\documentclass[fleqn,usenatbib]{mnras}

\usepackage{savesym}
\usepackage{amsmath}
\savesymbol{iint}
\savesymbol{iiint}
\savesymbol{iiiint}
\usepackage{txfonts}
\restoresymbol{TXF}{iint}
\restoresymbol{TXF}{iiint}
\restoresymbol{TXF}{iiiint}

\usepackage[T1]{fontenc}
\usepackage{ae,aecompl}

\usepackage{graphicx}	
\usepackage{amssymb}	
\usepackage{soul}
\usepackage{array}
\usepackage{mathptmx}
\usepackage{mathtools}
\usepackage[usenames]{color}
\usepackage{pdflscape}
\usepackage{hyperref}
\usepackage[all]{hypcap}
\hypersetup{colorlinks=true,citecolor=blue,linkcolor=purple,filecolor=black,runcolor=black,breaklinks=true}
\usepackage{etoolbox}
\makeatletter
\patchcmd\@combinedblfloats{\box\@outputbox}{\unvbox\@outputbox}{}{%
}%
 \makeatother
\DeclareMathAlphabet{\mathcal}{OMS}{cmsy}{m}{n}
\SetMathAlphabet{\mathcal}{bold}{OMS}{cmsy}{b}{n}

\newcommand{\comment}[1]{}
\definecolor{purple}{RGB}{160,32,240}

\newcommand{\Msun}{\;\mathrm{M}_{\odot}}

\newcommand{\vmax}{v_\mathrm{max}}

\newcommand{\vmp}{v_\mathrm{Mpeak}}

\renewcommand{\S}{Section}

\title[Measuring Halo Properties Beyond Mass]{Observational Measures of Halo Properties Beyond Mass}

\author[P. Behroozi et al.]{Peter Behroozi,$^{1}$\thanks{E-mail: behroozi@arizona.edu} Andrew Hearin,$^2$ Benjamin P.\ Moster$^3$
\\
$^{1}$ Department of Astronomy and Steward Observatory, University of Arizona, Tucson, AZ 85721, USA\\
$^{2}$ High-Energy Physics Division, Argonne National Laboratory, Argonne, IL 60439, USA\\
$^{3}$ Universit{\"a}ts-Sternwarte, Ludwig-Maximilians-Universit{\"a}t M{\"u}nchen, Scheinerstr. 1, 81679 M{\"u}nchen, Germany\\
}

\pubyear{2020}

\begin{document}
\label{firstpage}
\pagerange{\pageref{firstpage}--\pageref{lastpage}}
\maketitle

\begin{abstract}
Different properties of dark matter haloes, including growth rate, concentration, interaction history, and spin, correlate with environment in unique, scale-dependent ways.  While these halo properties are not directly observable, galaxies will inherit their host haloes' correlations with environment. In this paper, we show how these characteristic environmental signatures allow using measurements of galaxy environment to constrain {\em which} dark matter halo properties are most tightly connected to observable galaxy properties. We show that different halo properties beyond mass imprint distinct scale-dependent signatures in both the galaxy two-point correlation function and the distribution of distances to galaxies' $k$th nearest neighbours, with features strong enough to be accessible even with low-resolution (e.g., grism) spectroscopy at higher redshifts.  As an application, we compute observed two-point correlation functions for galaxies binned by half-mass radius at $z=0$ from the Sloan Digital Sky Survey, showing that classic galaxy size models (i.e., galaxy size being proportional to halo spin) as well as other recent proposals show significant tensions with observational data. We show that the agreement with observed clustering can be improved with a simple empirical model in which galaxy size correlates with halo growth.
\end{abstract}




\section{Introduction}

\label{s:intro}

In Lambda Cold Dark Matter ($\Lambda$CDM) cosmologies, dark matter dominates both the mass budget and gravitational potential.  Hence, galaxies form in the centres of virialized, overdense regions of dark matter (known as \textit{haloes}), and the properties of galaxies are expected to be strongly influenced by the properties of their host dark matter haloes \citep[see][for reviews]{Silk12,Somerville15}.  Multiple  types of observations provide evidence for a robust connection between galaxy mass and host halo mass; these include clustering, weak lensing, X-ray mass measurements, and satellite kinematics \citep[see][for a review]{Wechsler18}.  More recently, flexible forward models of the galaxy--halo connection (i.e., empirical models) have provided evidence based on galaxy clustering that galaxy growth rates correlate with host dark matter halo accretion rates (e.g., \citealt{Moster17,BWHC19}, cf.\ \citealt{ODonnell20}).  However, because dark matter haloes are not directly observable, it has so far proved difficult to test additional claimed links between halo and galaxy properties.

As an example, it has long been suspected that galaxy sizes relate to the angular momenta of their host dark matter haloes \citep{Fall80,Mo98}.  Gas and dark matter are evenly mixed throughout the Universe, so they are expected to have the same specific angular momenta at first infall into a halo.  Whereas the dark matter cannot radiate away energy, the gas can radiatively cool, with the result that it will continue shrinking in extent until it becomes a rotationally-supported disc at the centre of the halo.  With the classical assumption that gas angular momentum is conserved, one finds that
\begin{eqnarray}
    R_\mathrm{disc} \propto \lambda R_\mathrm{h},
\end{eqnarray}
where $R_\mathrm{disc}$ is the gas disc radius, $R_\mathrm{h}$ is the halo radius and $\lambda$ is the dimensionless spin parameter of the halo, e.g.,
\begin{eqnarray}
    \lambda_\mathrm{P} \equiv \frac{J_\mathrm{h}|E_\mathrm{h}|^{1/2}}{GM_\mathrm{h}^{5/2}},
\end{eqnarray}
where $\lambda_\mathrm{P}$ is the \cite{Peebles69} spin, $J_\mathrm{h}$ is the halo's angular momentum, $E_\mathrm{h}$ is the sum of the halo's potential and kinetic energies, and $M_\mathrm{h}$ is its mass.  As stars form in the densest regions of the gas disc \citep[e.g.,][]{Kennicutt98}, the stellar disc size ($R_\mathrm{stars}$) is also expected to be proportional to $\lambda R_\mathrm{h}$.

Often, the halo radius is defined in terms of the halo mass, e.g., $R_\mathrm{h} = \sqrt[3]{\frac{M_\mathrm{h}}{\frac{4}{3}\pi \rho_\mathrm{vir}}}$, where $\rho_\mathrm{vir}$ is the virial overdensity \citep{mvir_conv}.  Halo radii are hence available whenever halo masses are known, and so observations have shown that $R_\mathrm{stars}\propto R_\mathrm{h}$ \citep[e.g.,][]{Kravtsov13,Huang17,Somerville18}.  The same observations have suggested that the scatter in $\lambda$ is similar to the scatter in galaxy size at fixed halo radius (cf.\ \citealt{Desmond15}).  However, it has been difficult to show actual proportionality (i.e., $R_\mathrm{stars}\propto \lambda$) because of the difficulty in measuring halo spins.  At the same time, alternate models have suggested different galaxy size--halo connections, including \cite{Jiang19} (i.e., $R_\mathrm{stars}$ correlated with halo concentration but not halo spin) and \cite{Desmond17} (i.e., $R_\mathrm{stars}$ correlated only with stellar mass, and not with halo radius, concentration, or spin).  Clarifying the physics responsible for galaxy sizes is hence a strong motivation for developing observational probes of halo properties beyond mass and growth rates.

In this context, it has been known for a long time that halo properties correlate with the larger-scale environment \citep{Gao05,Wechsler06,Gao07,Hahn07,Hahn07b,Wang11}.  The correlation between halo clustering and properties beyond halo mass is known as assembly bias or \textit{secondary bias}, the term we use here \citep{Salcedo18,Mao18}.  Secondary bias exists for halo growth rates, assembly times, spins, and concentrations, among many other properties \citep[see previous references, as well as][]{Lazeyras17,Villarreal17,Salcedo18}.  Critically, secondary bias is \textit{scale-dependent} \citep{Gao05,Wechsler06,Sunayama16}, so the correlation between halo properties and clustering depends on the distance scale especially in the 1-2 Mpc regime.  While this has been presented as a problem---i.e., that no halo definition exists for which secondary bias vanishes on all scales \citep{Villarreal17}---here, we recast it as an opportunity.  Different halo properties have unique scale-dependent clustering \citep{Gao05,Wechsler06,Sunayama16}, and these clustering differences are inherited by galaxies.  As a result, galaxy clustering offers unique insights into host halo properties beyond halo mass.

Recent work on the galaxy-halo connection motivates considering summary statistics beyond two-point clustering \citep[e.g.,][]{wang_etal19,banerjee_abel20}. With correlation functions, galaxies in dense environments receive much higher relative weights compared to galaxies in sparse environments, because the latter contribute fewer pair counts.  Typical halo properties change in unique ways between median-density and low-density environments \citep[e.g.,][]{Lee16,Goh19}.  This suggests that sensitivity to low-density environments could provide additional halo property information beyond galaxy clustering.  

In this paper, we show that correlations between halo properties and galaxy properties would imprint strong, observable signatures in galaxy projected correlation functions that have unique stellar mass and redshift dependencies for each halo property considered.  We show that this is also the case for an alternate environment measure sensitive to a wide range of local densities (i.e., distance to the $k^\mathrm{th}$ nearest neighbour).  As a proof-of-concept, we show that spin-based and concentration-based galaxy size models both struggle to reproduce observed clustering from the Sloan Digital Sky Survey at $z\sim 0.1$, and provide a simple empirical model based on halo growth rates that yields a better match to observational data.

Section \ref{s:methods} details the dark matter simulation that we use and mock catalogue construction, and Section \ref{s:results} demonstrates that environmental measures (two-point correlation functions and $k^\mathrm{th}$ nearest neighbour distances) have unique scale-dependent clustering for different halo properties. Section \ref{s:sizes} compares these results with observed two-point correlation functions for galaxies split by size.  Sections \ref{s:discussion} and \ref{s:conclusions} offer discussion and conclusions, respectively.  Appendix \ref{a:alternate} presents scale-dependent clustering ratios for additional halo properties; Appendix \ref{a:obs_5nn} provides observed $5^\mathrm{th}$ nearest neighbour distributions;  Appendices \ref{a:discs} and \ref{a:rthen} provide tests of additional variations for spin-based size models; Appendix \ref{a:halo_growth} describes our empirical model for galaxy sizes; Appendix \ref{a:clustering} compares overall galaxy clustering (as opposed to clustering ratios) between our mock catalogues and observations; and Appendix \ref{a:orphans} provides a simple test of the effects of orphans on galaxy clustering ratios.  We assume a flat, $\Lambda$CDM cosmology ($h=0.68$, $\Omega_M=0.307$, $\Omega_\Lambda=0.693$, $n_s=0.96$, $\sigma_8 = 0.823$) that is consistent with \textit{Planck} constraints \citep{Planck18}.  Halo masses are defined using the virial spherical overdensity definition of \cite{mvir_conv}.  Stellar masses assume a \cite{Chabrier03} initial mass function (IMF).

\section{Correlations between Halo Properties and Environment}

\label{s:methods}

Here, we describe the dark matter simulation and halo finder used (Section \ref{s:dm_sim}), the approach for building a mock galaxy catalogue (Section \ref{s:mock}), and the approach for splitting this catalogue based on halo properties (Section \ref{s:hprops}).

\begin{figure*}
\vspace{-8ex}
\phantom{\hspace{-6ex}}\includegraphics[width=1.1\columnwidth]{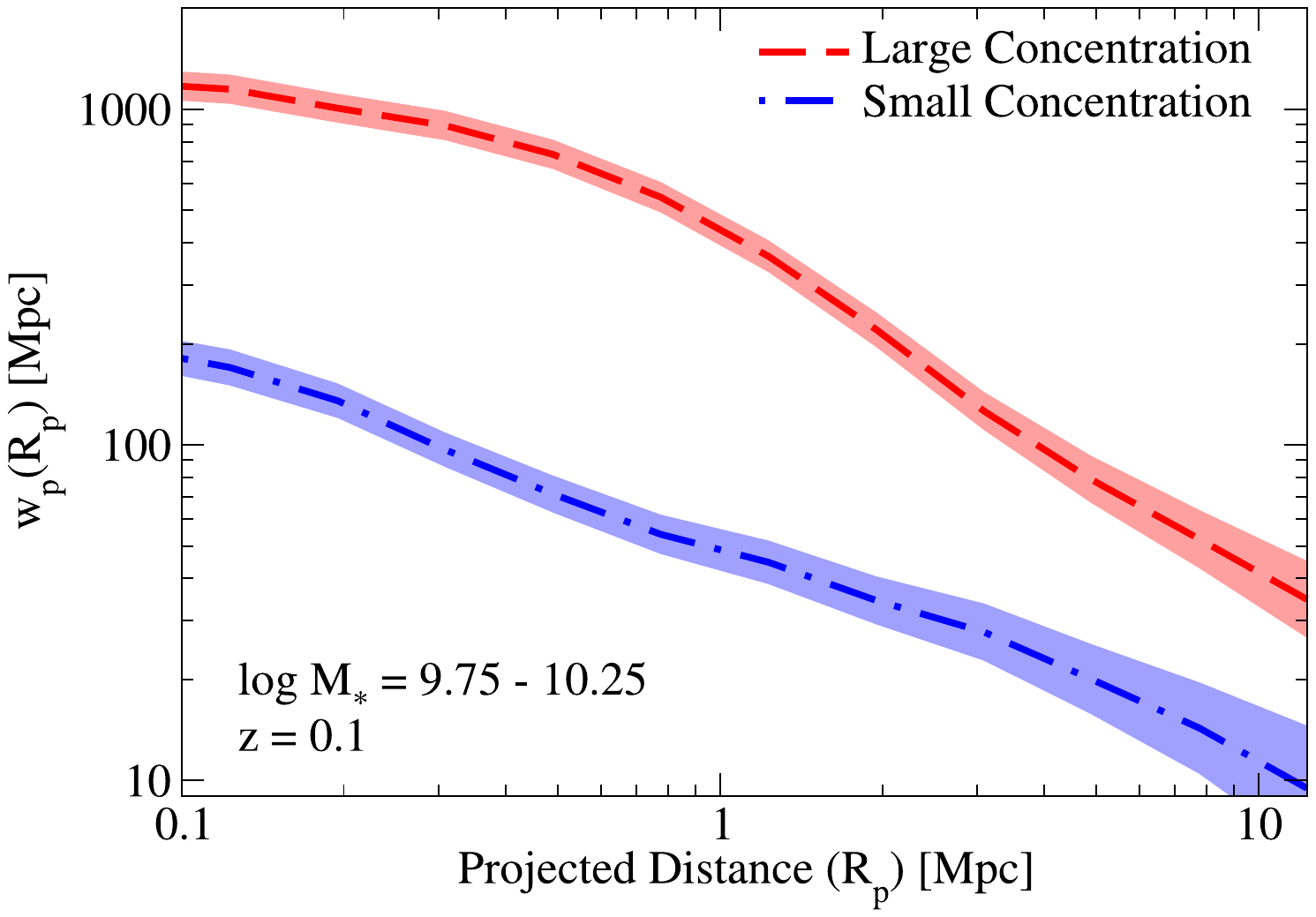}
\phantom{\hspace{-3ex}}\includegraphics[width=1.1\columnwidth]{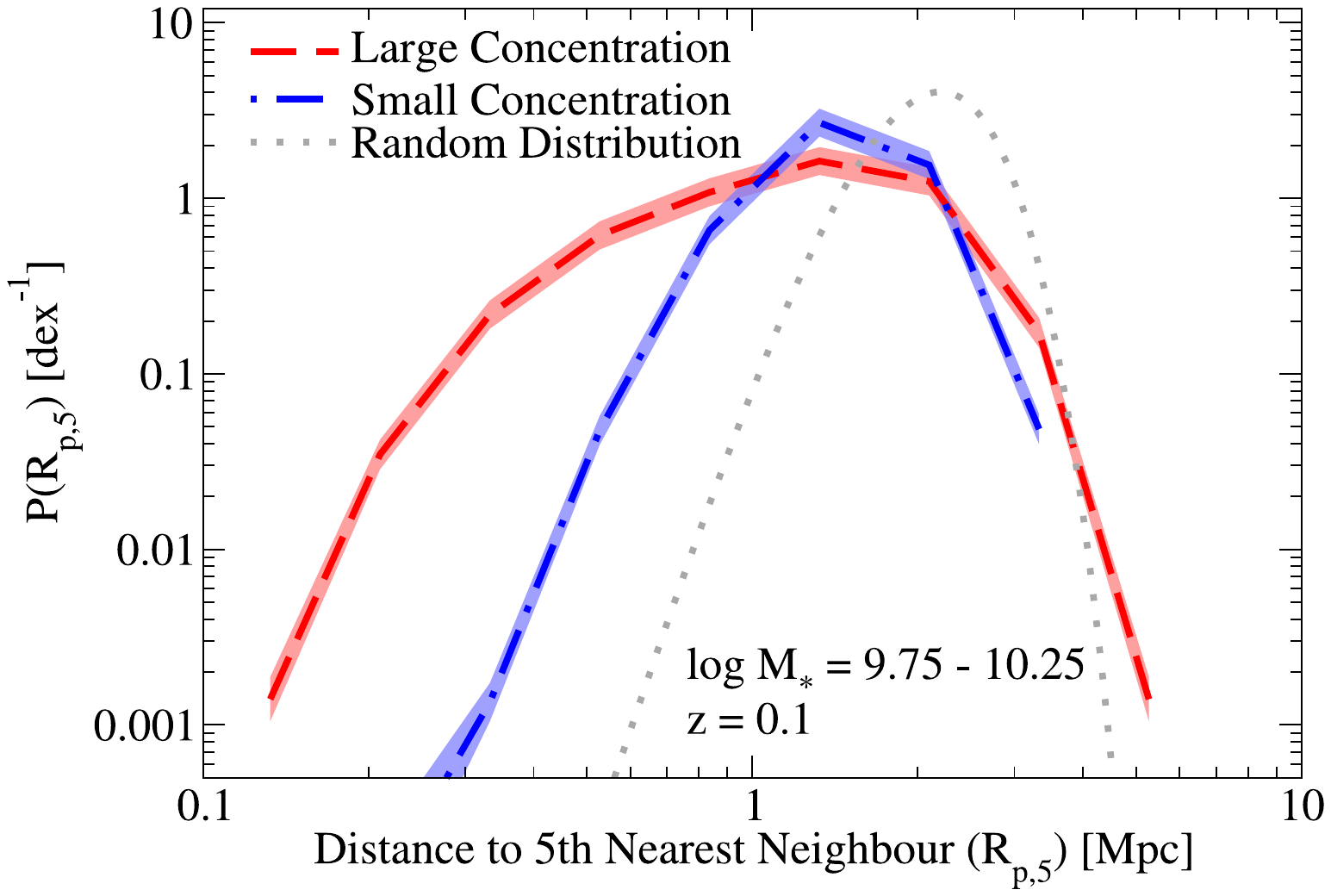}\\[-5ex]
\caption{High- and low-concentration haloes have dramatically different correlations with environment.  \textbf{Left} panel: Projected two-point correlation functions ($w_\mathrm{p}(R_\mathrm{p})$) for galaxies in our mock catalogue split by their halo concentrations.  Each sample has 50\% of the population, so ``Large Concentration'' refers to galaxies with above-median halo concentrations, and ``Small Concentration'' refers to galaxies with below-median halo concentrations.  \textbf{Right} panel: probability distributions of the projected distance to the 5$^\mathrm{th}$ nearest within-sample neighbour ($P(R_\mathrm{p,5})$) for galaxies split into the same samples as for the left panel.  For comparison, $P(R_\mathrm{p,5})$ for randomly-distributed points with the same number density as the galaxy samples is shown by the \textit{dotted line}.  \textbf{Both panels} are for galaxies with $10^{9.75} < M_\ast / \Msun < 10^{10.25}$ from an abundance-matched mock catalogue based on the \textit{SMDPL} dark matter simulation and the SDSS DR16 stellar mass function at $z=0.1$.  Shaded regions show $1\sigma$ jackknife errors added in quadrature to 10\% systematic errors (for correlation functions) or to 20\% systematic errors (for distance to the 5$^\mathrm{th}$ nearest neighbour).  Both panels count neighbouring galaxies within a maximum line-of-sight distance $\pi_\mathrm{max}=125$ Mpc $h^{-1}$.}
\label{f:examples}
\end{figure*}

\subsection{Dark Matter Simulation and Halo Finder}

\label{s:dm_sim}

Throughout this work, we use the public \textit{SMDPL}  simulation,\footnote{\url{https://www.cosmosim.org/cms/simulations/smdpl/}} which follows a periodic comoving cube of side length $400$ Mpc $h^{-1}$ from $z=120$ to $z=0$ with 3840$^3$ particles.  This gives \textit{SMDPL} both high particle mass resolution (1.4$\times 10^8 \Msun$) and force resolution (1.5 $h^{-1}$ kpc), allowing it to resolve $\sim 10^{10} \Msun$ haloes. \textit{SMDPL} was run with the \textsc{GADGET}-2 code \citep{Springel05}, with a flat, $\Lambda$CDM cosmology ($h=0.68$, $\Omega_M=0.307$, $\Omega_\Lambda=0.693$, $n_s=0.96$, $\sigma_8 = 0.823$).

Haloes were identified using the \textsc{Rockstar} phase-space halo finder \citep{Rockstar}.  A key advantage of \textsc{Rockstar} is its well-tested ability to accurately recover properties of satellite haloes regardless of distance to their host haloes' centres \citep{Knebe11,Onions12,Onions13,BehrooziNotts}.  This is beneficial for this work, as it implies that the measured dependencies of halo properties on environment are due to actual dark matter dynamics in the simulation, as opposed to biases from the halo finder.  Merger trees were constructed using the \textsc{Consistent Trees} code \citep{BehrooziTree}.

\subsection{Mock Galaxy Catalogues}

\label{s:mock}

To demonstrate that the environmental dependence of halo properties is observable, we must first build a mock galaxy catalogue.  To assign stellar masses to haloes in \textit{SMDPL}, we use abundance matching.  With no scatter, this process would assign the most-massive galaxies in rank order to the most-massive haloes in rank order within the same volume.  However, we here assume a log-normal scatter of 0.22 dex between stellar mass and our halo mass proxy, with is consistent with two-point correlation functions and other observables \citep{Reddick12}.  We implement scatter using the deconvolutional approach of \cite{Behroozi10}.  For our halo mass proxy, we use $\vmp$  \citep{BWHC19}, i.e., $\vmax$ ($\equiv\max\left[\sqrt{GM(<R)/R}\right]$) at the redshift of peak halo mass.  Notably, $\vmp$ is immune to the pseudo-evolution that affects halo masses \citep[e.g.,][]{Diemer13} and is also less sensitive to transient $\vmax$ spikes during mergers \citep[e.g.,][]{BehrooziMergers}.

At $z=0.1$, we abundance-match the stellar mass function from the Sloan Digital Sky Survey (SDSS) Data Release (DR) 16 \citep{DR16} to haloes in \textit{SMDPL}.  Specifically, we use redshifts from the DR16 catalogue and stellar masses from \citet{Brinchmann04}, which were based on DR7 photometry.  See \citet{BehrooziMM} for details on the stellar mass function calculation at this redshift.  At $z=1$ and $z=2.5$, we use the observed stellar mass functions from the best-fitting model of the \textsc{UniverseMachine}, which in effect interpolates between the relevant redshift bins in the stellar mass functions from PRIMUS, ULTRAVISTA, and ZFOURGE \citep{Moustakas13,Muzzin13,Ilbert13,Tomczak15}.

\begin{figure*}
\begin{center}
\vspace{-8ex}
\phantom{\hspace{-5ex}}\includegraphics[width=1.6\columnwidth]{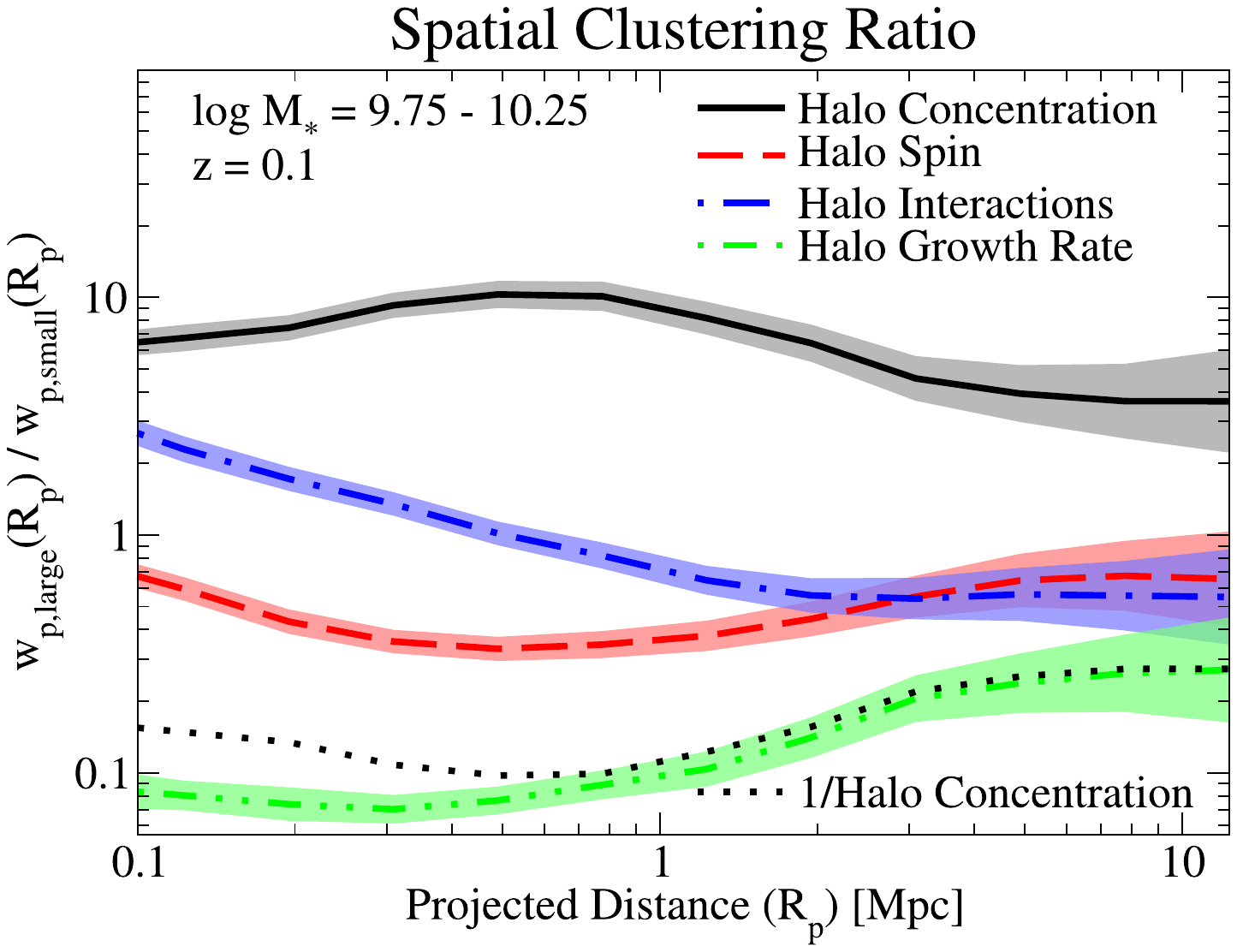}\\[-6ex]
\includegraphics[width=\columnwidth]{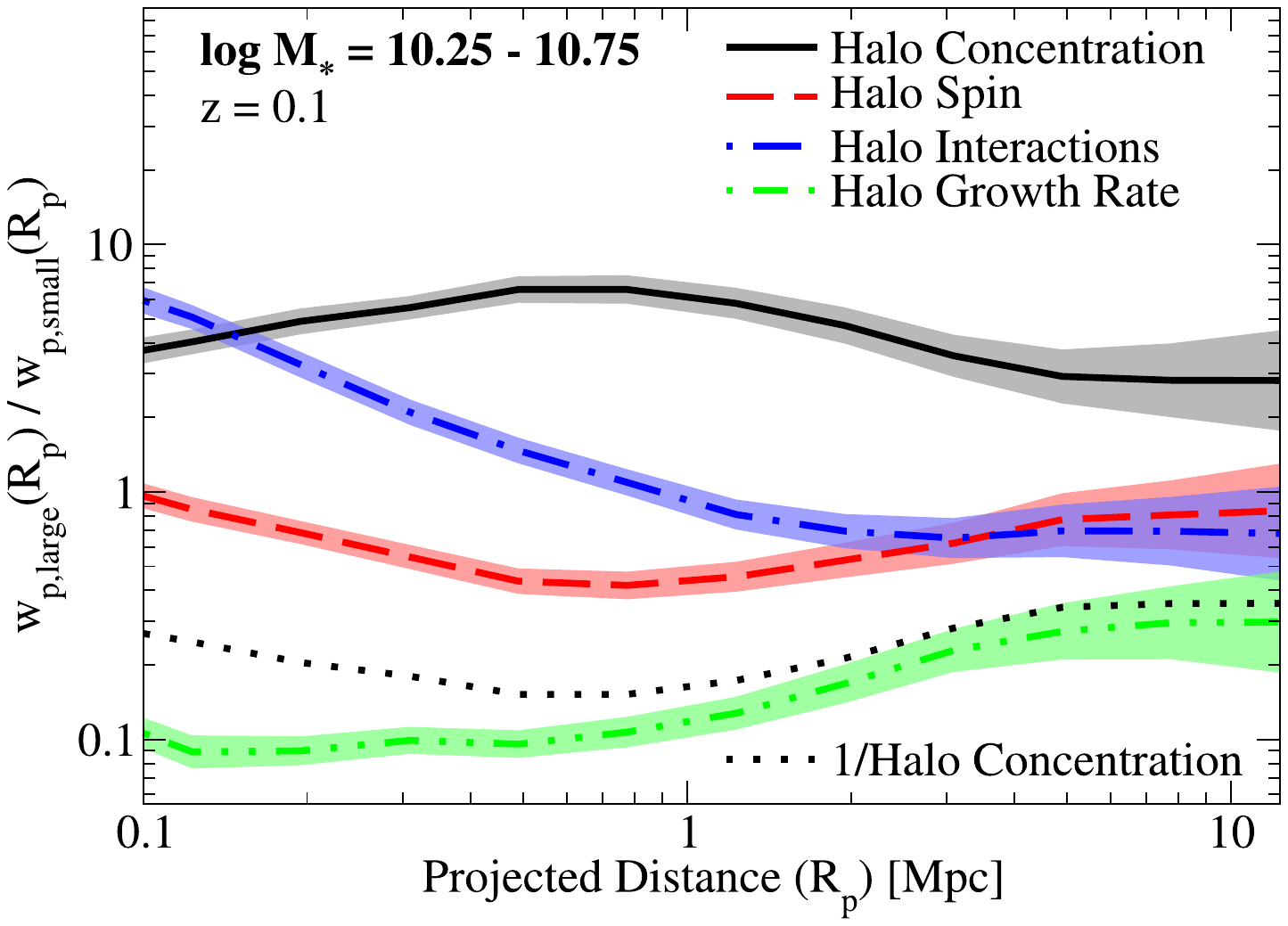}\hspace{-3ex}\includegraphics[width=\columnwidth]{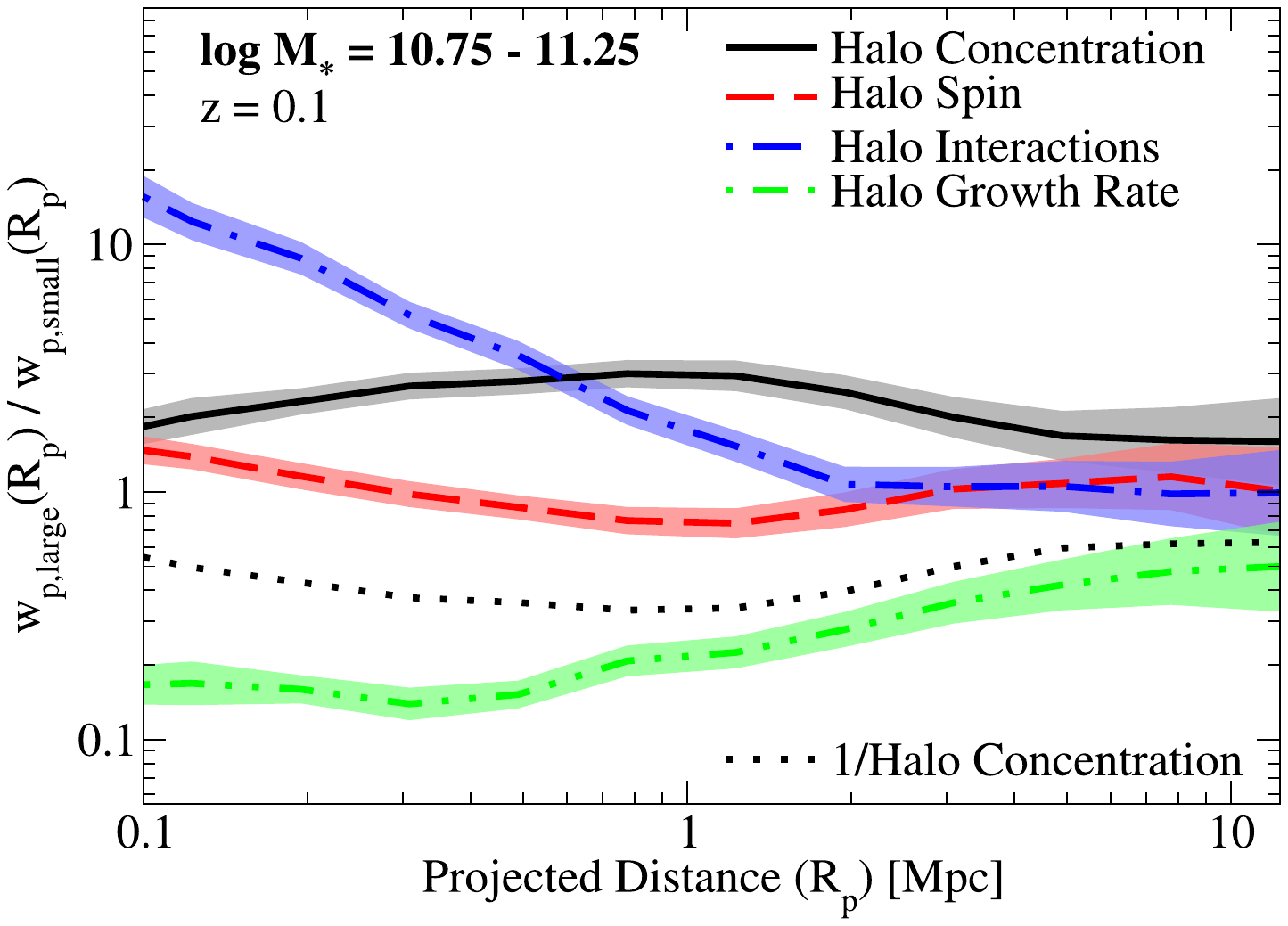}\\[-5ex]
\includegraphics[width=\columnwidth]{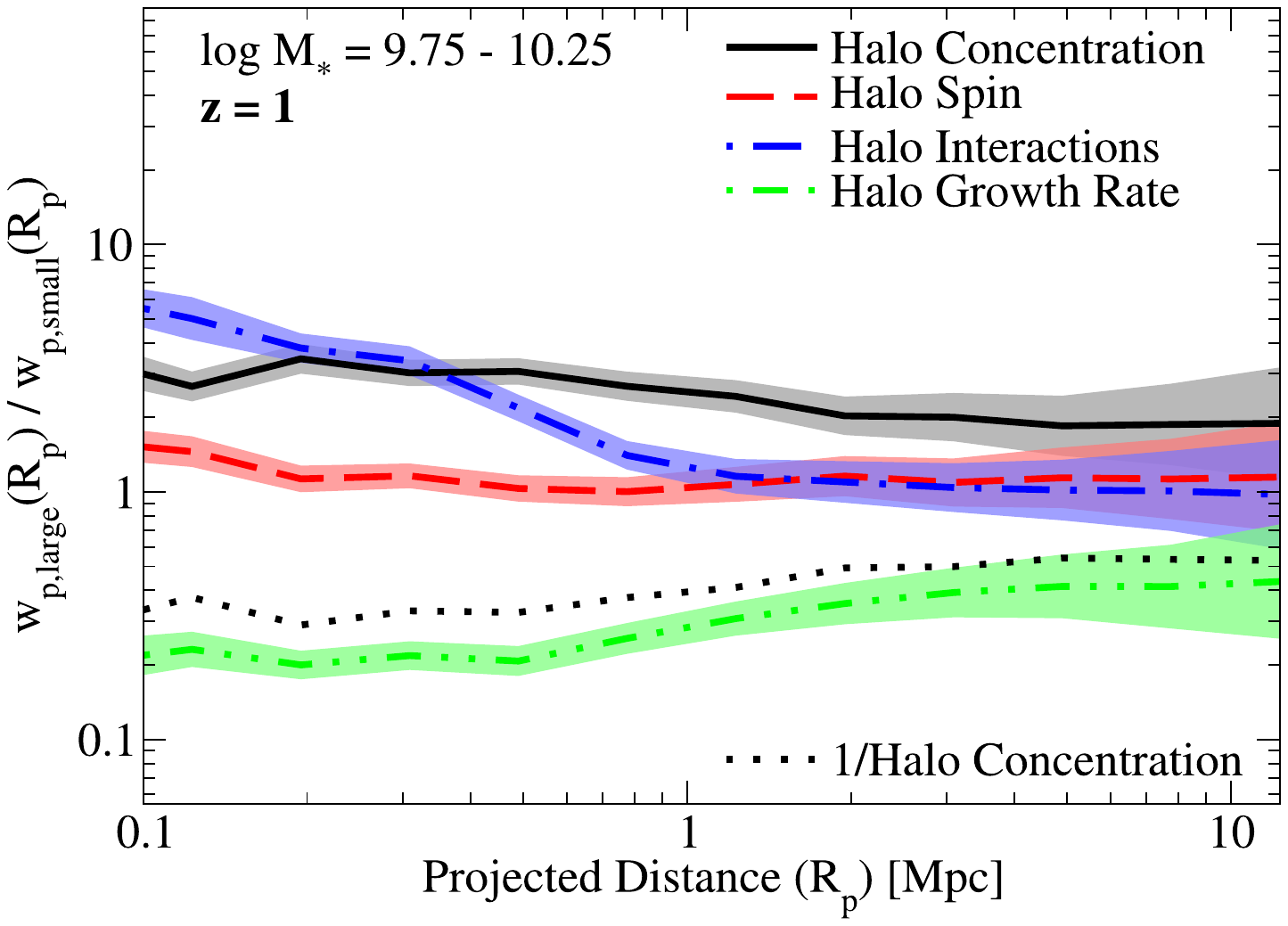}\hspace{-3ex}\includegraphics[width=\columnwidth]{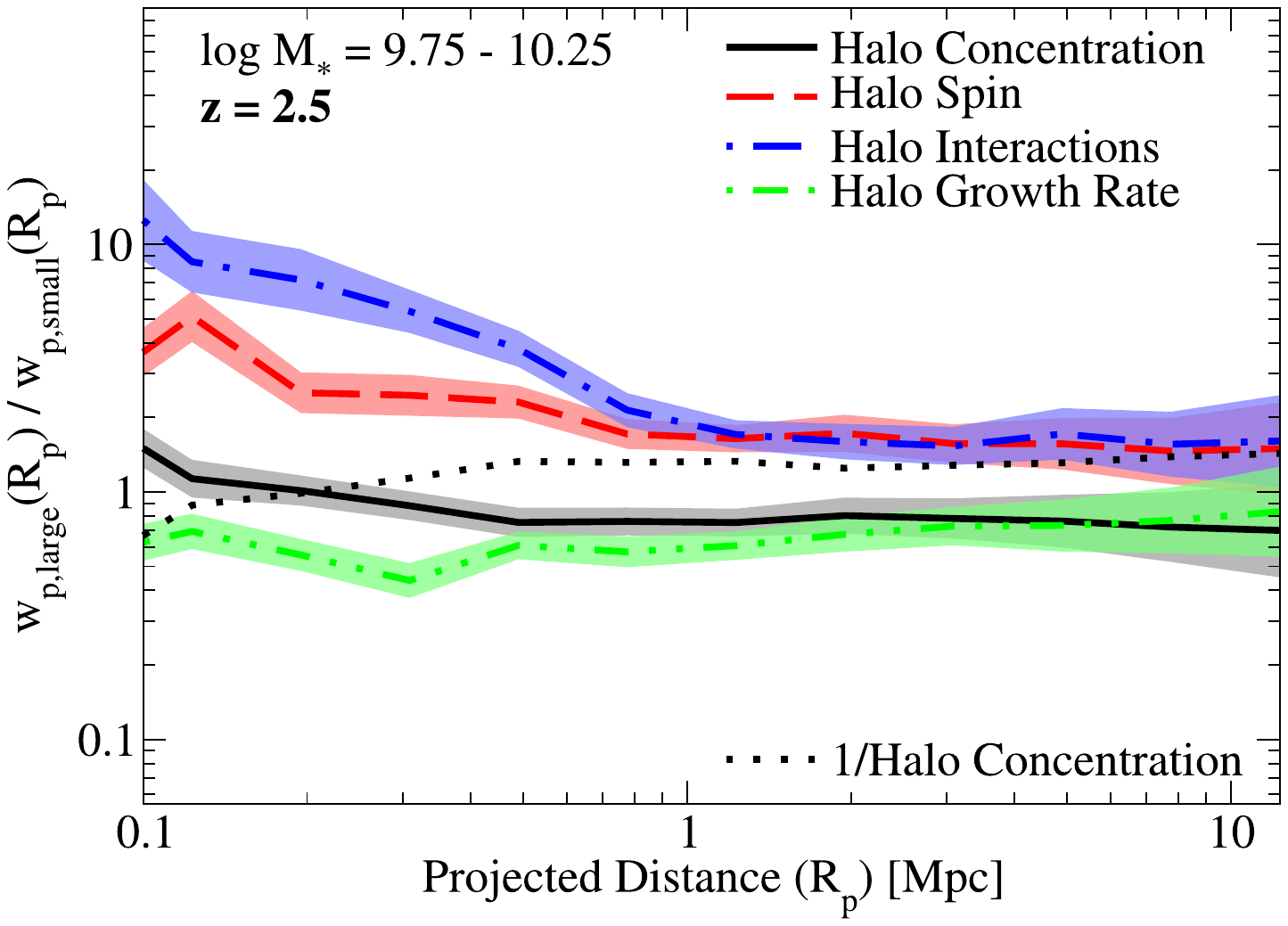}\\[-5ex]
\end{center}
\caption{Ratios of projected two-point correlation functions for galaxies split into two equal bins: 1) ``large,'' with above-median halo properties, and 2) ``small,'' with below-median halo properties.  Halo properties considered here include halo concentration, \citet{Bullock01} spin, ratio of kinetic to potential energy (\textit{Halo Interactions}), and average accretion rate over the past dynamical time (\textit{Halo Growth Rate}).  Since halo concentration and growth rate are anti-correlated, the \textit{dotted line} shows the reciprocal of the clustering ratio for halo concentration for ease of comparison.  In all cases shown, the scale-dependent clustering ratios are unique for each halo property.  The \textbf{top panel} shows simulated clustering ratios for galaxies with $10^{9.75} <  M_\ast / \Msun < 10^{10.25}$ from an abundance-matched catalogue at $z=0.1$ (Section \ref{s:mock}).  The \textbf{middle panels} show clustering at $z=0.1$ for higher-mass galaxies.  The \textbf{bottom panels} show clustering at $z=1$ and $z=2.5$ for galaxies with $10^{9.75} <  M_\ast / \Msun < 10^{10.25}$.  In all panels, shaded regions correspond to $1\sigma$ jackknife uncertainties convolved with 10\% assumed systematic errors; pairs are counted to a maximum line-of-sight distance of $\pi_\mathrm{max}=125$ Mpc/h.}
\label{f:wp_ratios}
\end{figure*}

\subsection{Splitting Galaxies Based on Host Halo Properties Beyond Mass}

\label{s:hprops}

In this section, we partition galaxies in the mock catalogue (Section \ref{s:mock}) into two equal bins for each of the following secondary halo properties:
\begin{enumerate}
    \item Halo growth rate, as measured by the average rate of change in halo mass over the past dynamical time ($\tau_\mathrm{dyn} \equiv (G \rho_\mathrm{vir})^{-1/2}$).
    \item Halo concentration, as measured by fitting a \cite{NFW97} density profile to the halo.
    \item Halo interaction history, as quantified by the ratio of kinetic to potential energy ($T/|U|$) in the halo's bound particles.
    \item Halo spin, using the \cite{Bullock01} definition ($\lambda_\mathrm{B}\equiv J/\sqrt{2GM^3R}$).
\end{enumerate}
The above properties were chosen primarily because of their relevance for models of galaxy size and morphology.  We test additional halo properties in Appendix \ref{a:alternate}, including half-mass scale factor ($a_\mathrm{1/2}$), scale factor of last major merger ($a_\mathrm{lmm}$), \cite{Peebles69} spin ($\lambda_P$), and halo shape ($\frac{c}{a}$).  As expected, similar measures have similar qualitative scale-dependent clustering ratios.

Observationally, one would like to be able to measure how galaxy properties (e.g., galaxy size) correlate with halo properties.  Typically, galaxy properties vary with stellar mass (and therefore host halo mass).  To remove this dependence for a given galaxy property, a common method is to measure the median of that property as a function of stellar mass, and then to split the sample into two bins: 1) galaxies with above-median properties, and 2) galaxies with below-median properties.  For example, \cite{Hearin19} split galaxies into those with above-median and below-median sizes to compare their relative clustering.

Here, we mimic that process.  We first divide haloes into narrow (0.1 dex) bins of $M_\ast$ and calculate the median of each of the above secondary halo properties (growth rate, concentration, interaction history, and spin).  For each of these secondary halo properties, we split galaxies into two groups, each containing half of the total population: 1) galaxies whose haloes have above-median values of the property, and 2) galaxies whose haloes have below-median values of the property.  Satellite galaxies are partitioned according to the properties of their enclosing subhalo.  

In the following sections, we compute the ratios of 2PCFs and $k$NN distributions for galaxies split by the secondary halo properties above.  This would correspond most straightforwardly to models where a galaxy property would correlate perfectly with a single halo property.  For example, in the classic galaxy size model, above- vs.\ below-median galaxy sizes would correlate perfectly with above- vs.\ below-median galaxy spins.

Nonetheless, the measures of environment for above-median vs.\ below-median haloes have a much broader application: they can be thought of as \textit{basis vectors} for more complicated models in which galaxy properties have imperfect correlations with multiple halo properties.  As we show, the shape of each halo property's environmental dependence is distinct from that of other halo properties.  As a result, one may jointly constrain the multivariate correlation of any galaxy property to multiple halo properties, because a unique set of correlations with halo properties corresponds to a unique predicted environmental dependence.

\begin{figure}
    \centering
\vspace{-8ex}
\phantom{\hspace{-5ex}}\includegraphics[width=1.1\columnwidth]{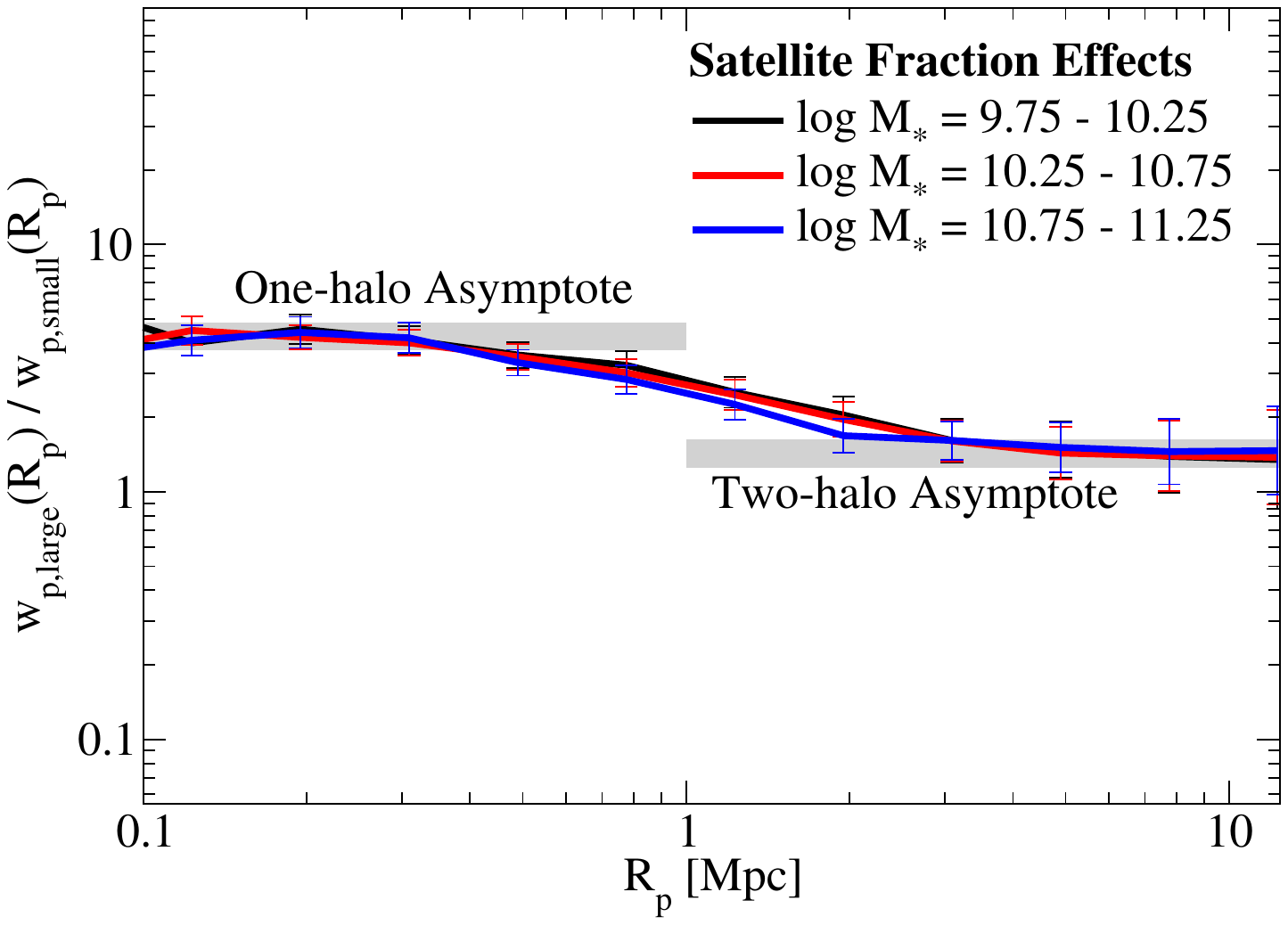}\\[-7ex]
\phantom{\hspace{-5ex}}\includegraphics[width=1.1\columnwidth]{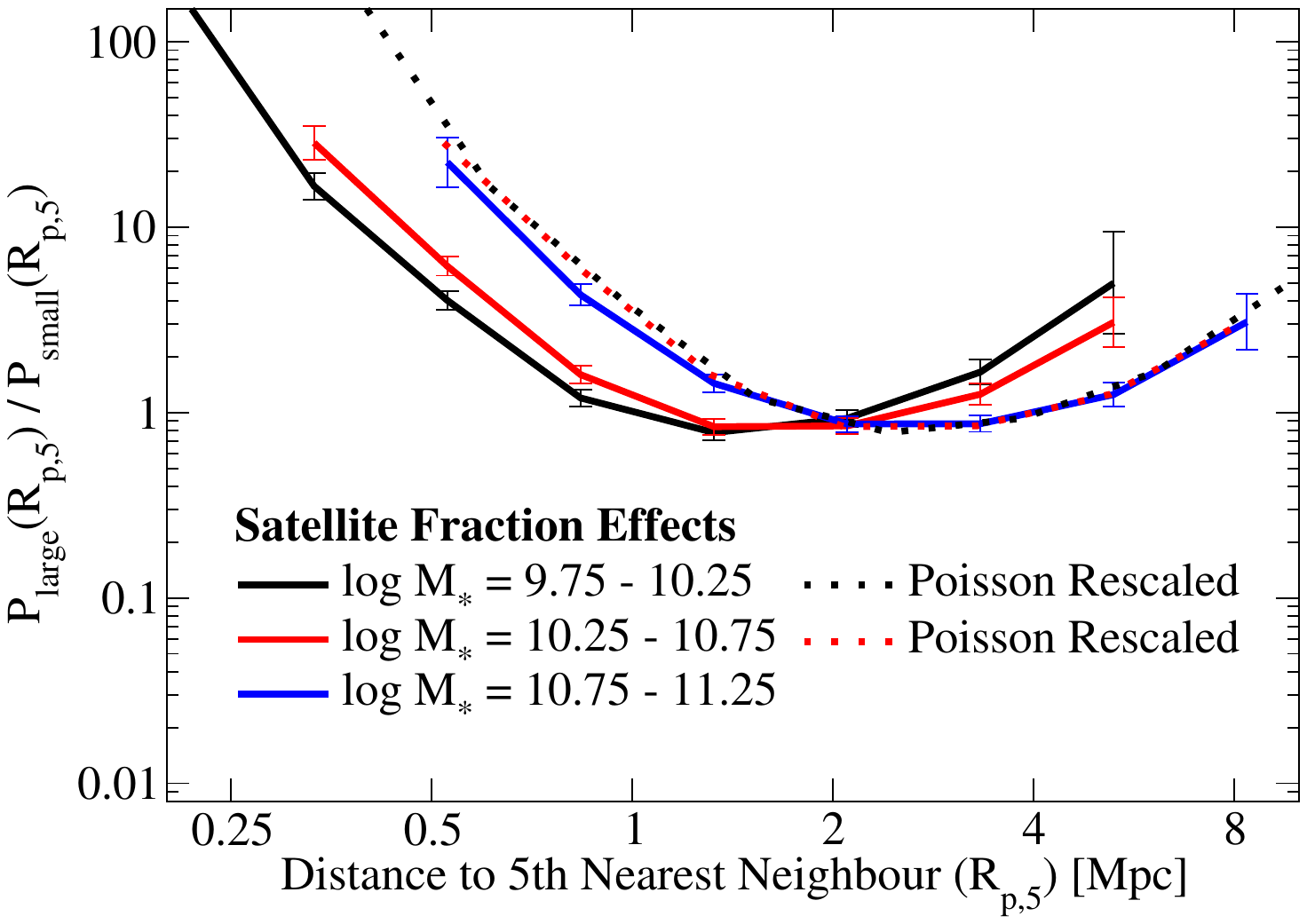}\\[-4.2ex]
\caption{\textbf{Top panel}: The effect on clustering ratios for a toy model with only satellite fraction differences and no central halo assembly/secondary bias.  Specifically, this figure shows ratios of projected two-point halo correlation functions at $z=0.1$, where both central and satellite haloes are randomly assigned labels of ``large'' or ``small,'' but satellites have a larger chance of being assigned a ``large'' label than centrals (as shown in Fig.\ \ref{f:sat_fracs}; see Section \ref{s:phys_overview} for details).  These clustering ratios all asymptote to a fixed value on one-halo scales (below 1 Mpc) and to a different fixed value on two-halo scales (above 2 Mpc), with a smooth, monotonic transition between them.  Of note, satellite fractions affect clustering on two-halo scales as well as one-halo scales, since haloes that host satellite galaxies are more massive (and more biased) than those that host only central galaxies.  \textbf{Bottom panel}: Ratios of probability distributions of distances to 5$^\mathrm{th}$ nearest within-sample neighbours (5NN PDFs) for galaxies in the same toy model.  The shapes of these ratios have little mass-dependence, as shown by the ``Poisson Rescaled'' versions of the two smaller mass bins.  For these rescaled curves, distances (i.e., the $x$-axes) have been multiplied by a constant factor to correct for sample number density differences (see Section  \ref{s:knn_physical_overview}).  As in Fig.\ \ref{f:wp_ratios}, error bars correspond to $1\sigma$ jackknife uncertainties convolved with 10\% assumed systematic errors for correlation functions and 20\% assumed errors for 5NN PDFs; both panels assume $\pi_\mathrm{max}=125$ Mpc/h.}
\label{f:sat_ratios}
\end{figure}

\begin{figure}
    \centering
\vspace{-8ex}
\phantom{\hspace{-5ex}}\includegraphics[width=1.1\columnwidth]{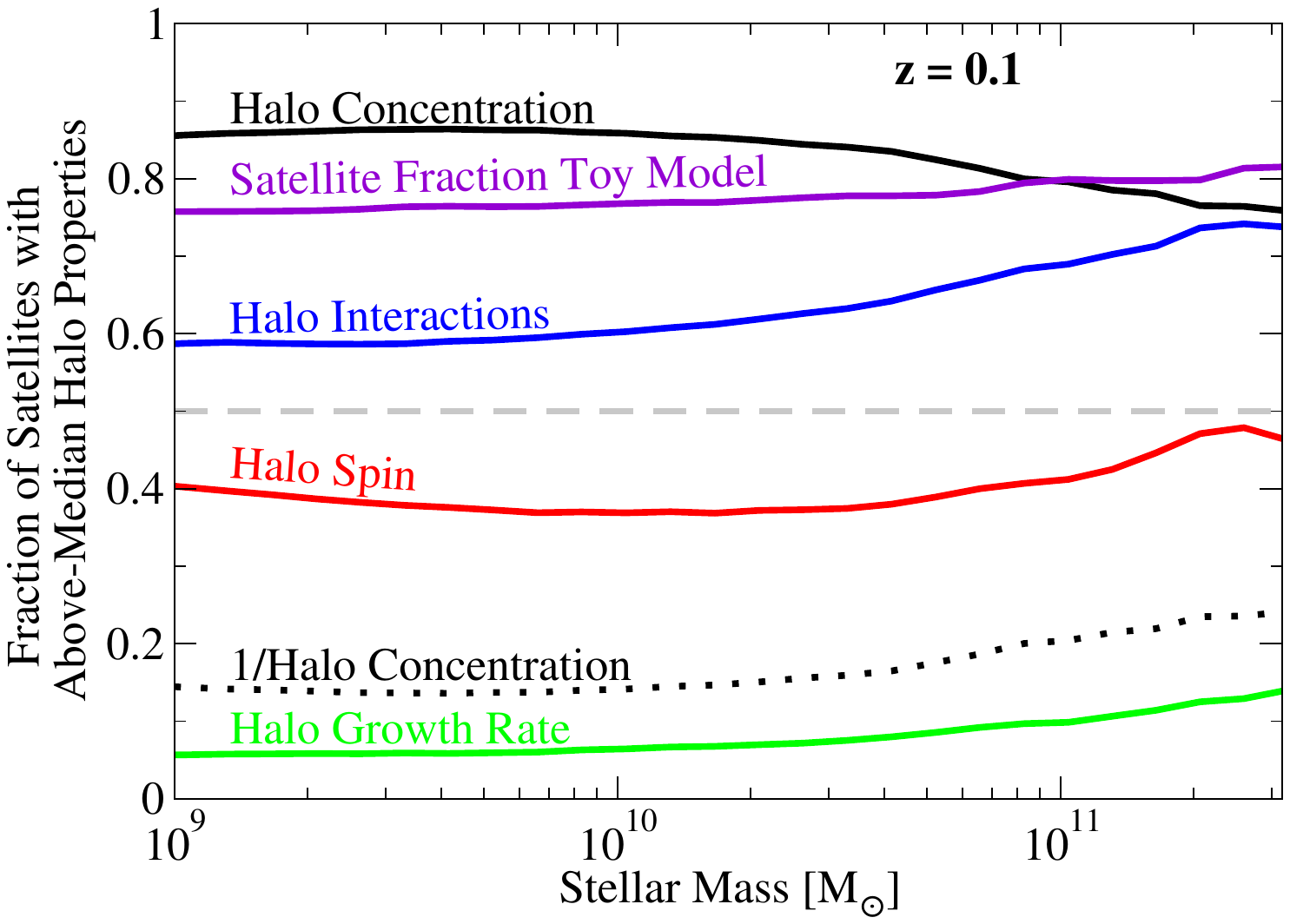}\\[-7ex]
\phantom{\hspace{-5ex}}\includegraphics[width=1.1\columnwidth]{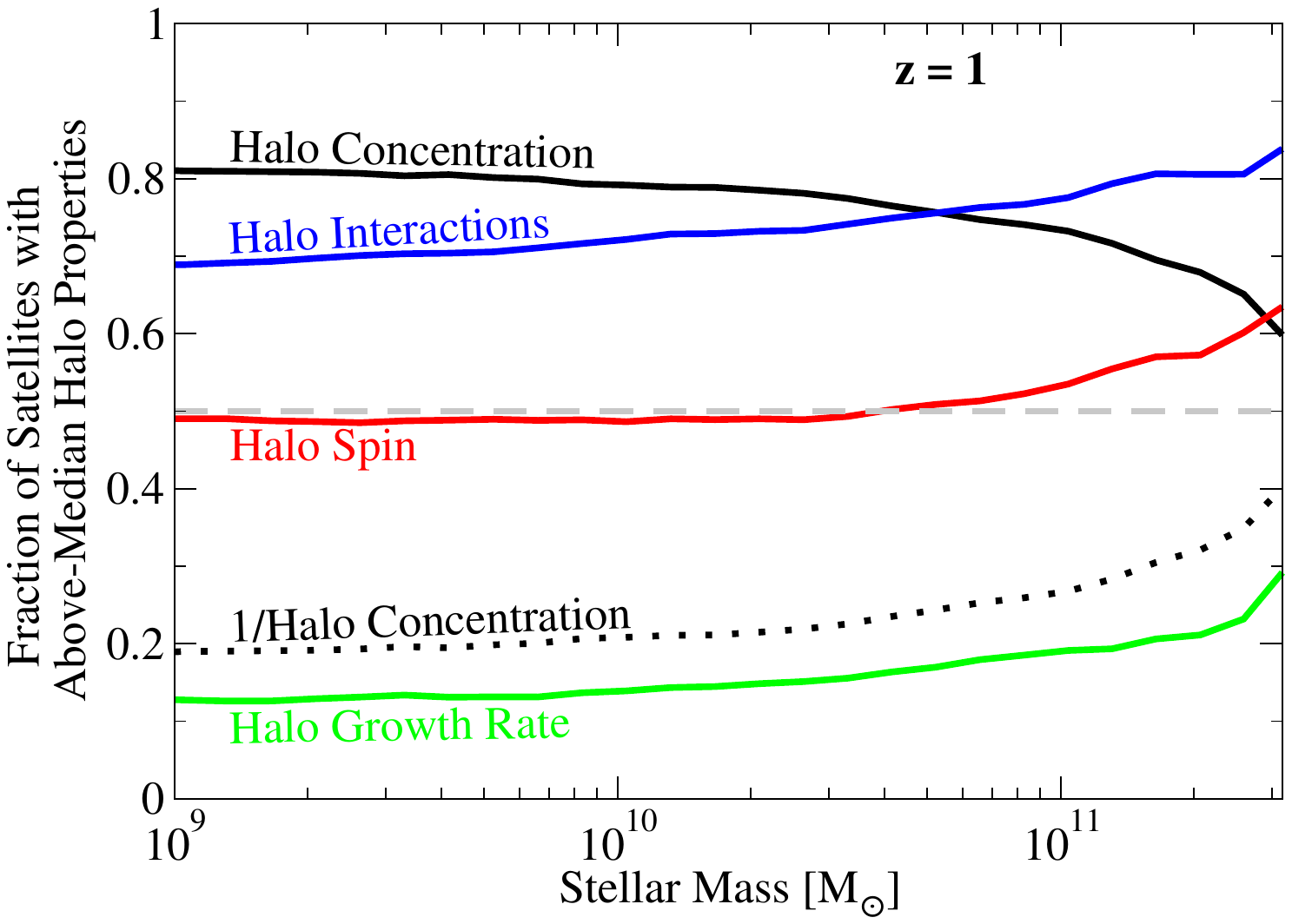}\\[-7ex]
\phantom{\hspace{-5ex}}\includegraphics[width=1.1\columnwidth]{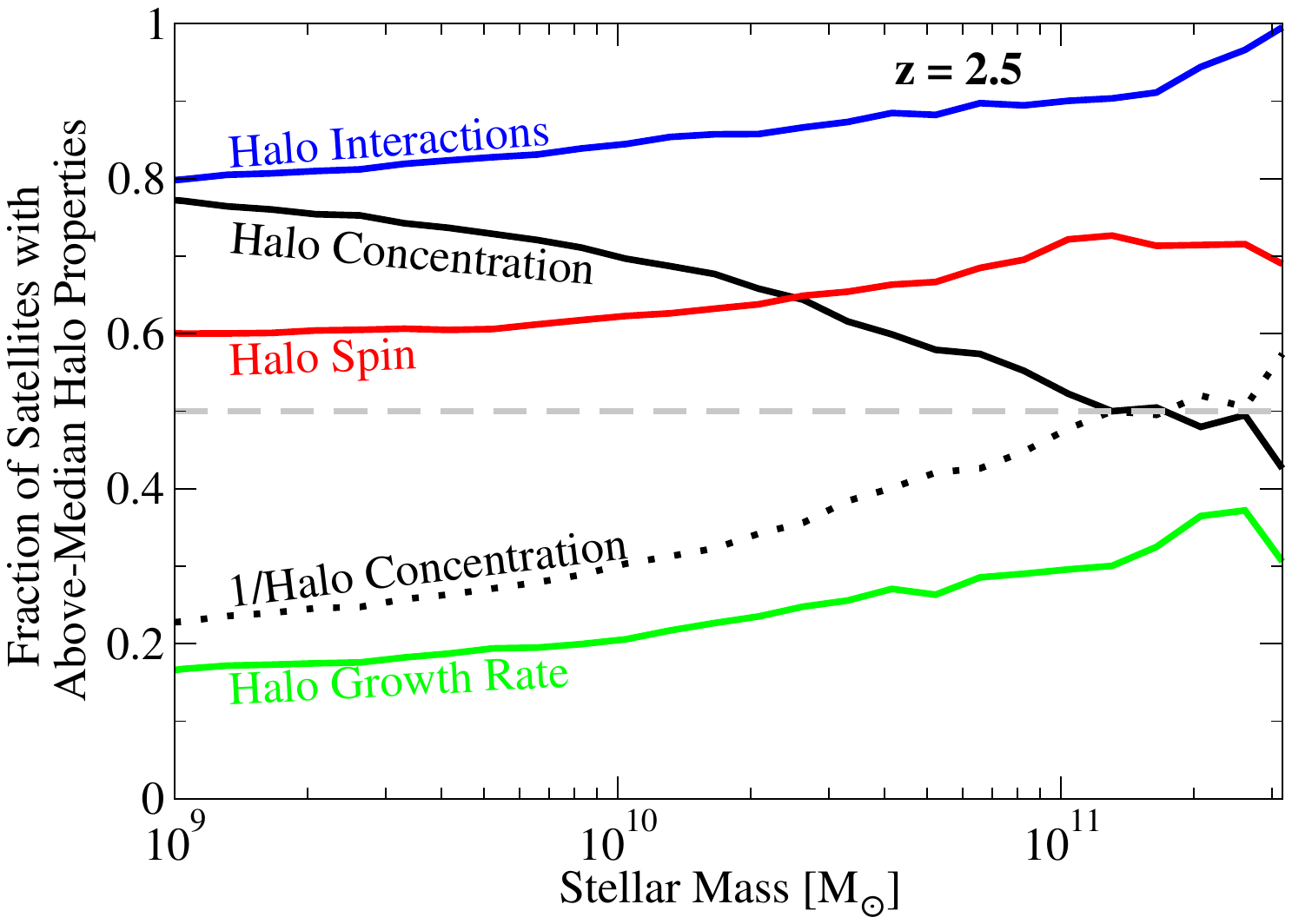}\\[-4ex]
\caption{The fraction of satellite galaxies in our mock catalogues (Section \ref{s:mock}) whose halo properties have above-median values, as a function of redshift and stellar mass.  A higher value for this fraction (in absence of differences in central halo bias) would result in higher ratios for $w_\mathrm{p,large}/w_\mathrm{p,small}$ on both one-halo and two-halo scales, as in Fig.\ \ref{f:sat_ratios}.}
\label{f:sat_fracs}
\end{figure}

\begin{figure*}
\begin{center}
\vspace{-8ex}
\phantom{\hspace{-5ex}}\includegraphics[width=1.6\columnwidth]{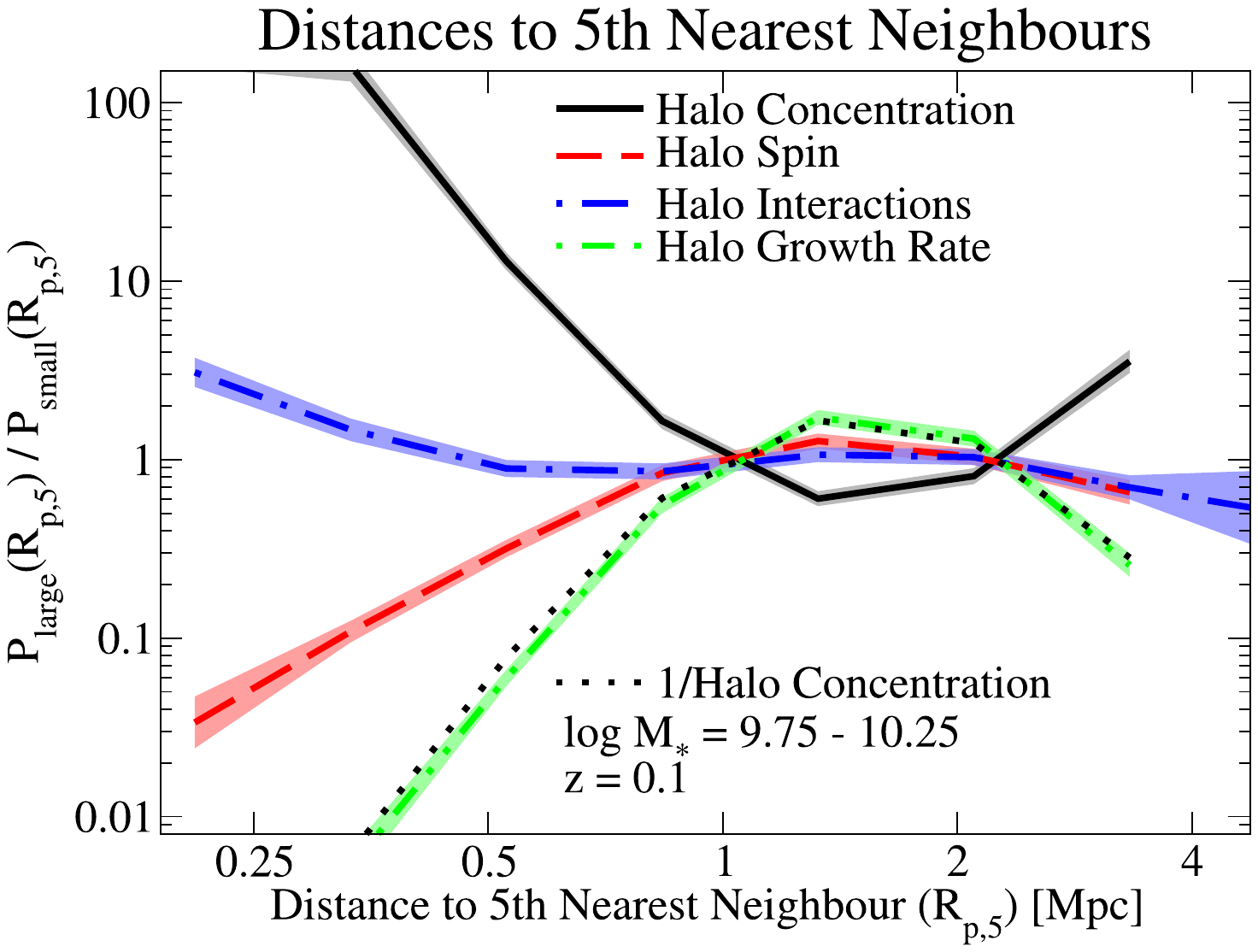}\\[-6ex]
\includegraphics[width=\columnwidth]{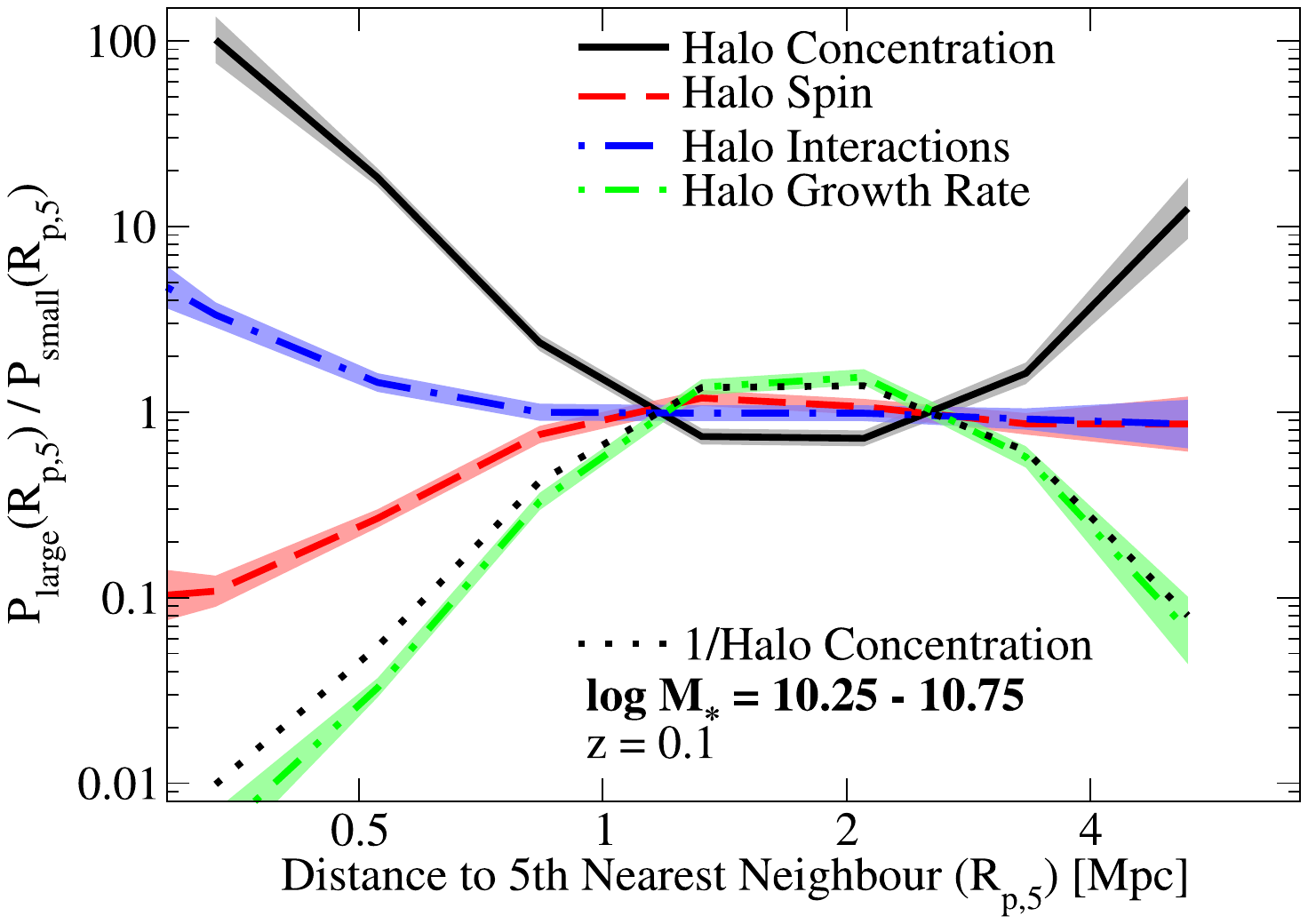}\hspace{-3ex}\includegraphics[width=\columnwidth]{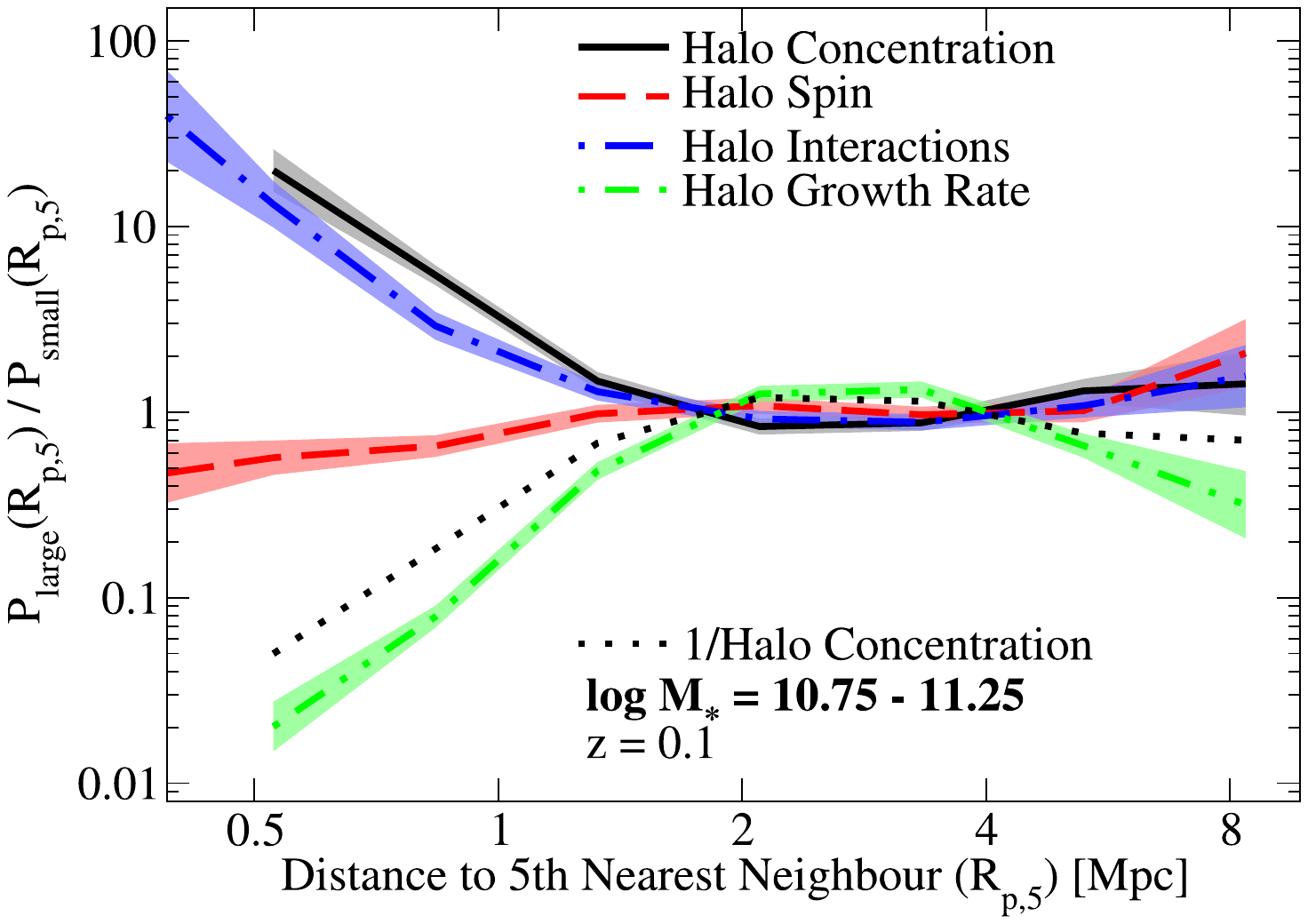}\\[-5ex]
\includegraphics[width=\columnwidth]{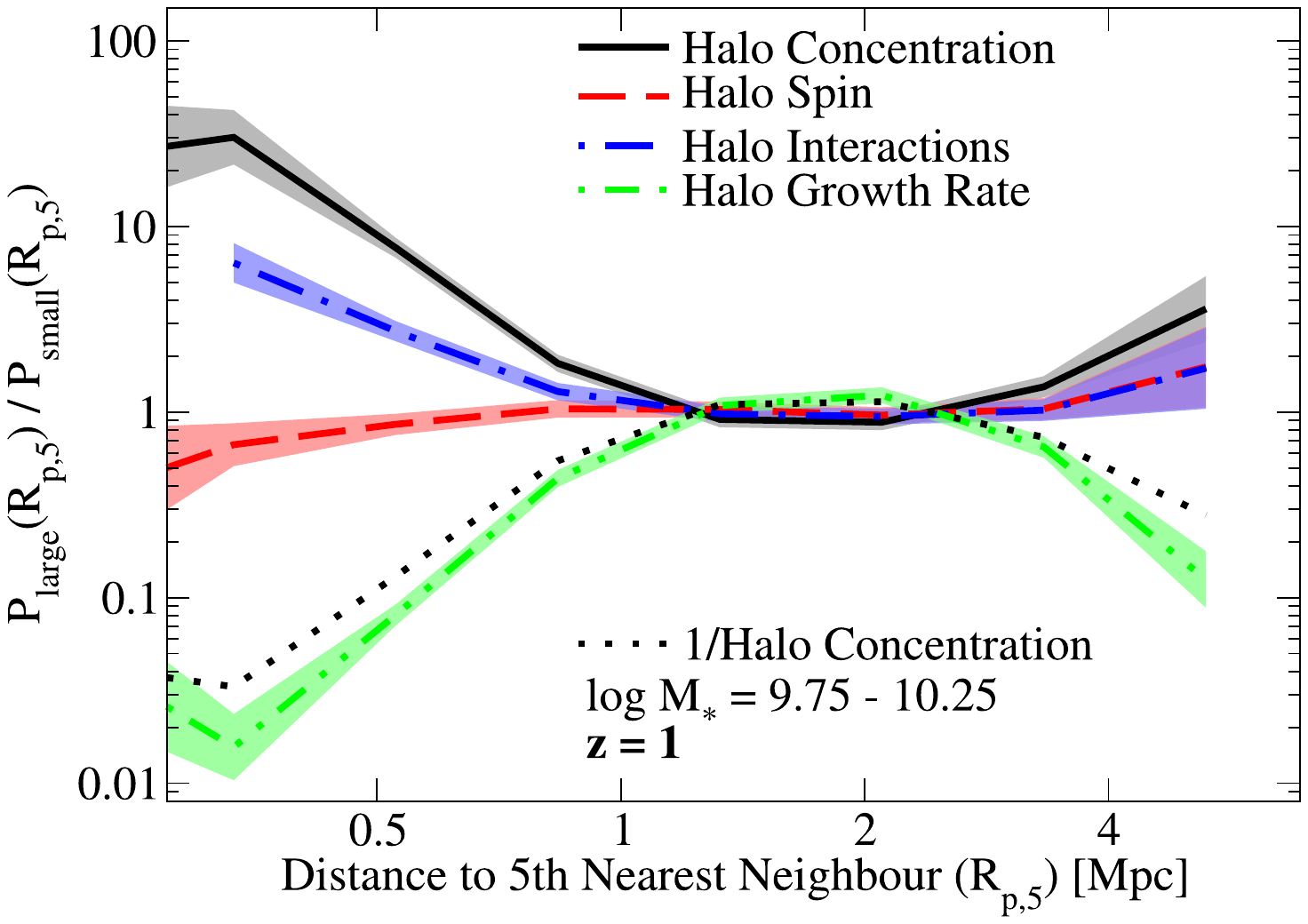}\hspace{-3ex}\includegraphics[width=\columnwidth]{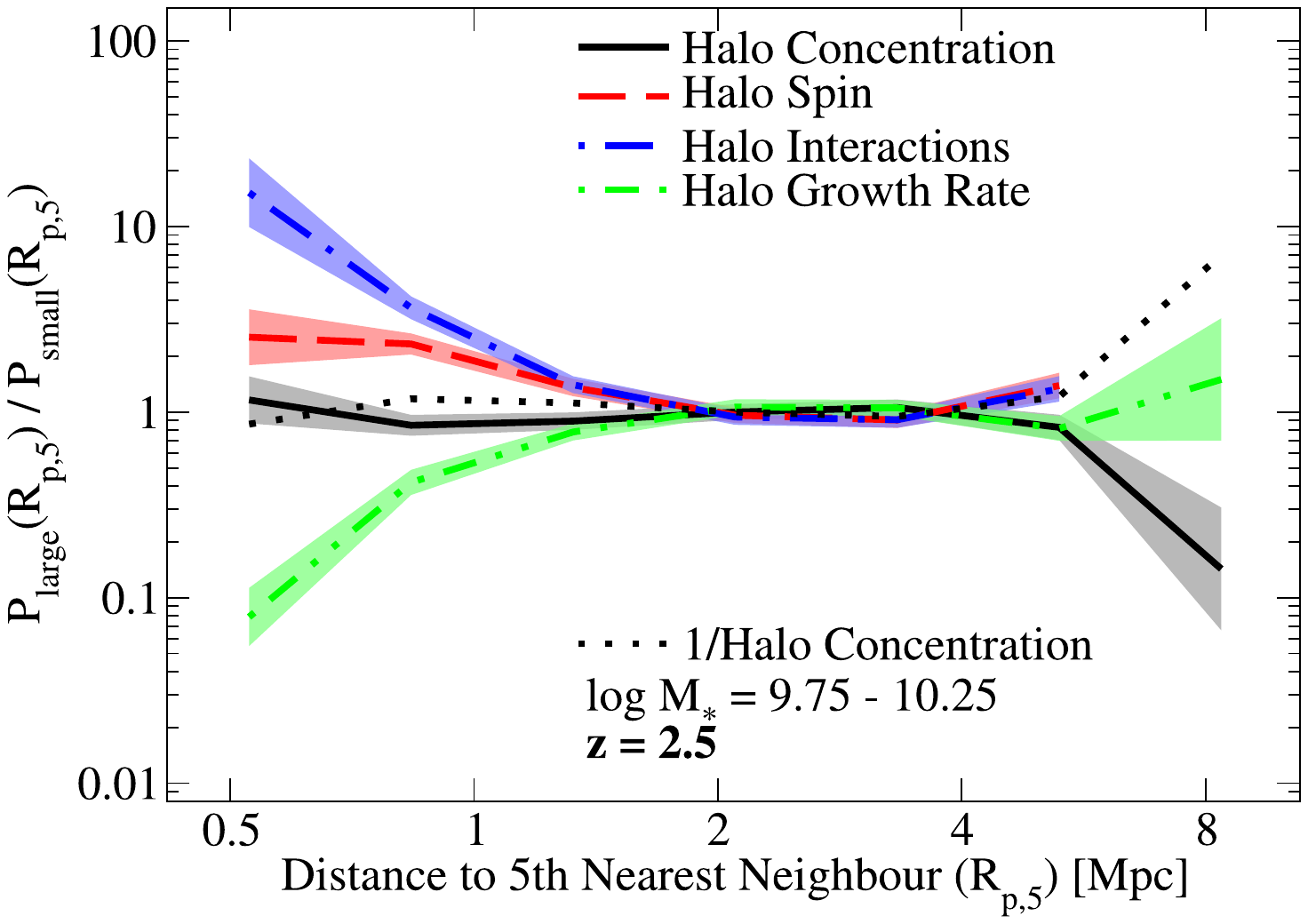}\\[-5ex]
\end{center}
\caption{Ratios of probability distributions of distances to 5$^\mathrm{th}$ nearest within-sample neighbours (5NN PDFs) for galaxies split by their halo properties.  As in Fig.\ \ref{f:wp_ratios}, galaxies are split into two equal bins: 1) ``large,'' with above-median halo properties, and 2) ``small,'' with below-median halo properties.  Halo properties considered here include halo concentration, \citet{Bullock01} spin, ratio of kinetic to potential energy (\textit{Halo Interactions}), and average accretion rate over the past dynamical time (\textit{Halo Growth Rate}).  Since halo concentration and growth rate are anti-correlated, the \textit{dotted line} shows the reciprocal of the probability ratio for halo concentration for ease of comparison.  In all cases shown, the scale-dependencies are unique for each halo property.  The \textbf{top panel} shows simulated 5NN PDF ratios for galaxies with $10^{9.75} <  M_\ast/\Msun < 10^{10.25}$ from an abundance-matched catalogue at $z=0.1$ (Section \ref{s:mock}).  The \textbf{middle panels} show 5NN PDF ratios at $z=0.1$ for higher-mass galaxies.  The \textbf{bottom panels} show 5NN PDF ratios at $z=1$ and $z=2.5$ for galaxies with $10^{9.75} <  M_\ast/\Msun < 10^{10.25}$.  In all panels, shaded regions correspond to $1\sigma$ jackknife uncertainties convolved with 20\% assumed systematic errors; neighbours are counted to a maximum line-of-sight distance of $\pi_\mathrm{max}=125$ Mpc/h.}
\label{f:5nn_ratios}
\end{figure*}

\section{Results}

\label{s:results}

\subsection{Two-Point Correlation Functions}

\label{s:2pcf}

\subsubsection{Results} 

We compute projected two-point auto-correlation functions (2PCFs) for galaxies in our mock catalogue using the standard integral of the 3D (projected $+$ redshift-space) correlation function $\xi(R_p, \pi)$:
\begin{eqnarray}
    w_p(R_p) = \int_{-\pi_\mathrm{max}}^{\pi_\mathrm{max}} \xi(R_p, \pi) d\pi, \label{e:wp}
\end{eqnarray}
where $R_p$ is the projected distance, and $\pi$ is the line-of-sight distance including redshift-space distortions from peculiar motion.  For this section, we choose $\pi_\mathrm{max}=125$ Mpc $h^{-1}$ (comoving).  This is motivated by the fact that, at $z>0$, many redshifts are obtained with low-resolution prism or grism spectroscopy; these instruments have typical redshift errors of $\sigma_z \lesssim 0.005 (1+z)$ \citep{Coil11,Momcheva16}.  At $z\sim 2$, these errors would correspond to $\sigma_\mathrm{\pi}$ of $45$ Mpc $h^{-1}$ (comoving), well within our chosen $\pi_\mathrm{max}$.  In Section \ref{s:sizes}, we use a smaller value of $\pi_\mathrm{max}$ that is better-matched to the smaller SDSS redshift errors.

To compute uncertainties, we subdivide \textit{SMDPL} into 16 jackknife regions, each $100\times100$  Mpc$^2$ $h^{-2}$ along the $X$ and $Y$ axes of the simulation volume.  We assume that the line-of-sight direction $\pi$ is parallel to the $Z$ axis of the box; hence, our large $\pi_\mathrm{max}$ does not allow independent divisions of the box volume along the $Z$ axis.  We also assume minimum systematic uncertainties of 10\% (e.g., from fibre collision corrections, masking errors, etc.).

The left panel of Fig.\ \ref{f:examples} shows an example of 2PCFs for galaxies split by host halo concentration.  There is a large, scale-dependent bias on non-linear and quasi-linear scales that results in large-concentration haloes clustering 3--10$\times$ more strongly than small-concentration haloes.  This effect is much larger than typical observational uncertainties.  Hence, if a given galaxy property correlated strongly with halo concentration, it would be straightforward to measure large clustering differences for galaxies with above-median and below-median values of that property.

To simplify our presentation, we show \textit{ratios} of 2PCFs in Fig.\ \ref{f:wp_ratios}.  For example, the ratio of large-concentration to small-concentration 2PCFs in Fig.\ \ref{f:examples} is shown as the black solid line in the top panel of Fig.\ \ref{f:wp_ratios}.  The top panel of Fig.\ \ref{f:wp_ratios} also shows clustering differences for galaxies split by host halo spin, interaction history, and growth rate.  All properties considered show unique, scale-dependent clustering differences of 0.3--1 dex on non-linear and quasi-linear scales ($<10$ Mpc).

\subsubsection{Physical overview}

\label{s:phys_overview}

Correlation functions like those in Fig.\ \ref{f:examples} are traditionally divided into two regimes: 1) one-halo scales ($\lesssim 1-2$ Mpc) where most pairs are within the same halo, and 2) two-halo scales ($\gtrsim 1-2$ Mpc), where most pairs are in different haloes \citep{Ma00,Seljak00,Scoccimarro01,Sheth01}.  On one-halo scales, a greater fraction of halo pairs are in bound orbits with each other than on two-halo scales, causing a distinct feature in the spatial clustering (e.g., on 1-2 Mpc scales in Fig.\ \ref{f:examples}).

In this traditional interpretation, central haloes' properties correlate with the large-scale environment, so that haloes with above- or below-median properties have different large-scale secondary biases \citep[hereafter called ``central halo bias''; see, e.g.,][]{Gao05,Wechsler06,Gao07}.  Since satellite haloes' properties are influenced by interactions with their host haloes, halo populations with above- or below-median properties will also have different satellite fractions.  This in turn influences clustering on both one- and two-halo scales.  On one-halo scales, increased satellite fractions boost satellite-satellite and central-satellite pairs.  On two-halo scales, more massive central haloes tend to host more satellite haloes, and so a higher satellite fraction preferentially up-weights higher-mass haloes and hence increases the large-scale bias.

To help illustrate satellite fraction effects, we construct a toy model in which haloes are randomly assigned labels of ``large'' and ``small,'' with satellites being more likely to receive the ``large'' label than centrals.  In this toy model, haloes were assigned stellar masses per Section \ref{s:mock}.  Each halo was then assigned a unit-variance Gaussian-distributed random number $R$, centered at $0$ for centrals and at $1$ for satellites.  Haloes were then split into ``large'' and ``small'' samples based on whether they received an above- or below-median value of $R$ in a given bin of stellar mass.   As shown in Fig.\ \ref{f:sat_ratios}, constructing the toy model in this way results in clustering ratios that are very nearly independent of mass.\footnote{We have also constructed toy models where the fraction of satellites in each sample is independent of stellar mass, but this makes it more difficult to see that a given one-halo effect size corresponds to a similar two-halo effect size regardless of mass.}  Across the mass range of $\log(M_*/\Msun)=9.75 - 11.25$, the fraction of satellites assigned to the ``large'' sample was nearly constant, ranging from 75-80\% (Fig.\ \ref{f:sat_fracs}).  As noted above, the largest clustering differences are seen on one-halo scales, but satellite fractions also impact clustering on two-halo scales.

The combination of scale-independent central halo bias and different satellite fractions would lead to a particularly simple form for clustering ratios.  The effect of scale-independent central halo bias would be (by definition) to shift clustering ratios by a constant factor on all distance scales.  As in Fig.\ \ref{f:sat_ratios}, adding the effect of different satellite fractions would result in clustering ratios that showed: 1) an asymptotic value within one-halo scales, 2) an asymptotic value on two-halo scales, and 3) a smooth, monotonic transition between the two on 1-2 Mpc scales. 

The clustering ratios in Fig.\ \ref{f:wp_ratios} do not match this simple expectation.  In the lowest-mass $z=0$ bin in Fig.\ \ref{f:wp_ratios}, clustering ratios for halo concentration, spin, and interactions do not asymptote to a fixed small-scale value.  As discussed in the next section, this arises due to satellite-satellite interactions, which become important on smaller scales than the traditional two-halo to one-halo transition.  Second, the transition between two-halo to one-halo scales does not occur at the same distance for all halo properties.  For example, halo concentration clustering ratios do not reach an asymptotic value until 5-6 Mpc for the lowest-mass $z=0$ bin, whereas clustering ratios for halo interactions are constant beyond 2-3 Mpc (Fig.\ \ref{f:wp_ratios}).  Changes in the clustering ratios beyond 2-3 Mpc at $z=0$ indeed correspond to scale-dependent secondary biases, since changes in satellite fraction alone cannot result in scale-dependent clustering ratios beyond 2-3 Mpc (Fig.\ \ref{f:sat_ratios}).

With the above said, asymptotic clustering ratios on two-halo scales do at least match expectations from the combination of central halo bias and different satellite fractions \citep[see also][]{Wechsler06,Gao07,Salcedo18,Contreras19}.  For reference, satellite fractions for all halo properties, stellar mass ranges, and redshifts are shown in Fig.\ \ref{f:sat_fracs}.  We note that even in cases where central halo bias can be expressed as a function of peak height alone (e.g., for concentration as in \citealt{Wechsler06}), this is no longer the case for clustering ratios including satellites.  For example, at $z=0$, concentration and halo growth rate have opposite correlations with large scale clustering regardless of halo mass (Fig.\ \ref{f:wp_ratios}).  At $z=2.5$, the correlation sign for large-scale clustering switches for concentration alone.  Hence, the relationship between peak height and large-scale clustering cannot be redshift independent for both concentration and halo growth rates.

\subsubsection{Physical interpretation for specific halo properties}

Halo concentration is very density-dependent \citep[e.g.][]{Lee16}, with higher-concentration haloes much more likely to be found in dense environments.  Halo growth rates are also density-dependent, and anti-correlate with concentration \citep{Wechsler02}.  However, the anti-correlation is not perfect: increased rates of interactions and fly-by encounters in dense environments (\textit{blue dash-dotted line} in Fig.\ \ref{f:wp_ratios}) can decrease concentration without increasing halo growth rates.  As a result, clustering differences between high- and low-concentration haloes show a shallower scale-dependence than halo growth rates (compare \textit{black dotted line} vs.\ \textit{green dash double-dotted line} in Fig.\ \ref{f:wp_ratios}). A significant majority of the haloes whose growth has been prematurely arrested due to a flyby event are found in the densest environments \citep{Mansfield_Kravtsov20}, and these haloes exhibit a characteristic scale-dependent feature in their correlation function at $\sim1-2$ Mpc \citep{Sunayama16}. Finally, halo spins tend to be higher in mid-density environments (as compared to either high-density or low-density; \citealt{Lee16}), but they also show twin bias \citep{Johnson19,SatoPolito19}, in that two haloes passing near each other can increase both haloes' angular momenta.  This results in a U-shaped ratio of separations between high-spin and low-spin halo clustering.

Fig.\ \ref{f:wp_ratios} also shows the mass dependence of these clustering ratios (middle panels).  Broadly speaking, clustering ratios for higher-mass haloes show largely self-similar behaviour as lower-mass haloes, but there are some mild differences that we describe for completeness.  At the higher masses considered here, halo concentration and spin have a weaker density-dependence \citep{Lee16}, resulting in reduced clustering differences.  On the other hand, higher density becomes a stronger predictor of halo interactions, leading to increased clustering differences for galaxies split by host halo interaction history.

Finally, Fig.\ \ref{f:wp_ratios} also shows the redshift dependence of these clustering ratios (bottom panels).  Even at $z=2.5$, there are still clear clustering differences (and different shapes) between galaxies split by different halo properties.  However, there are important qualitative differences.  The density-dependence of halo concentration reverses sign at high redshift for low-mass haloes \citep{Lee16}, which causes a large drop in the clustering ratio in the two-halo regime for galaxies split by halo concentration.  In addition, the density-dependence of halo spin changes sign near $z\sim 1$ \citep{Lee16}, so that large-spin haloes cluster more strongly than small-spin haloes at $z\sim 2.5$.

\subsection{Distances to $k$th Nearest Neighbours}

\label{s:knn}

\subsubsection{Results} 

\label{s:knn_results}

We also compute projected distances to the $k$th nearest neighbour ($k$NN distances), denoted as $R_{p,k}$.  Traditionally, $R_{p,k}$ is used as a density measure, so it is common to compute $k$NN distances among all observed galaxies.  Here, we instead compute the \textit{within-sample} $k$NN distance.  E.g., for low-concentration haloes, we compute the distance to the $k$th nearest neighbour within the set of other low-concentration haloes.

The key advantage of this approach is that it maximizes the strength of the environmentally-dependent signal.  For example, low-concentration haloes are rare in high-density environments \citep{Lee16}.  Hence, it becomes exponentially less likely to find $k$ low-concentration neighbours within a given distance rather than $k$ high-concentration neighbours.  As a result, the fraction of low-concentration haloes with low $R_{p,k}$ is dramatically suppressed relative to the fraction of high-concentration haloes with low $R_{p,k}$.  Conversely, high-concentration haloes are relatively rare in low-density environments.  This means that the average distances between high-concentration haloes in low-density environments are larger than those for low-concentration haloes in low-density environments.  As a result, high-concentration haloes are \textit{more} likely to have very high values of $R_{p,k}$ than low-concentration haloes.  Stated another way, the relationship between $R_{p,k}$ and local matter density is different for high-concentration vs.\ low-concentration haloes using within-sample $k$NN distances, which enhances the differences in their $R_{p,k}$ distributions.

Because of the strong dependence of $R_{p,k}$ on environment, $k$NN distances offer greater sensitivity to halo property dependence than 2PCFs in very high-density environments.  $k$NN distances also provide more information than 2PCFs about very low-density environments.  2PCFs effectively weight galaxies by density, as the number of pairs of galaxies is proportional to density squared.  As a result, galaxies in low-density environments can be assigned very different behaviour without any measurable effect on the 2PCF.  In contrast, galaxies with large $k$NN distances live exclusively in low-density environments, so galaxy properties that change only in very low-density environments are still accessible.

As with Section \ref{s:2pcf}, we compute distances to neighbours that are within a line-of-sight separation of $\pi_\mathrm{max}=125$  Mpc $h^{-1}$ (comoving), assumed to be parallel to the $Z$ axis of the box.  To compute uncertainties, we subdivide \textit{SMDPL} into the same 16 jackknife regions.  We assume minimum systematic errors of 20\%, reflecting the greater difficulty of correcting for observational systematics than in 2PCFs.  The right panel of Fig.\ \ref{f:examples} shows an example of 5$^\mathrm{th}$ nearest neighbour distances (5NN distances) for galaxies split by host halo concentration.  There is an even larger scale-dependent ratio difference than for the 2PCF, which results in up to $\sim 100 \times$ differences between the probability distributions.  As with 2PCFs, this is much larger than typical observational uncertainties.

We show ratios of $P(R_{p,5})$ (the distribution of projected 5NN distances) for galaxies split by secondary halo properties in Fig.\ \ref{f:5nn_ratios}.  These ratios may be interpreted as the relative probability of a halo being above-median vs.\ below-median for a given $R_{p,5}$, since the above- and below-median bins have equal numbers of haloes by definition. As with 2PCFs in Fig.\ \ref{f:wp_ratios}, each halo property considered (concentration, spin, interaction history, and growth rate) shows unique, scale-dependent differences in these ratios.  As with concentration, the magnitudes of the effects are significantly larger than for two-point correlation functions, with probability ratios of 0.7--2 dex.

\subsubsection{Physical overview}

\label{s:knn_physical_overview}

Unlike correlation functions, there is no natural division into one-halo and two-halo scales for $k$NN PDFs.  Instead, the distance scale corresponds to local density.  The smallest scales for $R_{p,5}$ are only achieved in dense clusters ($\lesssim 0.5$ Mpc for the samples considered here), whereas the largest scales are only achievable in voids.  Because galaxies have positive correlation functions, most galaxies live in denser-than-average environments, leading to systematically lower typical $k$NN separations compared to randomly-distributed points (Fig.\ \ref{f:examples}).

The $k$NN distance is a function of both the sample density and the sample clustering.  For example, the bottom panel of Fig.\ \ref{f:sat_ratios} shows the effect of satellite fractions on 5NN PDFs using the same toy model as in Section \ref{s:phys_overview}.  A galaxy sample with a higher satellite fraction (at fixed number density) will have more galaxies in dense environments, leading to higher 5NN PDFs on small scales.  Such a galaxy sample will also have fewer galaxies present in voids.  Recalling that lower densities in voids will increase the chances of finding large separations (see Section \ref{s:knn_results}), this leads to higher 5NN PDFs on large scales.  For the 5NN PDF ratios shown here, the largest effect of changing sample density will be a rescaling of the distance scale (i.e., the $x$-axis), with a much smaller second-order effect from the strength of clustering on different scales.  For example, if we multiplied $R_\mathrm{p,5}$ by the cube root of the ratio of sample densities for two different galaxy samples in the satellite fraction toy model (the ``Poisson Rescaled'' lines in Fig.\ \ref{f:sat_ratios}), the shapes of the 5NN PDF ratios converge, telling us that the underlying samples have similar satellite fractions as in Fig.\ \ref{f:sat_fracs}.  Because of the dependencies on sample density and clustering, there is no fixed 5NN scale that corresponds to clusters or voids.  Instead, the locations of the tails of the 5NN PDF (e.g., the lowest or highest 5\% of the galaxy sample) are often better indicators of where the cluster and void regimes are reached.

As discussed in Section \ref{s:knn_results}, $k$NN PDFs are more sensitive to galaxies' behaviour at very high densities (cluster regimes) and very low densities (void regimes) than 2PCFs.  Hence, 5NN PDFs have greater sensitivity than 2PCFs to effects on galaxy samples from the high tidal and ram-pressure forces in clusters.  At the same time, they also have greater sensitivity to the effects of more quiescent environments (e.g., much lower external tidal forces, pre-processing, and pre-heating) around haloes in voids.  As a result, they are sensitive to complementary physical processes as compared to 2PCFs.

\subsubsection{Physical interpretation for specific halo properties}

We note broadly similar trends in 5NN probability ratios as for the 2PCF ratios.  5NN distances also offer unique sensitivity to low-density environments.  High-concentration haloes are less common in low-density environments, which results in higher incidence of very large $R_{p,5}$ for these haloes.
Conversely, low accretion rates are less common in low-density environments, so the opposite behaviour is seen.  Notably, 5NN distances are sensitive to very high density, mid-density, and low-density regions separately, and so probability ratios at low $R_{p,5}$ do not predict probability ratios at high $R_{p,5}$.  This is especially evident for halo spin and interaction history, which have large differences in 5NN distance probability ratios only at very high or very low $R_{p,5}$, but not both.

\section{Application to Galaxy Sizes}

\label{s:sizes}
In this section, we apply the intuition gained from Section \ref{s:results} to test models for 3D half-mass galaxy sizes.  Section \ref{s:size_sims} describes how we generate galaxy sizes for the mock catalogues, section \ref{s:size_obs} describes how we measure two-point correlation functions for galaxies in observations, and \ref{s:size_results} compares the predicted vs.\ observed two-point correlation functions.  Although distances to 5$^\mathrm{th}$ nearest neighbours would represent important additional constraints, accurately comparing observations to simulations would require accurate forward modeling of fibre collisions and survey masking effects, which are beyond the scope of this paper.  As a result, we present comparisons to uncorrected 5$^\mathrm{th}$ nearest neighbour distributions in Appendix \ref{a:obs_5nn} and focus here only on two-point correlation functions.

\subsection{Adding Galaxy Sizes to Simulations}

\label{s:size_sims}

The classic galaxy size model, $R_\mathrm{stars} \propto \lambda R_\mathrm{h}$, where $R_h$ is the halo radius, is deceptively simple.  Several sources of ambiguity are present in this relation:
\begin{enumerate}
    \item Which galaxies the size model should apply to: i.e., all galaxies, or disc galaxies only. 
    \item The choice of $\lambda$: e.g., $\lambda_\mathrm{P}$ \citep{Peebles69} or $\lambda_\mathrm{B}$ \citep{Bullock01}.
    \item The choice of satellite mass definition: e.g., $M_\mathrm{peak}$ or $M_\mathrm{now}$.
    \item The choice of overdensity to use for $R_\mathrm{h}$: e.g., $\rho_\mathrm{vir}(a_\mathrm{now})$ or $\rho_\mathrm{vir}(a_\mathrm{peak})$, where $a_\mathrm{peak}$ is the scale factor at which the halo reaches peak mass.
\end{enumerate}
Past work has shown that the scatter in $\lambda$ is $\sim 0.25-0.3$ dex \citep[e.g.,][]{Bullock01,Onions13,Lee16}, independent of halo mass.  This is similar to the scatter in galaxy sizes across all galaxy types.  Hence, $R_\mathrm{stars} \propto \lambda R_\mathrm{h}$ predicts the correct level of scatter in galaxy sizes only if: 1) it were applied to \textit{all} galaxies, and 2) satellite masses were taken to be $M_\mathrm{peak}$ or $M_\mathrm{acc}$ instead of $M_\mathrm{now}$ \citep{Somerville18}.  We primarily consider the cases where these conditions hold, but for completeness, we also show clustering predictions for the case where only disc galaxies satisfy $R_\mathrm{stars} \propto \lambda R_\mathrm{h}$ in Appendix \ref{a:discs}.

We also consider the models of \cite{Jiang19} and \cite{Desmond17}.  \cite{Jiang19} found in zoom-in hydrodynamical simulations that 3D galaxy size for all types of galaxies is most correlated with $R_\mathrm{h}c^{0.7}$ (where $c$ is the present-day halo concentration), with 0.15 dex scatter.  \cite{Desmond17} investigated disc galaxies only in a large cosmological hydrodynamical simulation, finding no residual correlations between galaxy size and either halo spin, concentration, or radius at fixed galaxy stellar mass; comparisons with this model are shown in Appendix \ref{a:discs}.

For all tested models, we assign galaxy stellar masses to haloes as in Section \ref{s:mock}.  Then, in narrow (0.1 dex) bins of stellar mass, we assign galaxy sizes to haloes in rank order of the following quantities:
\begin{enumerate}
    \item \textbf{Classic Size Models}: $\lambda R_h$.  Separate predictions are shown for Peebles and Bullock spins.  In the main text, we take $R_h = R_\mathrm{peak,now} \equiv \sqrt[3]{M_\mathrm{peak}/(\frac{4}{3}\pi \rho_\mathrm{vir}(a_\mathrm{now}))}$; we show comparisons in Appendix \ref{a:rthen} for $R_h = R_\mathrm{peak,then} \equiv R_\mathrm{vir}$ at the scale factor of peak mass.\label{m:classic}
    \item \textbf{Jiang et al.\ (2019) Model}: $R_\mathrm{peak,now} \, c_\mathrm{now}^{0.7} \times 10^{G(0.15)}$, where $c_\mathrm{now}$ is the present-day concentration and $G(0.15)$ is a Gaussian random variable with standard deviation 0.15.
    \item \textbf{Halo Growth Model, with correlation $\mathbf{r}$}: galaxy size at fixed stellar mass is correlated with $\dot{M}_h$ (i.e., the average mass accretion rate over the past dynamical time), with rank correlation strength $r$.  This model corresponds to a correlation between galaxy SFR (which correlates with halo growth) and galaxy size as qualitatively suggested in \citet{ElBadry15}.  Implementation is described in Appendix \ref{a:halo_growth}.  Here, we show predictions for $r=0.2-0.4$.
\end{enumerate}

We note that the above analysis is insensitive to the exact choice of galaxy sizes to assign.  Specifically, when comparing the clustering of above-median galaxies vs.\ below-median galaxies, the only factor that matters is a halo's rank order---i.e., haloes that are above-median in rank within their stellar mass bin will be assigned above-median galaxy sizes, and vice versa for haloes that are below-median in rank.  Hence, we omit the step of assigning a physical size, and we compare the clustering of haloes with above-median ranks vs.\ those with below-median ranks after the above assignments.

\begin{table}
\caption{Summary of Observational Galaxy Sample.}
\begin{tabular}{cccccc}
$\log_{10}(M_{*,\mathrm{min}})$ & $\log_{10}(M_{*,\mathrm{max}})$ & $N$ & $f_\mathrm{discs}$ & $z_\mathrm{max}$ & $V$ (Mpc$^{3}$)\\
\hline
9.75 & 10.25 & 9762 & 70\% & 0.044 & 4.4$\times 10^6$\\
10.25 & 10.75 & 25011 & 60\% & 0.066 & 1.5$\times 10^7$\\
10.75 & 11.25 & 34339 & 44\% & 0.100 & 5.2$\times 10^7$\\
\hline
\end{tabular}
\parbox{\columnwidth}{\textbf{Notes.} Stellar masses are in units of $\Msun$.  Discs are defined as having bulge fractions of $B/T<0.5$ in the \citet{Meert15} SerExp profile fits.  For all samples, the minimum redshift is $z=0.01$; $V$ refers to the volume of the galaxy sample used.}  
\label{t:obs_summary}
\end{table}

\begin{figure}
    \centering
\vspace{-8ex}
\phantom{\hspace{-5ex}}\includegraphics[width=1.1\columnwidth]{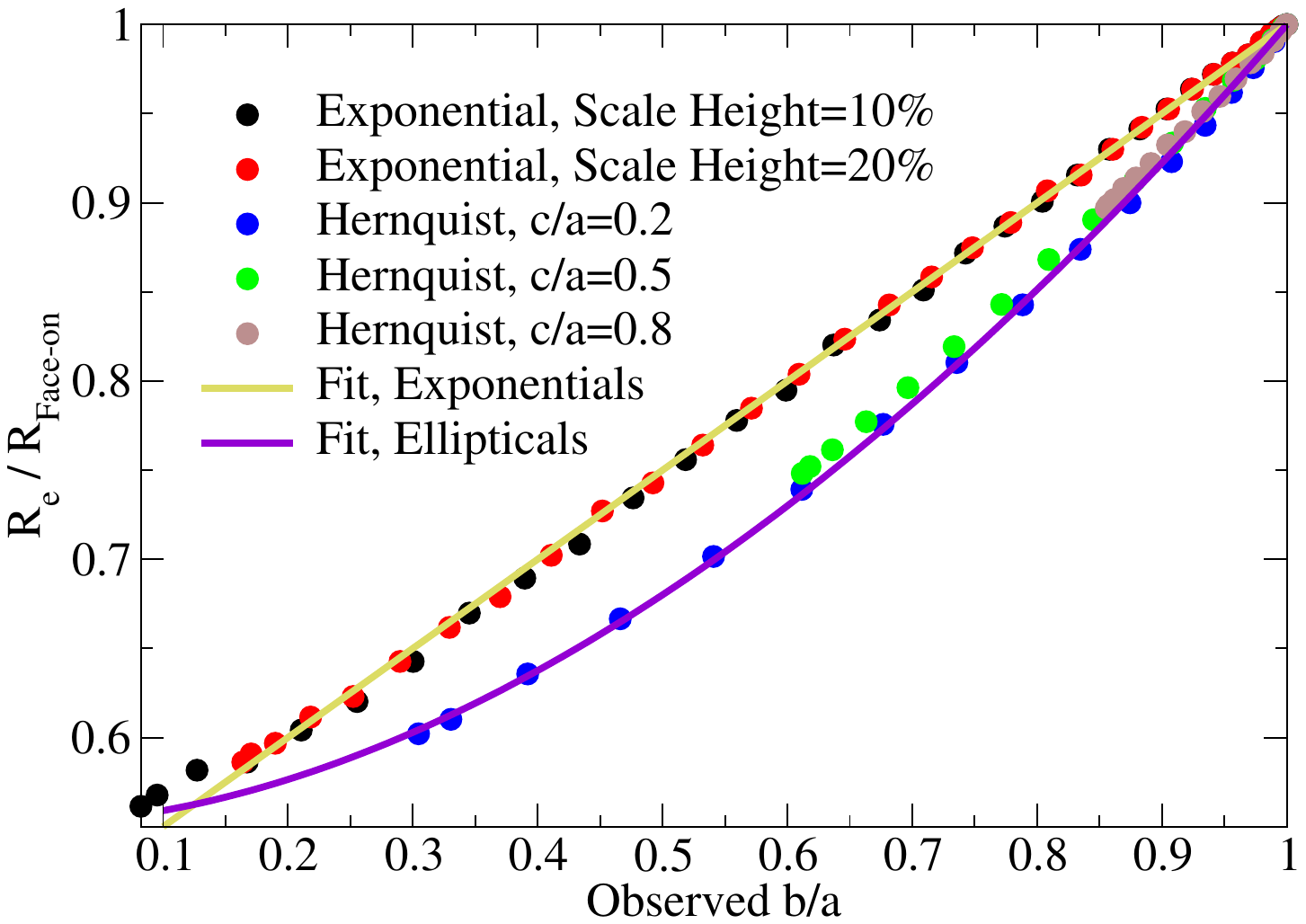}\\[-4.2ex]
\caption{The ratio of the projected half-mass radius ($R_e$) for a galaxy that is tilted with respect to the line of sight compared to the projected half-mass radius for a face-on galaxy with the same intrinsic mass profile ($R_\mathrm{Face-on}$).  This is a strong function of the observed axis ratio ($b/a$), but is a relatively weak function of the intrinsic profile shape.  Dots show numerical projections of common galaxy profiles (exponential and \citealt{Hernquist90}).  For discs (Exponential), profiles that have scale heights of $10$\% and $20$\% of the scale length are shown; for oblate spheroids (Hernquist), profiles that have major to (smallest) minor axis ratios of $c/a=0.2$, $0.5$, and $0.8$ are shown.  Fits for both mass distributions are in Section \ref{s:size_3d}.}
\label{f:ba_sims}
\phantom{\hspace{-5ex}}\includegraphics[width=1.1\columnwidth]{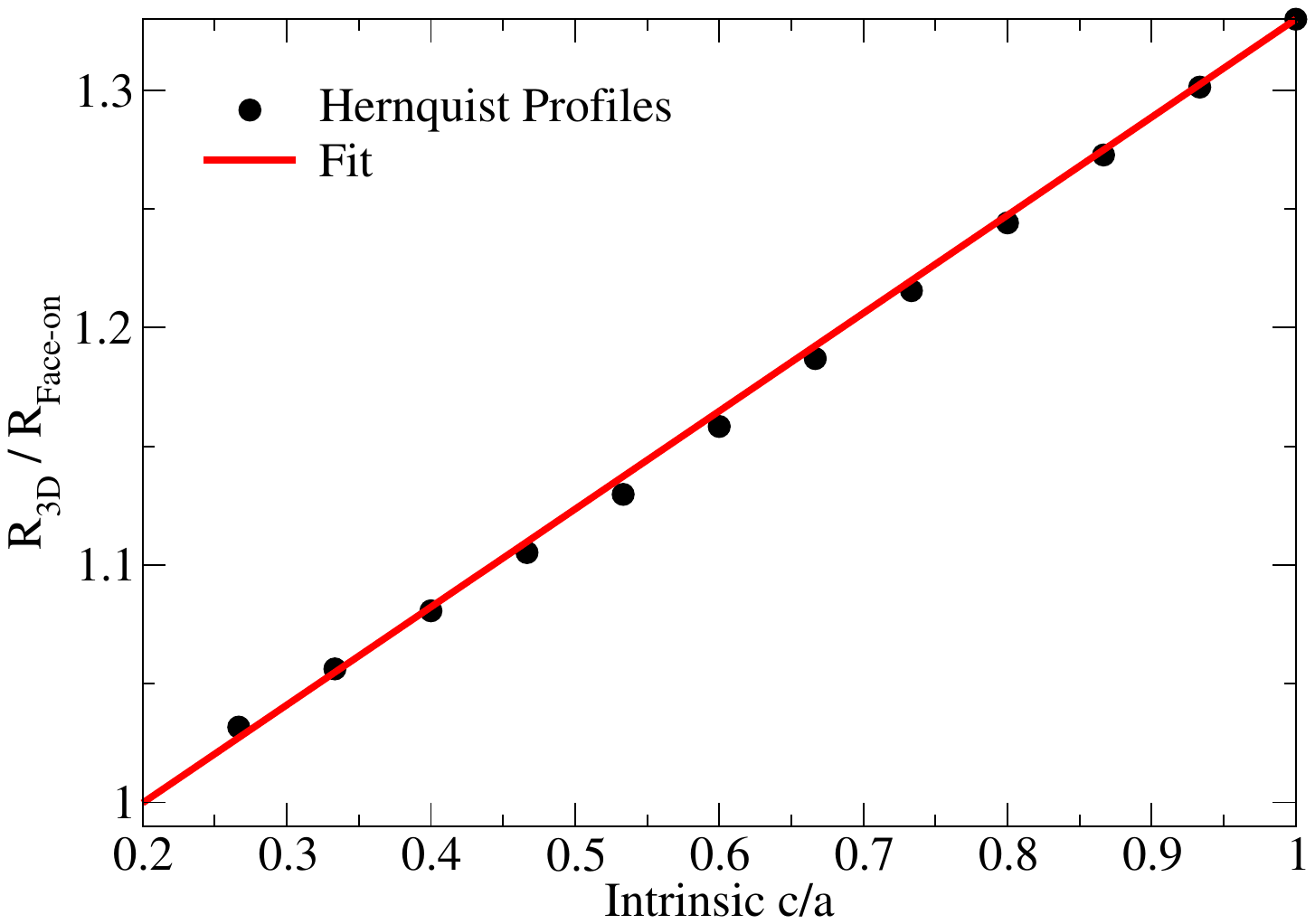}\\[-5ex]
\caption{The 3D half-mass radii for discs are essentially the same as the 2D half-mass radii when the discs are face-on.  However, the same is not true for spheroids.  This figure shows the ratio of the 3D half-mass radius to the 2D face-on half-mass radius for oblate spheroids with  \citet{Hernquist90} profiles as a function of the intrinsic major to (smallest) minor axis ratios $c/a$.  The dots show numerical projections, and the line shows the fit from Section \ref{s:size_3d}.}
\label{f:axis_sims}
\end{figure}

\begin{figure}
\vspace{-8ex}
\phantom{\hspace{-5ex}}\includegraphics[width=1.1\columnwidth]{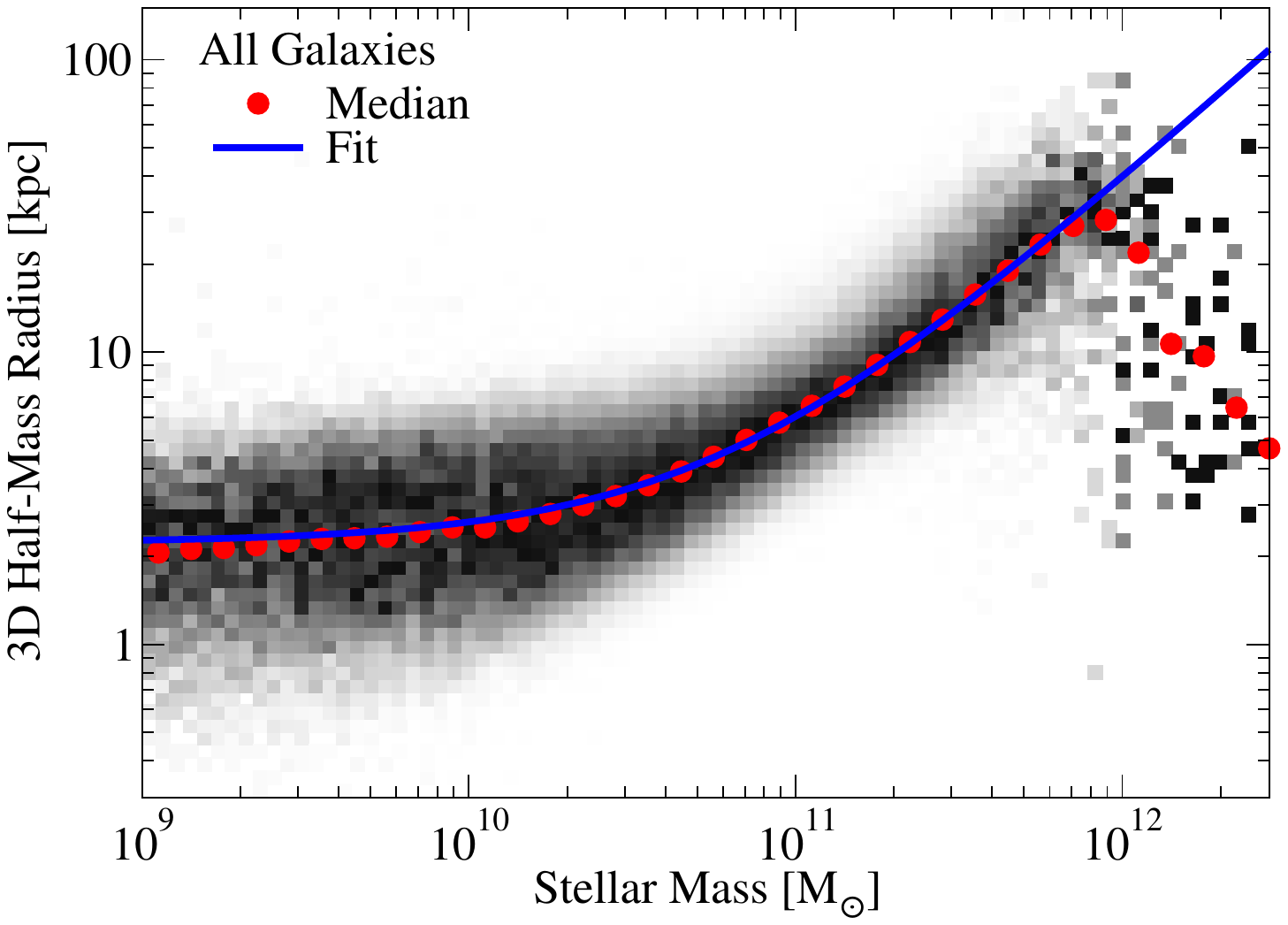}\hspace{-6ex}\\[-7ex]
\caption{Conditional probability distribution of 3D half-mass radii  as a function of stellar mass, based on \citet{Meert15} fits to SDSS galaxies at $z<0.2$, with 3D half-mass corrections from Section \ref{s:size_3d}.  Darker shades of gray indicate higher density on a linear scale.  The \textit{red dots} indicate the sample median, and the \textit{blue line} represents our fit (Eq.\ \ref{e:size_fit}).  Most observed objects with $M_\ast>10^{12}\Msun$ in the \citet{Meert15} catalogues are affected by artefacts such as cosmic rays or incorrect masking of surrounding objects; in addition, the SDSS fibre collision algorithm is biased against targeting brightest cluster galaxies \citep[e.g.,][]{BehrooziMM}.  Hence, the true size distribution is not expected to decrease at high masses.}
\label{f:size_dist}
\end{figure}

\subsection{Observations}
\label{s:size_obs}

We measure projected two-point correlation functions for observed galaxies from the SDSS DR16 \citep[DR16;][]{DR16}.  As in the mock catalogues (Section \ref{s:mock}), redshifts were taken from DR16, but stellar masses were taken from the \cite{Brinchmann04} catalogue (MPA-JHU), which is based on DR7 photometry; they were converted to a \cite{Chabrier03} initial mass function (IMF) by dividing the original \cite{Kroupa01} stellar masses by a factor 1.07.  Selection cuts are described in Section \ref{s:selection}.  Observed galaxy sizes were obtained from the \texttt{SerExp} fits in \cite{Meert15}, who measured projected half-light radii for galaxies from the SDSS DR7.  We convert these to 3D half-mass radii as described in Section \ref{s:size_3d}, and the correlation function measurements are described in Section \ref{s:correlation_functions}.

\subsubsection{Selection Cuts}

\label{s:selection}

We follow the same selection/completeness cuts as in \cite{BehrooziMM}.  The MPA-JHU catalogue includes 606,714 galaxy targets with positive stellar masses from the SDSS DR16, covering a sky area of 8032 deg$^2$.  This sample is $>90\%$ complete down to an apparent magnitude of $r=17.77$.  Duplicate targets within 2kpc projected distance are excluded, leaving 591,513 galaxies.  We also exclude galaxies with redshifts $<0.01$ due to the need for Hubble flow corrections, as well as galaxies within 2 Mpc of a survey boundary or region of significant incompleteness.  This leaves 479,285 galaxies over a sky area of 6262 deg$^2$.

These galaxies represent a flux-limited sample.  To convert to a stellar mass-limited catalogue, we exclude galaxies with stellar masses less than the limit from \cite{BehrooziMM}:
\begin{eqnarray}
    \log_{10}\left(\frac{M_{\ast,\mathrm{min}}}{\Msun}\right) = \frac{r_\mathrm{max} + 0.25 - \mu(z)}{-1.9},
\end{eqnarray}
where $r_\mathrm{max}=17.77$ is the survey's $r$-band limit and $\mu(z)$ is the distance modulus.  The 184,082 galaxies brighter than this limit are $>96\%$ complete at a given stellar mass and redshift \citep{BehrooziMM}.

In the \cite{Meert15} catalogue, about 5\% of these galaxies are not present, and an additional 3\% are flagged as having bad bulge/disc decompositions.  We have verified that there are $\lesssim 5\%$ differences in the overall correlation functions of all galaxies that pass the stellar mass completeness cut, all such galaxies in the \cite{Meert15} catalogue, and all such galaxies without bad fits in the \cite{Meert15} catalogue; these differences are well within the observational uncertainties.  After all cuts, there remain 159,706 galaxies.

\subsubsection{3D Galaxy Sizes}

\label{s:size_3d}
We converted projected half-light radii into projected half-mass radii by multiplying them by the average conversion factor from \cite{Somerville18} of $f_k = 0.85$, as the conversion factors for discs and ellipticals (0.83 and 0.87, respectively) were effectively identical.\footnote{Because we consider correlation function ratios within 0.5 dex bins in stellar mass, using a mass-dependent $f_k$ that varied from 0.83 for low-mass galaxies to 0.87 for high-mass galaxies would have minimal effect on our results.  Of note, \citet{Somerville18} adopt the inverse convention for $f_k$, giving values of $1.15-1.2$ for the factor of division.  The conversion we use neglects the galaxy-to-galaxy variations in star formation history and dust, which in practice will increase the amount of scatter when classifying galaxies as above- or below-median and reduce the observed clustering differences.}

Galaxies are oriented at random angles with respect to the line of sight, causing their projected half-mass radii to be correlated with their axial shape ratio $b/a$.  We perform several projections of common galaxy profiles to measure the importance of this effect.  Intrinsic galaxy shapes at $z\sim 0$ are primarily exponential discs and oblate spheroids (flattened spheres), per \citet{Padilla08}.  Hence, we numerically simulate projections of these two common galaxy profiles at angles from $0$ to $\frac{\pi}{2}$ with respect to the line of sight.  Specifically, we project both exponential discs:
\begin{eqnarray}
    \rho_\ast(R,z) \propto \exp\left(-\frac{R}{r_s} - \frac{|z|}{s r_s}\right),
\end{eqnarray}
where $r_s$ is the scale length and $s$ is the scale height to scale length ratio; as well as flattened \citet{Hernquist90} profiles:
\begin{eqnarray}
    \rho_\ast(R') \propto (R' (R'+a)^3)^{-1}\\
    R' = \sqrt{x^2 + y^2 + \left(\frac{za}{c}\right)^2},
\end{eqnarray}
where $a$ is the scale length and $c/a$ is the flattening ratio (i.e., the ratio of smallest to largest axes).

As shown in Fig.\ \ref{f:ba_sims}, the ratio of the half-mass radii for tilted galaxies to their face-on counterparts is a strong function of the observed shape ratio $b/a$, but a weak function of intrinsic mass distribution.  We find that exponential discs are well fit by:
\begin{eqnarray}
    \frac{R_{e\textrm{,Exp}}(b/a)}{R_\textrm{Face-on}} = 0.5 + 0.5\frac{b}{a},\label{e:face_discs}
\end{eqnarray}
and Hernquist profiles are well fit by:
\begin{eqnarray}
    \frac{R_{e\textrm{,Hern}}(b/a)}{R_\textrm{Face-on}} = (1-0.55)\left(\frac{b}{a}\right)^{1.8} + 0.55, \label{e:face_bulges}
\end{eqnarray}
where $R_e(b/a)$ is the projected half-mass radius as a function of observed shape ratio $b/a$, and $R_\textrm{Face-on}$ is the projected half-mass radius when the profile is viewed face-on.  Defining $f_{b/a}\equiv R_\textrm{Face-on}/R_e(b/a)$, the average value of $f_{b/a}$ across our sample of galaxies is 1.23.

For disc galaxies, the projected face-on half-mass radius is effectively the same as the 3D half-mass radius, but the same is not true for spheroids.  As shown in Fig.\ \ref{f:axis_sims}, the 3D half-mass radius $R_\textrm{3D}$ is larger than the 2D radius by a factor of
\begin{eqnarray}
    f_\mathrm{3D} = \frac{R_{\textrm{3D,Hern}}}{R_\mathrm{Face-on}} = 0.41\left(\frac{c}{a}-1\right)+1.33. \label{e:ca}
\end{eqnarray}
Because this depends on $c/a$, which is not directly observable, we use the average $c/a$ values from \cite{Padilla08} as a function of magnitude, which range from $\langle c/a \rangle =0.36$ ($R_{\textrm{3D}}/R_\mathrm{Face-on} = 1.07$) at $M_r = -19.5$ to $\langle c/a \rangle =0.76$ ($R_{\textrm{3D}}/R_\mathrm{Face-on} = 1.23$), linearly interpolating between magnitude bins.

The net conversion between projected half-light and 3D half-mass radii is given by:
\begin{eqnarray}
    r_{\textrm{3D,mass}} = f_\mathrm{3D}\cdot f_{b/a} \cdot f_k \cdot r_\textrm{2D,light},
\end{eqnarray}
where $f_k$ is 0.85, $f_{b/a}$ is given by the reciprocals of Eqs.\ \ref{e:face_discs}--\ref{e:face_bulges}, and $f_\mathrm{3D}$ is 1 for discs and given by Eq.\ \ref{e:ca} for ellipticals.  To decide which fits to use, we divide galaxies into discs ($B/T < 0.5$) and ellipticals $(B/T>0.5)$ based on their bulge fractions $B/T$ as measured in \cite{Meert15}.  The distribution of $r_{\textrm{3D,mass}}/r_\textrm{2D,light}$ is relatively narrow, even in different magnitude bins, with a mean of $0.07$ dex and a standard deviation of $0.06$ dex.  As this is small relative to the standard deviation of galaxy sizes at fixed mass ($\sim 0.25-0.3$ dex, \citealt{Somerville18}), we find that this results in relatively modest corrections to the size distribution.  Similarly, the clustering ratio  $w_\textrm{p,large}/w_\textrm{p,small}$ changes by $<10$\% between 2D half-light radii and 3D half-mass radii, which is well within observational errors.

\begin{figure*}
\begin{center}
\vspace{-8ex}
\phantom{\hspace{-5ex}}\includegraphics[width=1.1\columnwidth]{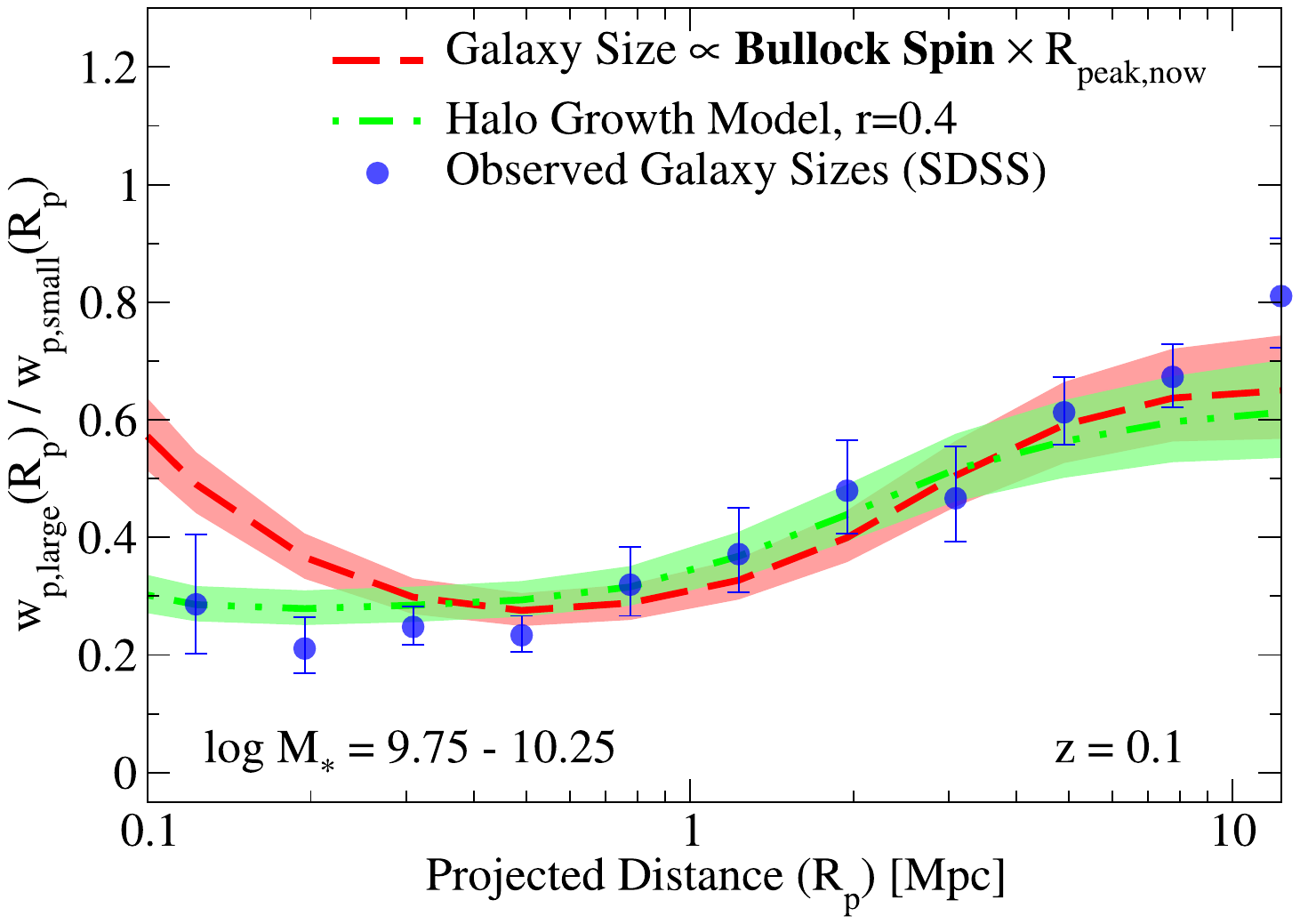}\hspace{-6ex}\includegraphics[width=1.1\columnwidth]{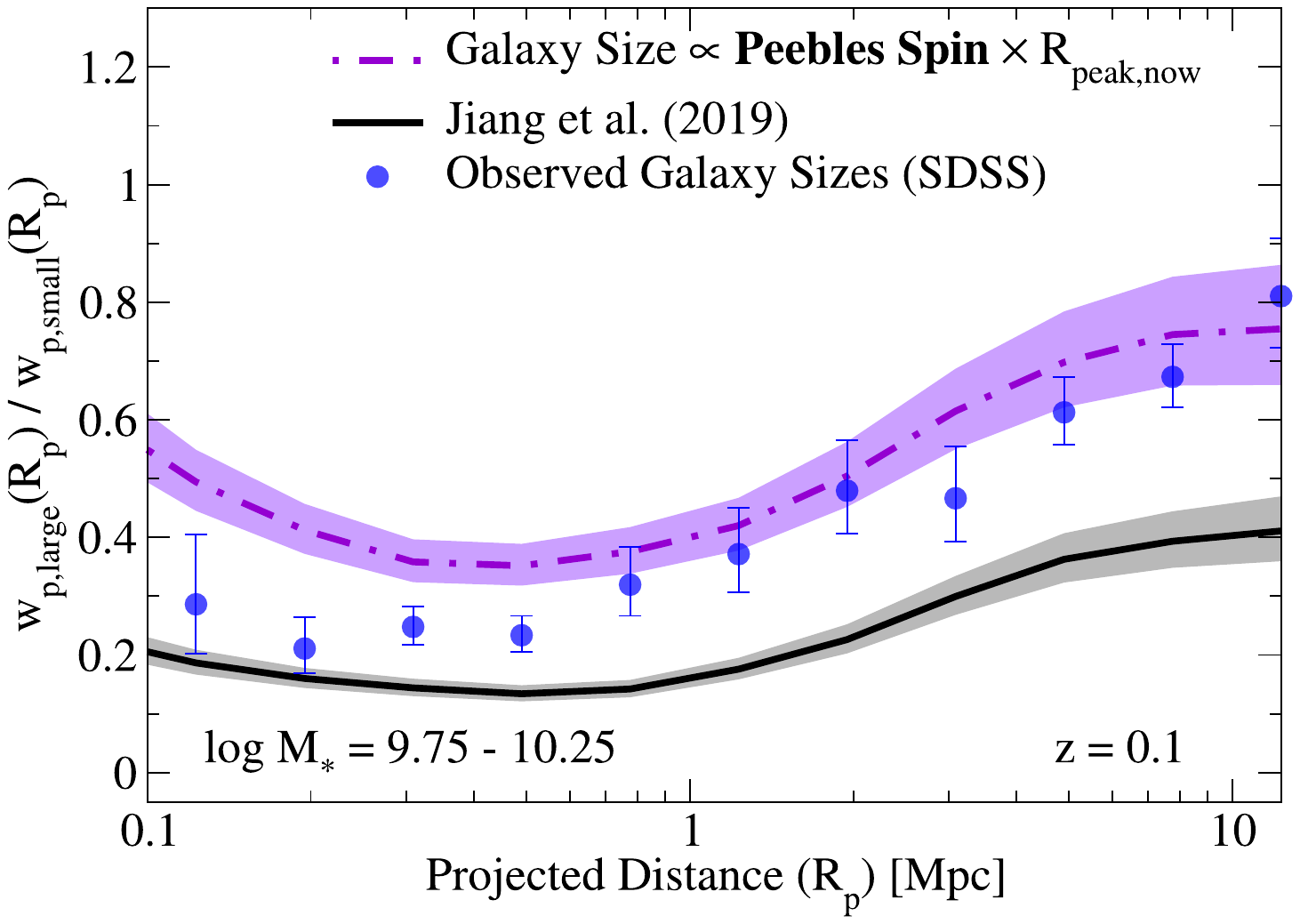}
\vspace{-3ex}
\phantom{\hspace{-5ex}}\includegraphics[width=1.1\columnwidth]{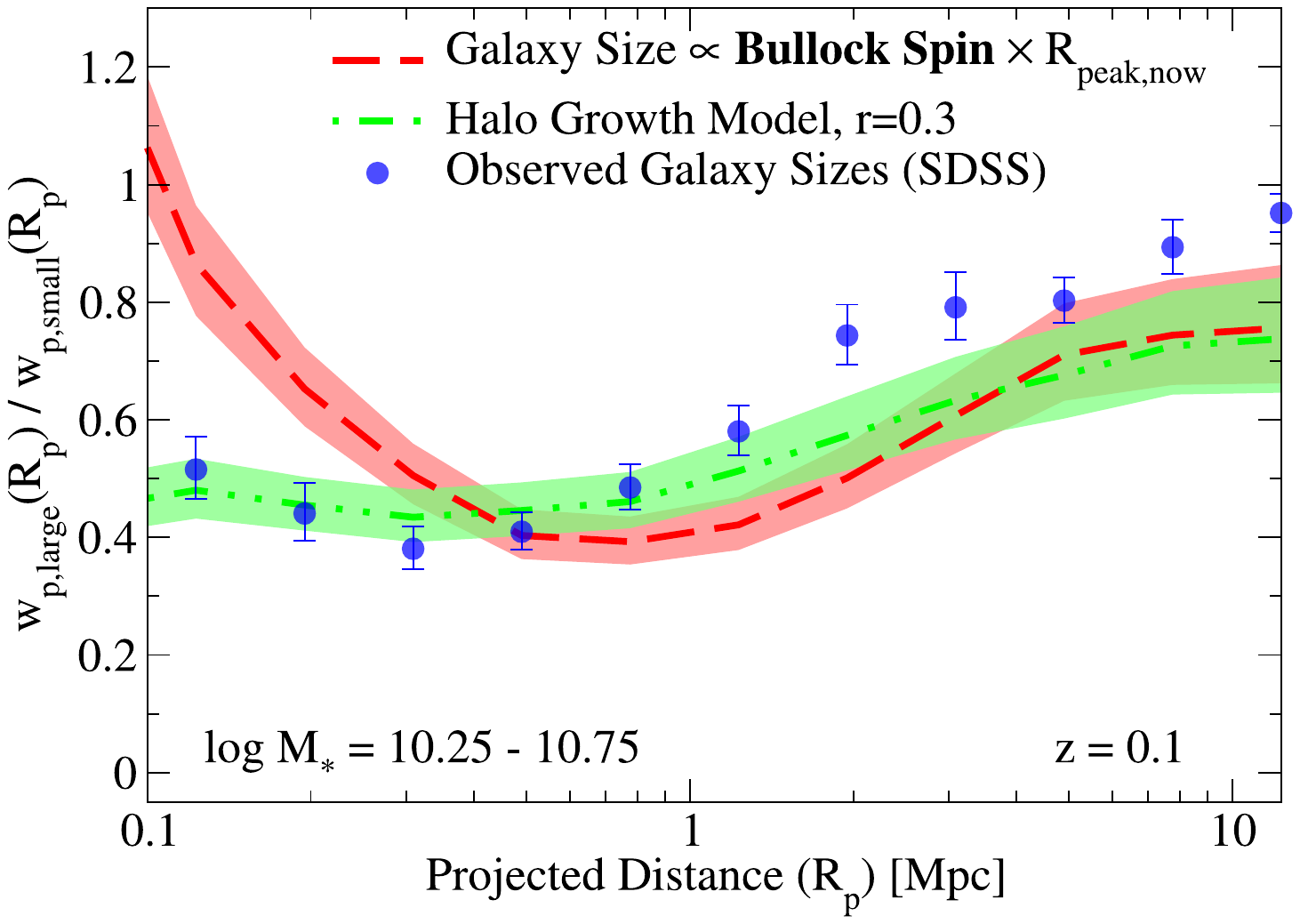}\hspace{-6ex}\includegraphics[width=1.1\columnwidth]{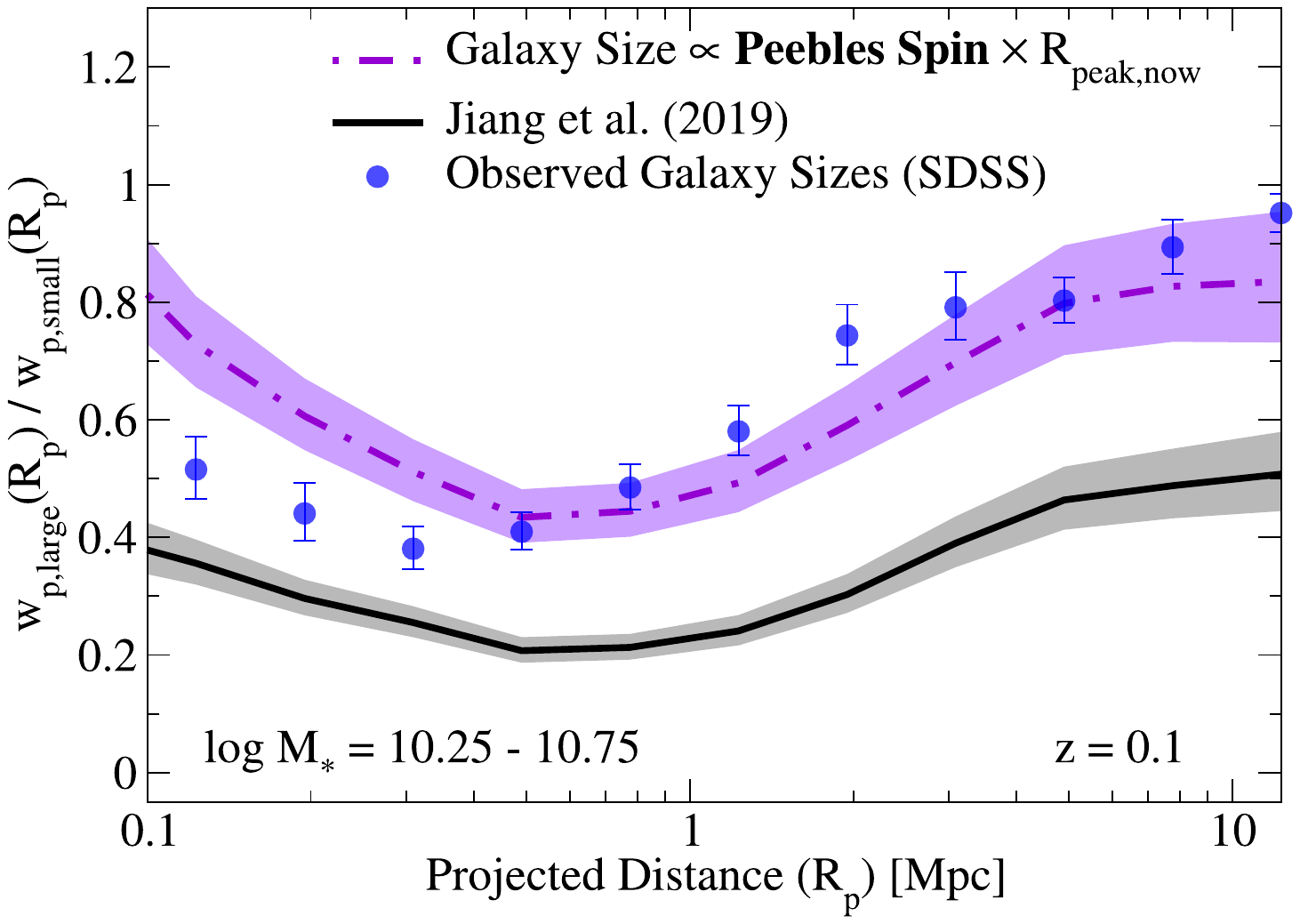}
\vspace{-3ex}
\phantom{\hspace{-5ex}}\includegraphics[width=1.1\columnwidth]{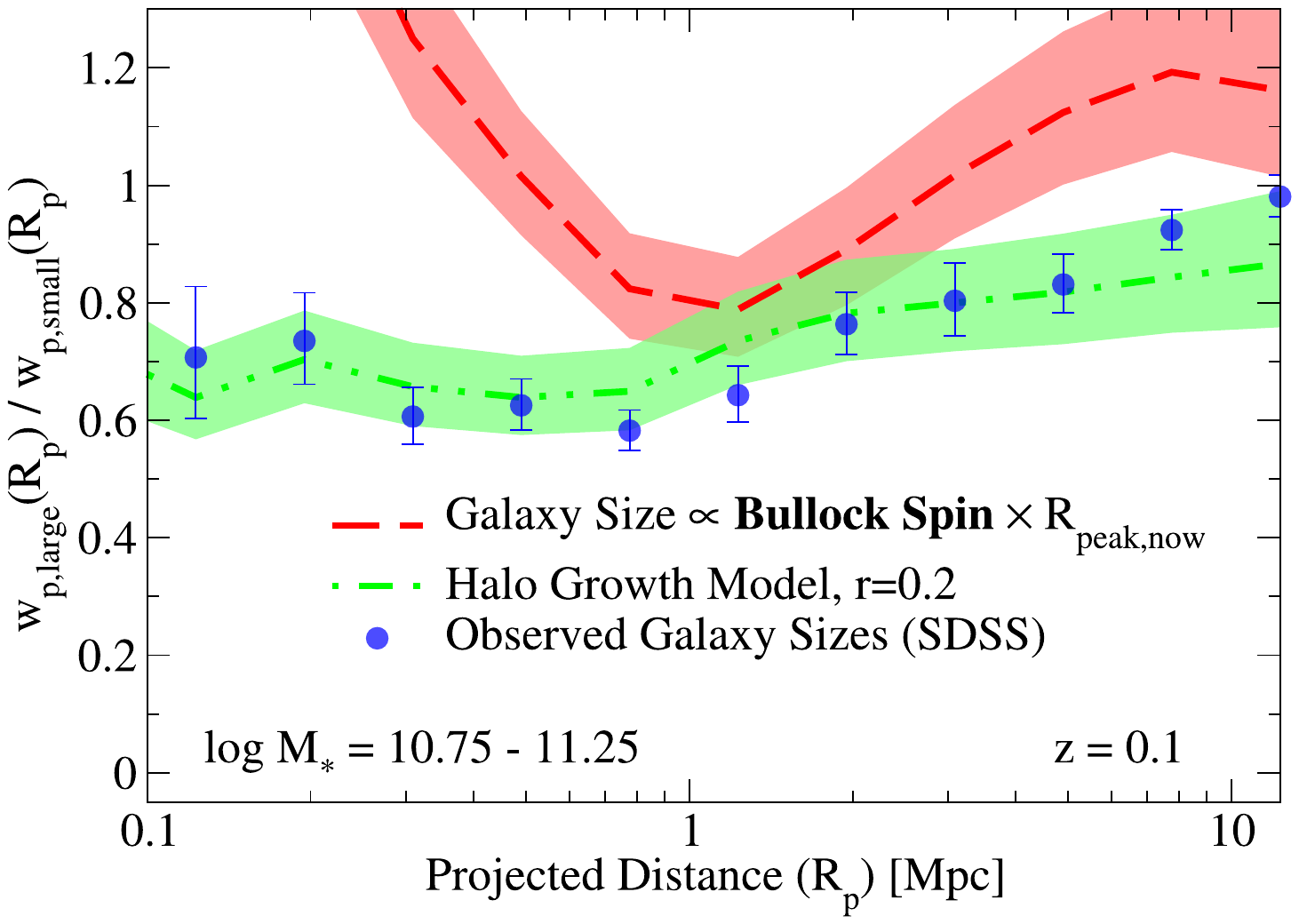}\hspace{-6ex}\includegraphics[width=1.1\columnwidth]{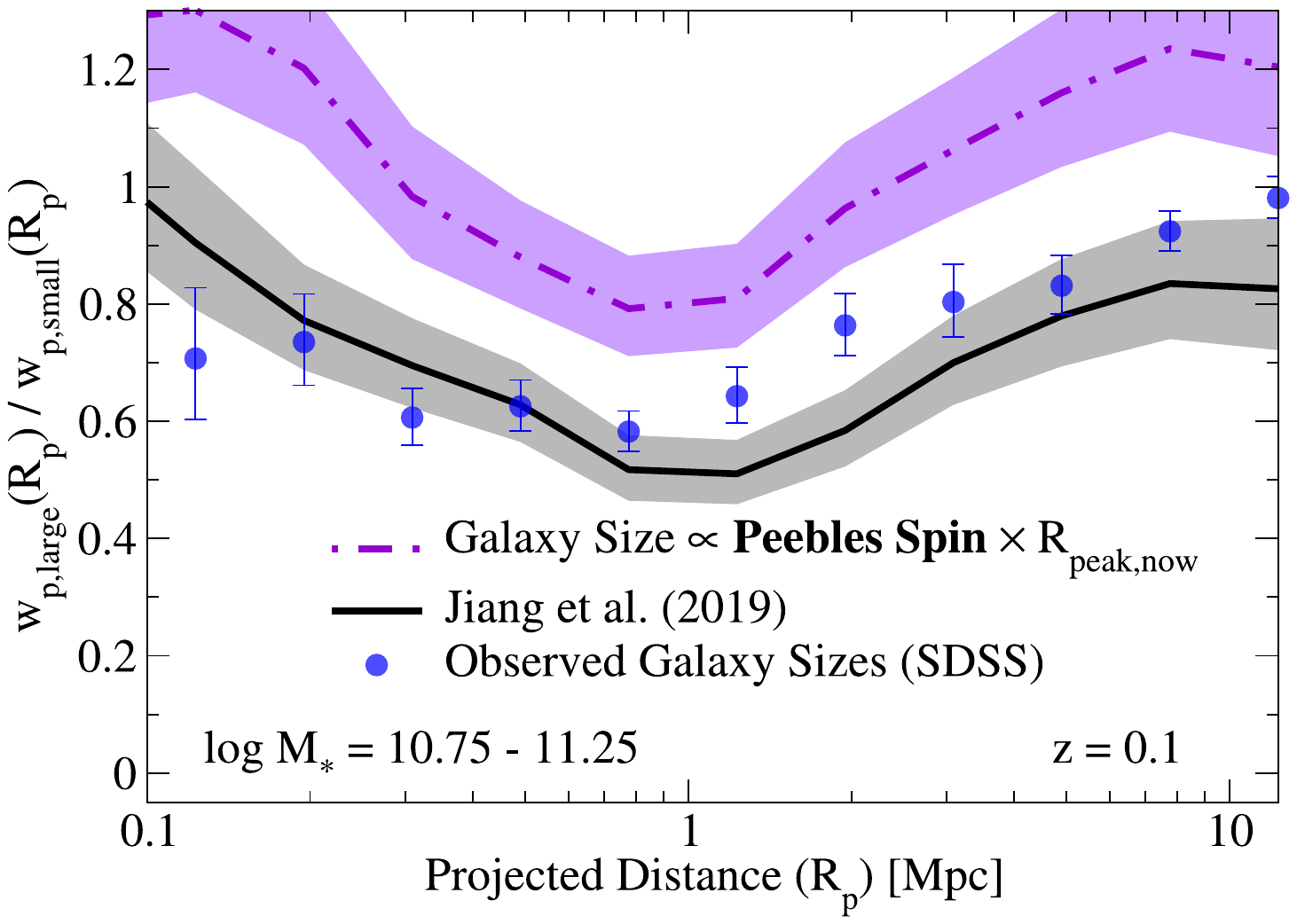}
\vspace{-3ex}
\end{center}
\caption{Ratios of projected two-point correlation functions for large vs.\ small galaxies in both observations and simulations, in several bins of stellar mass.  Galaxy projected half-light sizes in observations have had corrections applied from Section \ref{s:size_3d} to approximate 3D half-mass sizes.  \textbf{Left} panels show models where galaxy sizes are proportional to the Bullock spin (\textit{red} line) as well as to the halo growth rate (\textit{green line}); \textbf{Right} panels show a model where galaxy size is proportional to the Peebles spin (\textit{purple line}) as well as the \citet{Jiang19} concentration-based model (\textit{black line}).  The halo growth model gives the closest match to the SDSS observations (\textit{blue points}).  Other models have difficulties matching all the observational data.  
For both spin-based models, the halo radius used is the present-day radius of a halo with the same peak mass.  In all cases, error bars and shaded regions correspond to jackknife uncertainties.  In contrast to previous plots, two-point correlation functions here for both observations and simulations are computed with $\pi_\mathrm{max}=13.6$ Mpc $h^{-1}$.  See Appendices \ref{a:discs} and \ref{a:rthen} for additional model comparisons.}
\label{f:size_obs}
\end{figure*}

\subsubsection{Observed Projected Correlation Functions}

\label{s:correlation_functions}

We use the public \texttt{correl} utility distributed with the \textsc{UniverseMachine} to calculate auto-correlation functions.  The code computes the projected+redshift-space correlation function $\xi(R_p,\pi)$ using the \cite{Landy93} estimator:
\begin{eqnarray}
    \xi(R_p,\pi) = \frac{DD-2DR+RR}{RR} \; .
\end{eqnarray}
We use the default settings, where $DR$ and $RR$ are computed with 10$^6$ points distributed uniformly randomly, according to the same mask as for the data points.  For two targets at redshift-space positions $\mathbf{x}_1$ and $\mathbf{x}_2$, we define the mean position to be $\mathbf{l} = 0.5(\mathbf{x}_1+\mathbf{x}_2)$ and the separation to be $\mathbf{s} = \mathbf{x}_1-\mathbf{x}_2$.  We then take the usual definitions of $\pi \equiv (\mathbf{s}\cdot\mathbf{l})/ |\mathbf{l}|$ and $R_p = \sqrt{s^2 - \pi^2}$.  Fibre collisions are corrected to first order by increasing the weights of small-separation galaxy pairs.  Because about one-third of the SDSS footprint was observed twice, many galaxy pairs separated by less than 55 arcsec are still present.  We assume that for each observed pair of galaxies separated by less than 55 arcsec, we would miss 2.08 such galaxy pairs due to fibre collisions in the rest of the survey.  Hence, we multiply such pair counts by a factor 3.08 to correct for this deficit \citep{Patton13}.  We also expect that 2.08 galaxies in similar environments will be absent from the survey, and so for each galaxy in each observed low-separation pair, we multiply its longer-separation pair counts by a factor 2.04 (mimicking the effect of having 2.04+2.04 - 2 = 2.08 extra galaxies in the survey).  As discussed in \cite{BWHC19}, this correction results in only percent-level differences at scales above 100 kpc, which are the only scales considered in this paper.

The \texttt{correl} code integrates $\xi(R_p,\pi)$ up to a user-specified $\pi_\mathrm{max}$ to obtain the projected correlation function $w_p(R_p)$ as in Eq.\ \ref{e:wp}.  Here, we choose $\pi_\mathrm{max}=13.6$ Mpc $h^{-1}$ (i.e., 20 Mpc for our adopted value of $h$).  This is low relative to typical values ($> 50$ Mpc) because our aim is \textit{not} to estimate the pure-3D correlation function $\xi(r)$, but instead to compare with the identical calculation of $w_p$ in our mock catalogue.  Choosing a smaller value of $\pi_\mathrm{max}$ is advantageous because the sample variance from galaxies randomly projected in the line of sight is significantly reduced, especially at large $R_p$.  As noted in Section \ref{s:2pcf}, we also use this lower value of $\pi_\mathrm{max}$ when calculating $w_p(R_p)$ from the mock catalogues for this Section.

For our main analysis, we split our observed galaxy sample into several bins of mass and two bins of size.  Since median galaxy size varies across our mass bins, we first compute the median 3D half-mass galaxy size as a function of mass.  As shown in Fig.\ \ref{f:size_dist}, this is fit well by:
\begin{equation}
    R_\mathrm{3D,all}(M_*) = 2.26\left[\left(\frac{M_*}{10^{10.78} \Msun}\right)^{1.00} + \left(\frac{M_*}{10^{10.78} \Msun}\right)^{0.003}\right]\;\mathrm{kpc}, \label{e:size_fit}
\end{equation}
down to stellar masses of $10^{9}\Msun$.

We compute error bars via jackknife sampling, subdividing each observed sample into 16 spatially coherent regions.  The number of galaxies in each observed sample, as well as redshift ranges, are shown in Table \ref{t:obs_summary}.

\subsection{Comparisons of Galaxy Size Models to Observations}

\label{s:size_results}

Fig.\ \ref{f:size_obs} shows two-point correlation function ratios for both galaxy size models and observations; overall clustering normalizations are shown in Appendix \ref{a:clustering}.  As in \citet{Hearin19}, we find scale-dependent clustering differences between large and small galaxies.  These clustering differences are larger for less-massive galaxies, indicating a larger net correlation between size and environment for lower-mass galaxies.

All the spin-based size models show an upturn in the clustering ratio at small radii that does not match the observed clustering.  This upturn is also seen in Fig.\ \ref{f:wp_ratios}; however, it is much more apparent in Fig.\ \ref{f:size_obs} due to the much smaller vertical axis range, and due to our choice to use a smaller value of $\pi_{\rm max}$ in Fig.\ \ref{f:size_obs}.  When the present-day overdensity is used, closely interacting galaxies tend to be assigned larger sizes than necessary to match the data.  Using the halo overdensity definition at the time of peak mass ($R_\mathrm{peak,then}$) results in systematically smaller satellite galaxy sizes (Appendix \ref{a:rthen}; Fig. \ref{f:rthen}).  This halo definition reduces the upturn in the clustering ratio at small radii but then increases the difference in the clustering ratio on two-halo scales beyond what the observations allow.  This same effect occurs for any attempt to change the relative sizes of satellites and central galaxies, since satellites are important for clustering on both one-halo and two-halo scales \citep[see discussion in, e.g.,][]{BWHC19}.  As a result, it may be difficult to ameliorate spin-based size models with another choice of halo radius.

In comparison, the \citet{Jiang19} model uniformly predicts much larger differences between the clustering of large vs.\ small galaxies relative to what is observed.  This discrepancy arises because satellites have significantly higher concentrations than central haloes \citep[e.g.,][]{Bullock01b}, so the assumptions underlying the \citet{Jiang19} model result in an excessively large ratio of satellites that have small sizes vs.\ large sizes (4:1). In the Peebles spin model in Fig.\ \ref{f:size_obs}, this ratio is only 2:1, as satellites have only modestly smaller spins than central haloes \citep{Onions13}.

As it may be useful to have an empirical model that is a closer match to the observed data than existing models, we note that the halo growth model described in Section \ref{s:size_sims} and Appendix \ref{a:halo_growth} performs reasonably well with a correlation coefficient $r$ of:
\begin{eqnarray}
    r(M_*) = 0.4 - 0.2\log\left(\frac{M_*}{10^{10}\Msun}\right).
\end{eqnarray}

\section{Discussion}

\label{s:discussion}

\subsection{Measuring Halo Properties Beyond Mass}

Correlations between galaxy properties and halo properties result in strong, unique, and observable environmental signatures.  We have shown in Section \ref{s:results} and Appendix \ref{a:alternate} that two-point correlation functions and $k$th nearest neighbour statistics contain information about host halo growth, concentration, spin, merger history, and shape.  This opens exciting new avenues for exploring how galaxy properties, such as size, morphology, metallicity, gas fraction, etc.\ depend on these halo properties.  Observable signatures of these halo properties are accessible in lower-resolution grism and prism redshift surveys, allowing measurements out to at least $z\sim 2.5$ in surveys such as PRIMUS and 3D-HST \citep{Coil11,Momcheva16}.

While the environmental signatures of halo-galaxy connections are present in two-point correlation functions, the signatures are even more clear for higher-order density statistics, such as the distance to the $k$th nearest neighbour (Section \ref{s:results}).  To date, two-point correlation functions have received much of the modeling effort, both in terms of developing fast halo occupation models and in correcting for systematic observational effects like fibre collisions \citep[e.g.,][]{Guo12,Burden17,Yang19}.  We have greatly benefited from this effort, in the sense that it is now straightforward to compare two-point correlation functions in simulations and observations on relatively equal footing.  Nonetheless, correcting for fibre collisions for $k$th nearest neighbour distances is much more difficult, and is strongly sensitive to the algorithm employed.  Given the potential to reveal key information about the galaxy--halo connection, we encourage readers involved with current and future spectroscopic surveys (e.g., the Dark Energy Spectroscopic Instrument Survey, the Subaru Prime Focus Spectrograph Survey, the Astrophysics Telescope for Large Area Spectroscopy Survey, etc.) to plan their fibre collision algorithms to make it straightforward to model the effects on higher-order environmental statistics.  For example, the lack of a transparent way to model fibre collisions for $k$th nearest neighbours with the Sloan Digital Sky Survey makes it difficult to perform further tests on size models here.  In contrast, this will be much easier when full public data for the Galaxy And Mass Assembly survey becomes available, due to their strategy of high spectroscopic completeness \citep{Robotham10}.

One may also ask whether the environmental statistics discussed here are optimal measures of host halo properties.  Almost certainly, they are not.  Machine learning offers an elegant solution to search for better statistical measures.  In H.\ Bowden et al., in preparation, we use machine learning to search the space of environmental measures to find one that is optimized for halo property separation.  After splitting haloes into those with high and low values of a given property, we generate a mock galaxy catalogue as in Section \ref{s:mock}.  For each galaxy in the mock, we compute the distances to its $k$ nearest neighbours, which is fed as an $k$-dimensional vector to a deep neural network with a single output statistic $S$.  This network is trained via reinforcement learning to maximize the difference between the probability distributions of $S$ for the two different halo samples, resulting in an improved observational measure for determining correlations between galaxy properties and the halo property in question.

\subsection{Interpreting Physical Scales in Correlation Functions}

Traditionally, correlation functions are interpreted as having a single important physical scale--e.g., the transition between the one- and two-halo regimes at 1-2 Mpc.  Nonetheless, halo properties are influenced by more than whether a halo is a satellite or not (see Section \ref{s:2pcf} and references therein).  On small scales, satellite-satellite interactions can increase spin and decrease concentration; on larger scales, tidal forces can influence growth rates well before haloes become satellites.  Because all of these effects scale with the strength of gravitational interactions, dimensional analysis would suggest that they should all occur on scales similar to the one- to two-halo transition.  However, there is no \textit{a priori} expectation that all the scales should be perfectly identical.

In this paper, we find that correlation function ratios for galaxies split by host halo properties continue changing well into the one-halo regime.  As discussed in Section \ref{s:2pcf}, this cannot arise due to satellite fraction differences, but instead is consistent with a satellite-satellite interaction scale becoming important for halo spins and concentrations at $\lesssim$ 500 kpc at $z=0$.

At larger distances, different halo properties have different scales over which the environment becomes important.  Halo concentrations are subject to environmental effects on scales as large as 5-6 Mpc, whereas other properties only begin to be affected on smaller scales of 2-3 Mpc (Section \ref{s:2pcf}).  We note that the different scales for environmental effects on different halo properties make it extremely difficult to develop a halo mass definition that is assembly-bias free \citep[see also][]{Villarreal17}.  At the same time, different relevant scales for different halo properties make it easier to disentangle the effects of each on galaxy formation, using both correlation functions and $k$NN statistics.

\subsection{Galaxy Sizes}

We have also shown that many past galaxy size models do not provide good matches to observed two-point correlation functions (Section \ref{s:size_results}).  Spin-based models predict that galaxy sizes should increase in very dense environments, which they do not appear to do in observations, both in this paper and in previous studies \citep[e.g.,][]{Spindler17,Hearin19,Rodriguez20}.  This applies regardless of whether one is considering all galaxies (Section \ref{s:size_results}) or discs only (Appendix \ref{a:discs}), although the modeling for disc galaxies is subject to additional theoretical uncertainties.  We have tested several alternate prescriptions for spin-correlated galaxy sizes beyond those considered here, but have not been able to find any that perform significantly better than those shown in this paper.  For example, we have tested taking both the Peebles spin and halo radius from the time of the halo's peak mass, and we have also tested using the \cite{Mo98} formula directly, which introduces a concentration dependence to galaxy sizes.  Both perform very similarly to the existing Peebles spin model shown in Section \ref{s:size_results}.

Hydrodynamical simulations have uniformly found relatively little correlation between halo spin and galaxy size \citep[e.g.,][]{Desmond17,Jiang19}.  We find that the alternate concentration-based model in \cite{Jiang19} results in clustering differences between large and small galaxies that are much more pronounced than allowed by observations.  This suggests that the \cite{Jiang19} model is incomplete.  We have briefly tested whether accounting for observational uncertainties in measuring galaxy sizes could improve the model's performance; however, no amount of scatter added to the model gave a match to clustering on both one-halo and two-halo scales.  We also test the null hypothesis in \cite{Desmond17} of no correlation between galaxy size and halo properties, finding that it predicts clustering differences much smaller than observed (Appendix \ref{a:discs}).

We have presented a simple empirical model that correlates galaxy size with halo growth at fixed stellar mass.  This model is inspired by the correlations between galaxy size and galaxy growth suggested in \cite{ElBadry15}, as well as correlations between galaxy growth and halo growth \citep{Moster17,BWHC19}.  It would also be consistent with the observational need for an anti-correlation between halo concentration and galaxy size \citep{Desmond17b,Desmond19}, as halo growth rates and concentrations are strongly anti-correlated.  Although the physical picture in \cite{ElBadry15} is qualitatively consistent with the observational measurements herein, we caution that (as with any empirical model) consistency does not automatically imply correctness.  In future work (H.\ Bowden et al., in preparation), we will provide a more advanced, self-consistent empirical model for galaxy sizes that will be integrated into the \textsc{Emerge} and \textsc{UniverseMachine} galaxy formation frameworks.

Finally, we note that we have not considered orphan galaxies (i.e., galaxies whose host haloes are no longer tracked by simulations; see, e.g., \citealt{MillOrphan,Moster17,Hearin19,BWHC19}) in the main body of this paper.  At present, there is no way to estimate how most halo properties should evolve for orphan galaxies--meaning that none of the existing spin-based or concentration-based models would have given clear predictions for orphan galaxies' sizes.  It is again beyond the scope of this paper to create models for orphan haloes' properties, which we hope will be investigated comprehensively in future work.  Nonetheless, we find that orphan models do not necessarily result in a large impact on the clustering ratios considered here, as shown in Appendix \ref{a:orphans}.

\section{Conclusions}
\label{s:conclusions}
In this paper, we investigated the observable consequences of correlations between halo properties and galaxy properties. Key findings include:
\begin{itemize}
    \item Correlations between galaxy properties and halo spin, concentration, growth rate, and interaction history produce unique, scale-dependent signatures in multiple observational statistics, including galaxies' two-point correlation functions and the distribution of distances to galaxies' $k^\mathrm{th}$ nearest neighbours (Sections \ref{s:2pcf} and \ref{s:knn}).  These signatures are most evident on distance scales of less than 10 Mpc. This finding is in accord with past studies that have shown unique scale-dependent clustering for haloes (as opposed to galaxies) split by different halo properties, as well as unique density-dependence for different halo properties.
    \item These scale-dependent signatures can be very strong, reaching factors of 10$\times$ differences in projected two-point correlation functions, and 100$\times$ differences in distributions of distances to galaxies' $5^\mathrm{th}$ nearest neighbours (Sections \ref{s:2pcf} and \ref{s:knn}).
    \item These signatures are observationally accessible even with lower-resolution grism and prism redshifts out at least $z\sim 2.5$ (Sections \ref{s:2pcf} and \ref{s:knn}).
    \item We provide a systematic approach to convert 2D projected half-light galaxy sizes to intrinsic 3D half-mass sizes based on the shape ratio and provide fits for the median 3D sizes of observed galaxies (Sections \ref{s:size_3d} and \ref{s:correlation_functions} for all galaxies, as well as Appendix \ref{a:discs} for disc galaxies).
    \item We measure galaxy two-point correlation functions split by galaxy size at fixed stellar mass in the Sloan Digital Sky Survey (SDSS).  As with previous studies, we find larger clustering differences at lower stellar masses (Section \ref{s:size_results}).
    \item All classic spin-based galaxy size models predict a small-scale upturn in the clustering of large galaxies compared to small galaxies, which is not present in our SDSS measurements (Section \ref{s:size_results}).
    \item The \cite{Jiang19} model predicts much larger clustering differences than observed between large and small galaxies (Section \ref{s:size_results}).
    \item We measure clustering differences between large and small disc galaxies at fixed stellar mass in the SDSS (Appendix \ref{a:discs}), which rules out the hypothesis in \cite{Desmond17} of there being no correlations between galaxy size and halo properties beyond a correlation with galaxy stellar mass.
    \item We provide a simple empirical model that closely matches observed galaxy clustering measurements based on correlating galaxy size with halo growth (Section \ref{s:size_sims}, Appendix \ref{a:halo_growth}).  We also confirm that the model in \citet{Hearin19} provides a reasonable match to our updated clustering data (Appendix \ref{a:orphans}).
\end{itemize}

\section*{Data Availability}

All data for this paper are available \href{https://peterbehroozi.com/data.html}{online}.

\section*{Acknowledgements}

We thank the anonymous referee for suggestions that improved the clarity and presentation of the paper.  We also thank Harry Desmond, Michael Fall, Andrey Kravtsov, and Ari Maller for insightful discussions.

PB was partially funded by a Packard Fellowship, Grant \#2019-69646.
Work done at Argonne National Laboratory was supported under the Department of Energy contract DE-AC02-06CH11357.
BPM acknowledges an Emmy Noether grant funded by the Deutsche Forschungsgemeinschaft (DFG, German Research Foundation) -- MO 2979/1-1.
This material is based upon High Performance Computing (HPC) resources supported by the University of Arizona TRIF, UITS, and RDI and maintained by the UA Research Technologies department.  The University of Arizona sits on the original homelands of Indigenous Peoples (including the \href{http://www.tonation-nsn.gov/history-culture/}{Tohono O'odham} and the Pascua Yaqui) who have stewarded the Land since time immemorial. 

The authors gratefully acknowledge the Gauss Centre for Supercomputing e.V.\ (\href{www.gauss-centre.eu}{www.gauss-centre.eu}) and the Partnership for Advanced Supercomputing in Europe (PRACE, \href{www.prace-ri.eu}{www.prace-ri.eu}) for funding the MultiDark simulation project by providing computing time on the GCS Supercomputer SuperMUC at Leibniz Supercomputing Centre (LRZ, \href{www.lrz.de}{www.lrz.de}).

Funding for the Sloan Digital Sky Survey IV has been provided by the Alfred P.\ Sloan Foundation, the U.S.\ Department of Energy Office of Science, and the Participating Institutions. SDSS-IV acknowledges support and resources from the Center for High Performance Computing  at the University of Utah. The SDSS website is \href{www.sdss.org}{www.sdss.org}.  SDSS-IV is managed by the Astrophysical Research Consortium for the Participating Institutions of the SDSS Collaboration including the Brazilian Participation Group, the Carnegie Institution for Science, Carnegie Mellon University, Center for Astrophysics | Harvard \& Smithsonian, the Chilean Participation Group, the French Participation Group, Instituto de Astrof\'isica de Canarias, The Johns Hopkins University, Kavli Institute for the Physics and Mathematics of the Universe (IPMU) / University of Tokyo, the Korean Participation Group, Lawrence Berkeley National Laboratory, Leibniz Institut f\"ur Astrophysik Potsdam (AIP),  Max-Planck-Institut f\"ur Astronomie (MPIA Heidelberg), Max-Planck-Institut f\"ur Astrophysik (MPA Garching), Max-Planck-Institut f\"ur Extraterrestrische Physik (MPE), National Astronomical Observatories of China, New Mexico State University, New York University, University of Notre Dame, Observat\'ario Nacional / MCTI, The Ohio State University, Pennsylvania State University, Shanghai Astronomical Observatory, United Kingdom Participation Group, Universidad Nacional Aut\'onoma de M\'exico, University of Arizona, University of Colorado Boulder, University of Oxford, University of Portsmouth, University of Utah, University of Virginia, University of Washington, University of Wisconsin, Vanderbilt University, and Yale University.

\bibliography{master_bib} 

\appendix

\section{Alternate Halo Properties}

\label{a:alternate}

\begin{figure*}
\begin{center}
\vspace{-8ex}
\phantom{\hspace{-5ex}}\includegraphics[width=1.6\columnwidth]{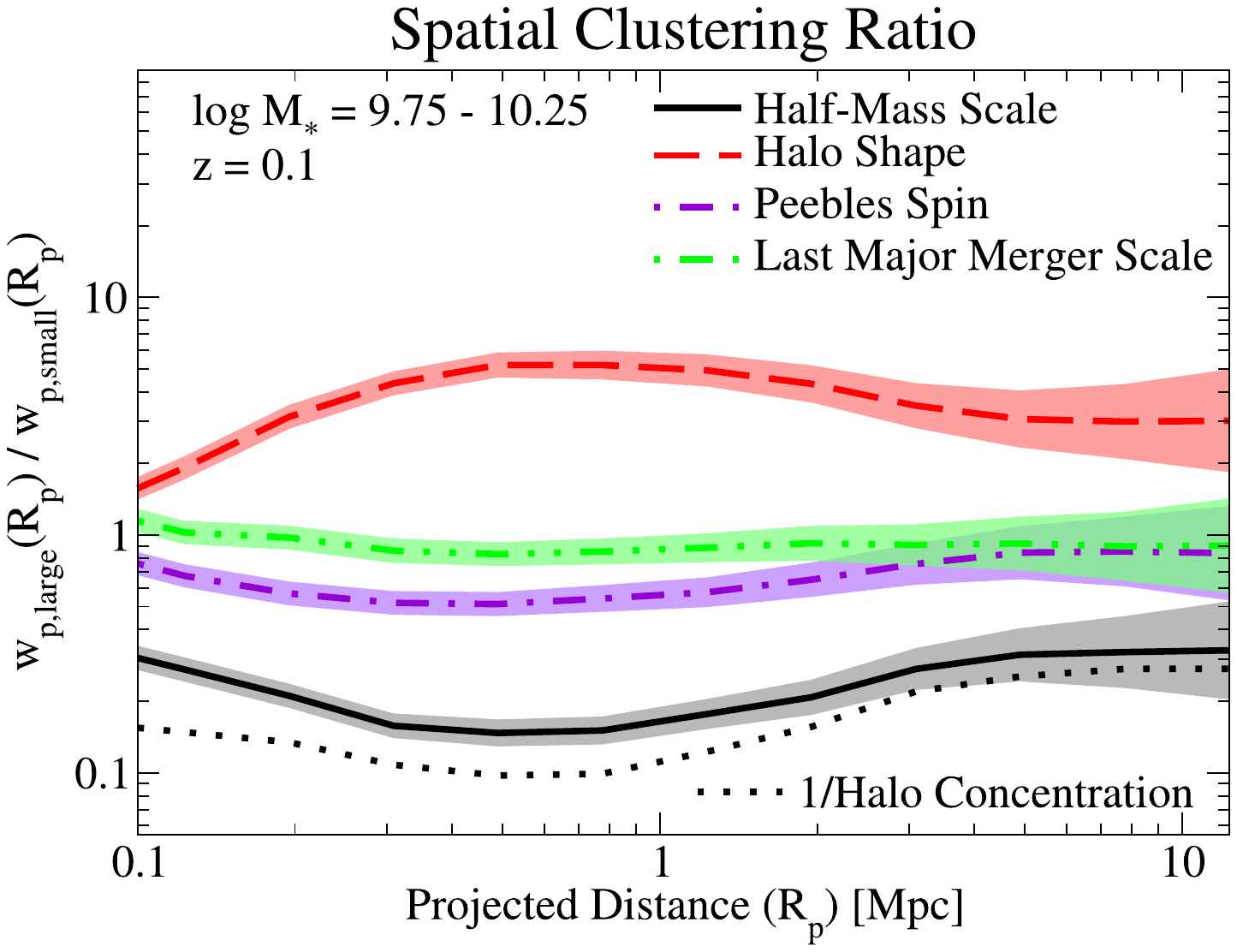}\\[-6ex]
\includegraphics[width=\columnwidth]{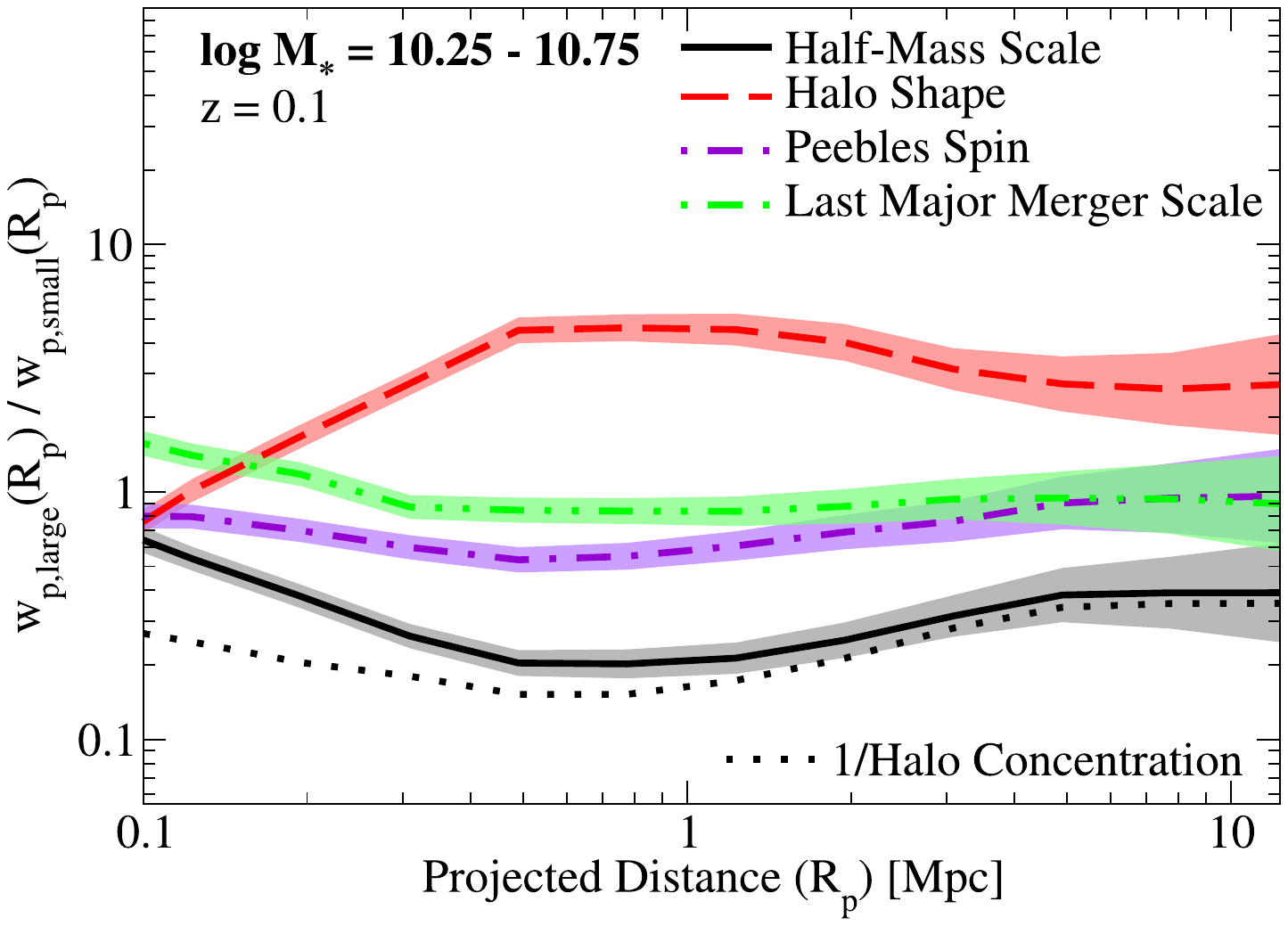}\hspace{-3ex}\includegraphics[width=\columnwidth]{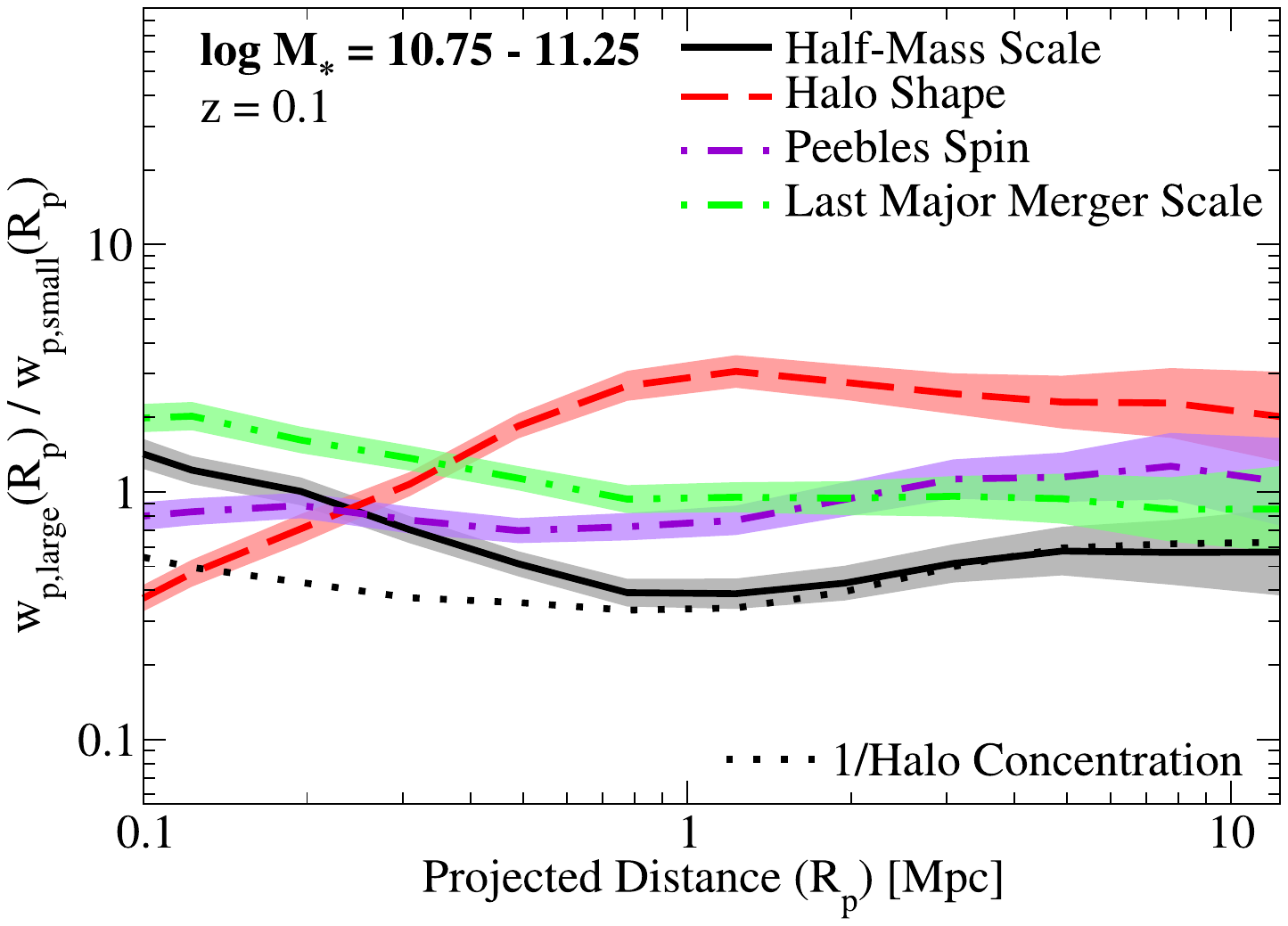}\\[-5ex]
\includegraphics[width=\columnwidth]{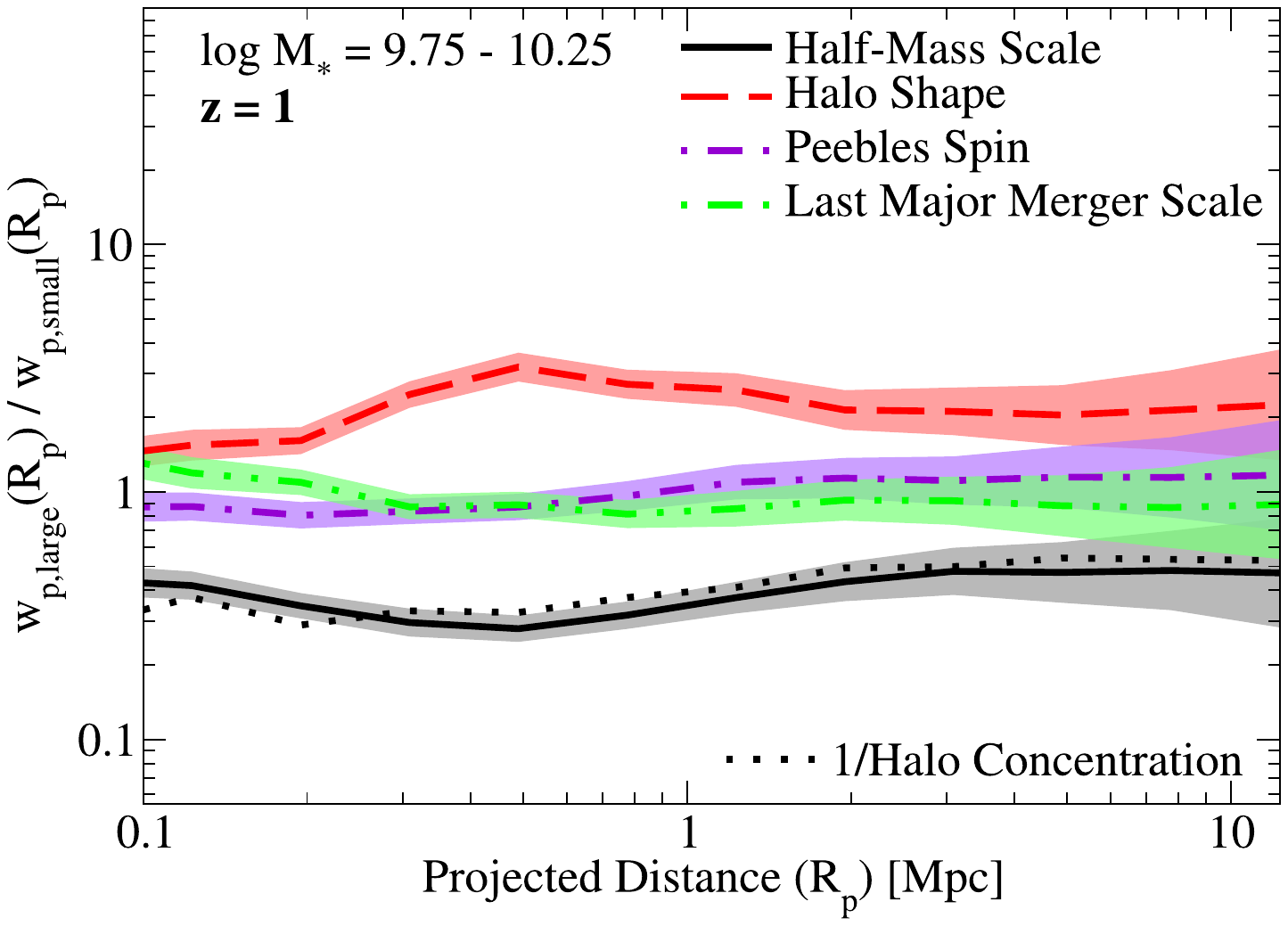}\hspace{-3ex}\includegraphics[width=\columnwidth]{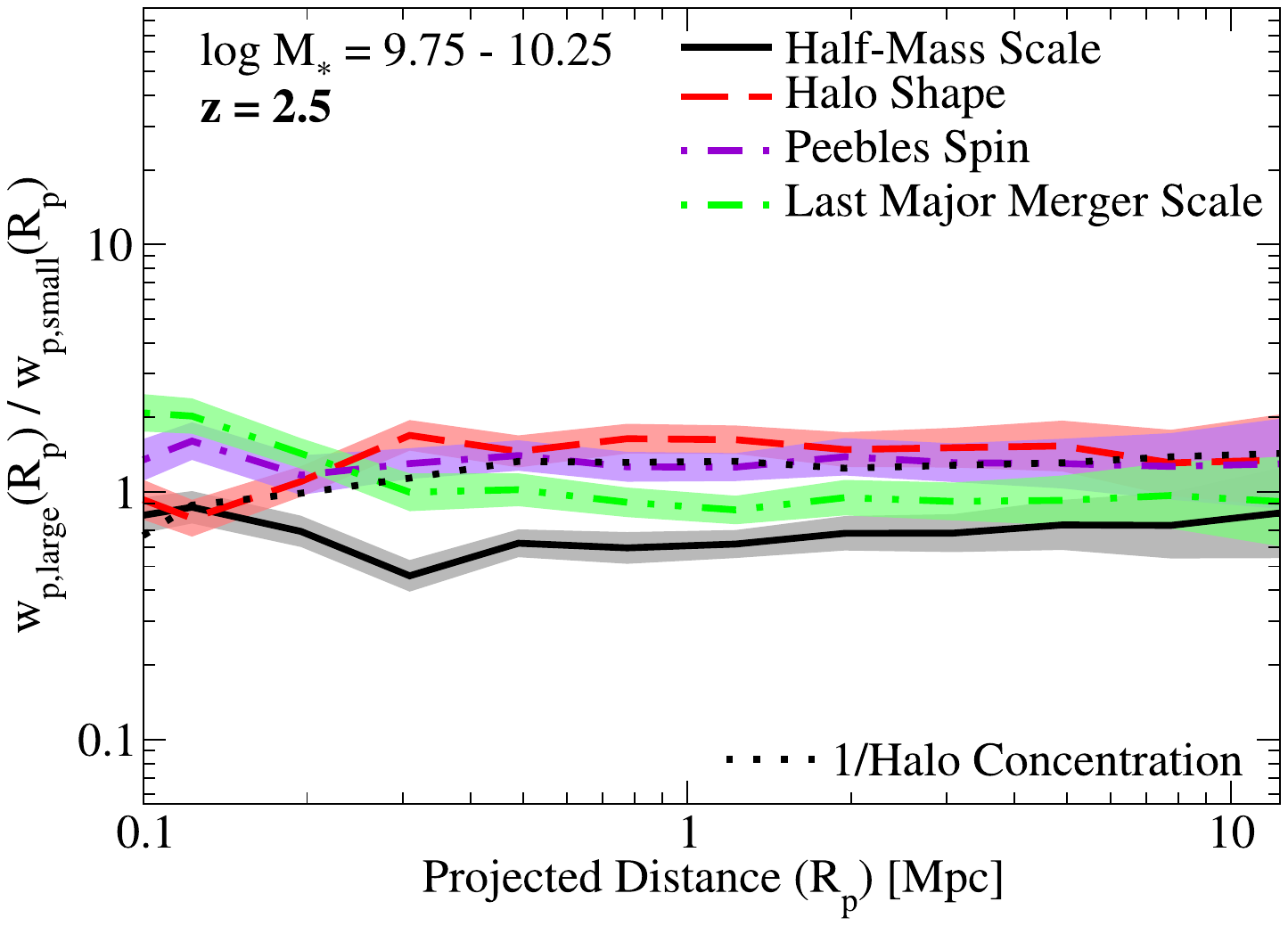}\\[-5ex]
\end{center}
\caption{Ratios of projected two-point correlation functions for galaxies split by host-halo properties. These halo properties include the scale factor at which the halo reached half its peak mass (\textit{Half-Mass Scale}), the shape ratio $\frac{c}{a}$ (\textit{Halo Shape}), \citet{Peebles69} spin, and the scale factor at which the halo had its last major merger (mass ratio of 1:3 or greater; \textit{Last Major Merger Scale}).  The \textit{dotted line} shows the reciprocal of the clustering ratio for halo concentration for ease of comparison with Fig.\ \ref{f:wp_ratios}.  There are scale-dependent clustering differences for each halo property except for the last major merger scale.  The \textbf{top panel} shows simulated clustering ratios for galaxies with $10^{9.75} <  M_\ast / \Msun < 10^{10.25}$ from an abundance-matched catalogue at $z=0.1$ (Section \ref{s:mock}).  The \textbf{middle panels} show clustering at $z=0.1$ for higher-mass galaxies.  The \textbf{bottom panels} show clustering at $z=1$ and $z=2.5$ for galaxies with $10^{9.75} <  M_\ast / \Msun < 10^{10.25}$.  In all panels, shaded regions correspond to $1\sigma$ jackknife uncertainties convolved with 10\% assumed systematic errors; all panels assume $\pi_\mathrm{max}=$125 Mpc/h.}
\label{f:alt_wp_ratios}
\end{figure*}

\begin{figure*}
\begin{center}
\vspace{-8ex}
\phantom{\hspace{-5ex}}\includegraphics[width=1.6\columnwidth]{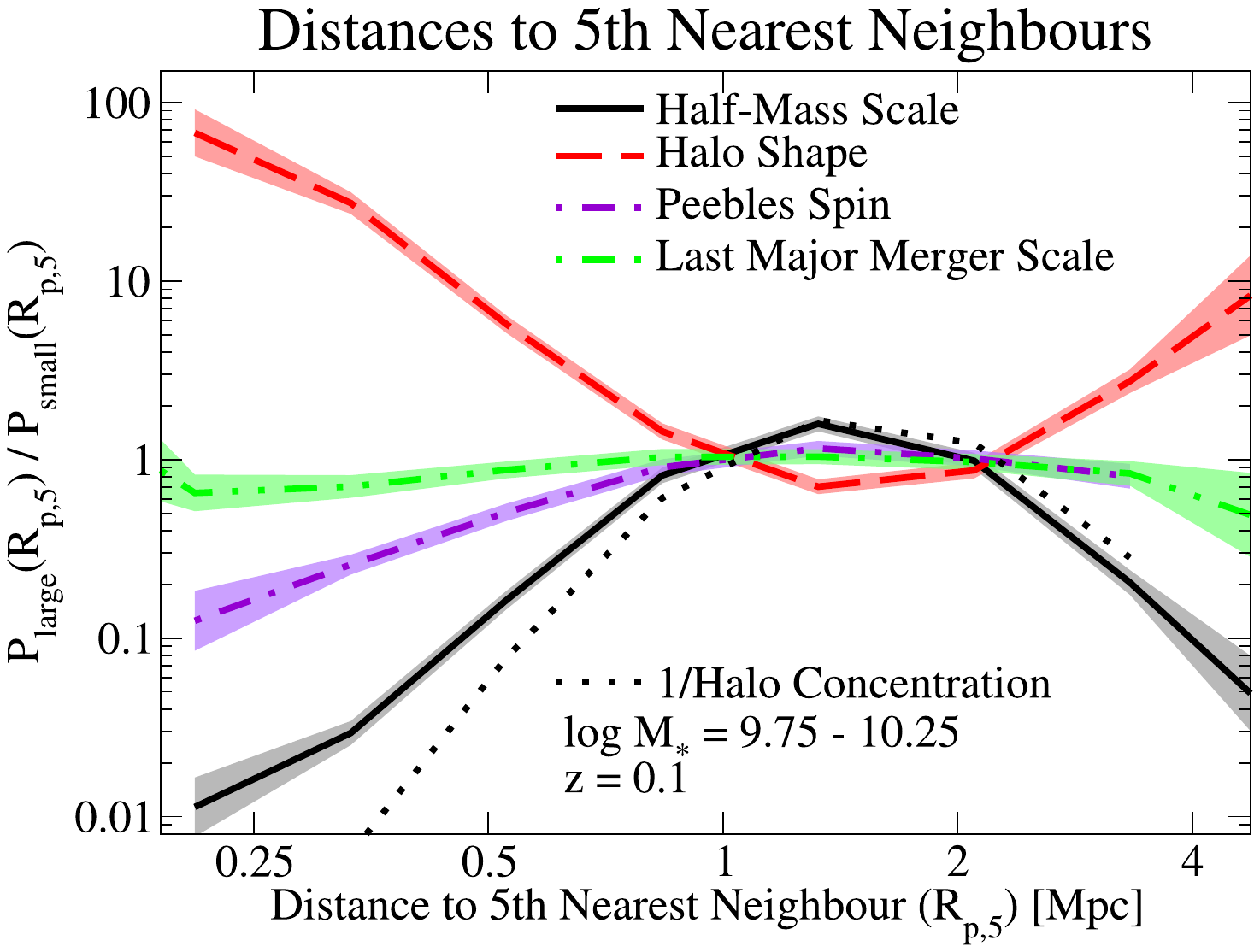}\\[-6ex]
\includegraphics[width=\columnwidth]{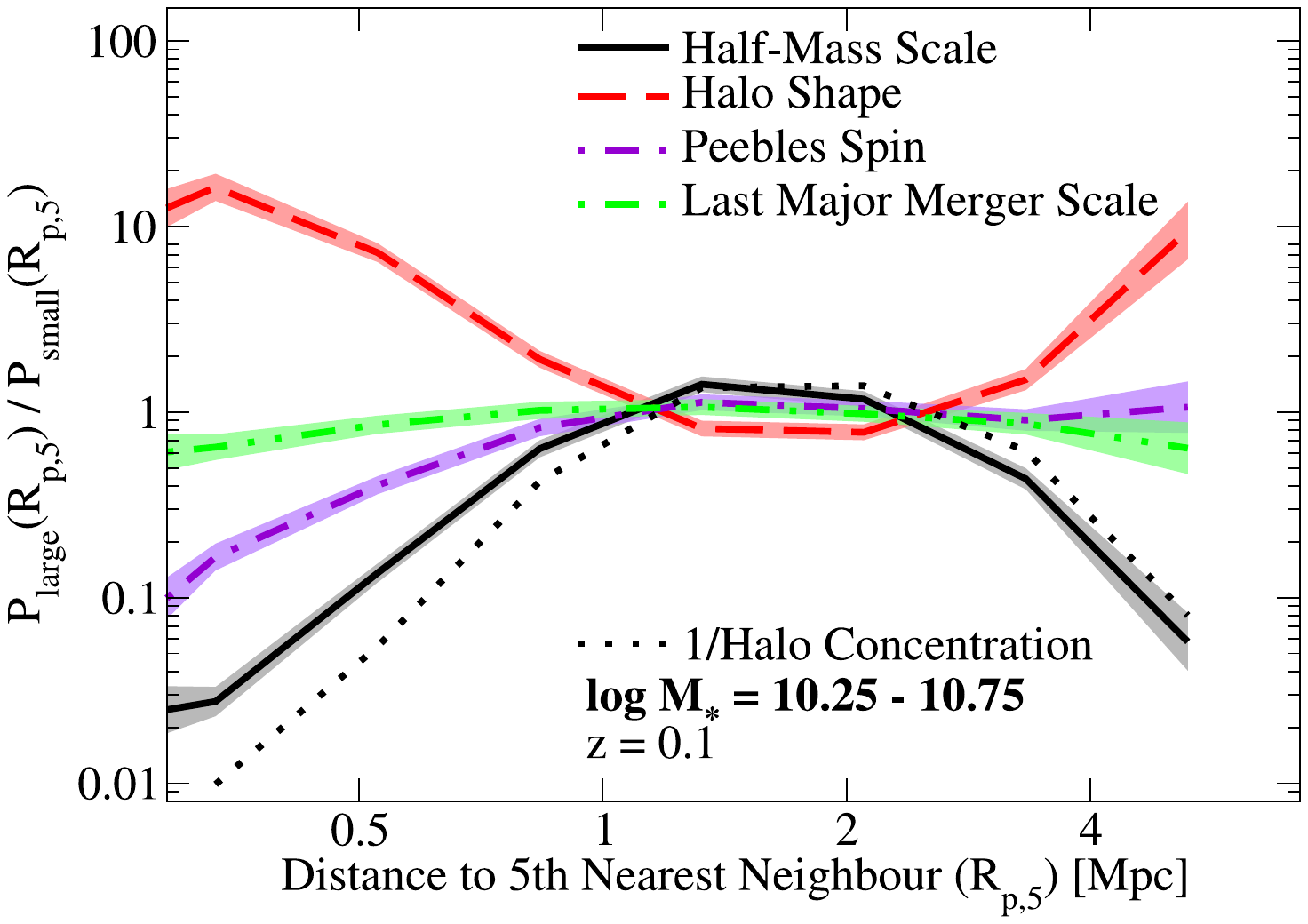}\hspace{-3ex}\includegraphics[width=\columnwidth]{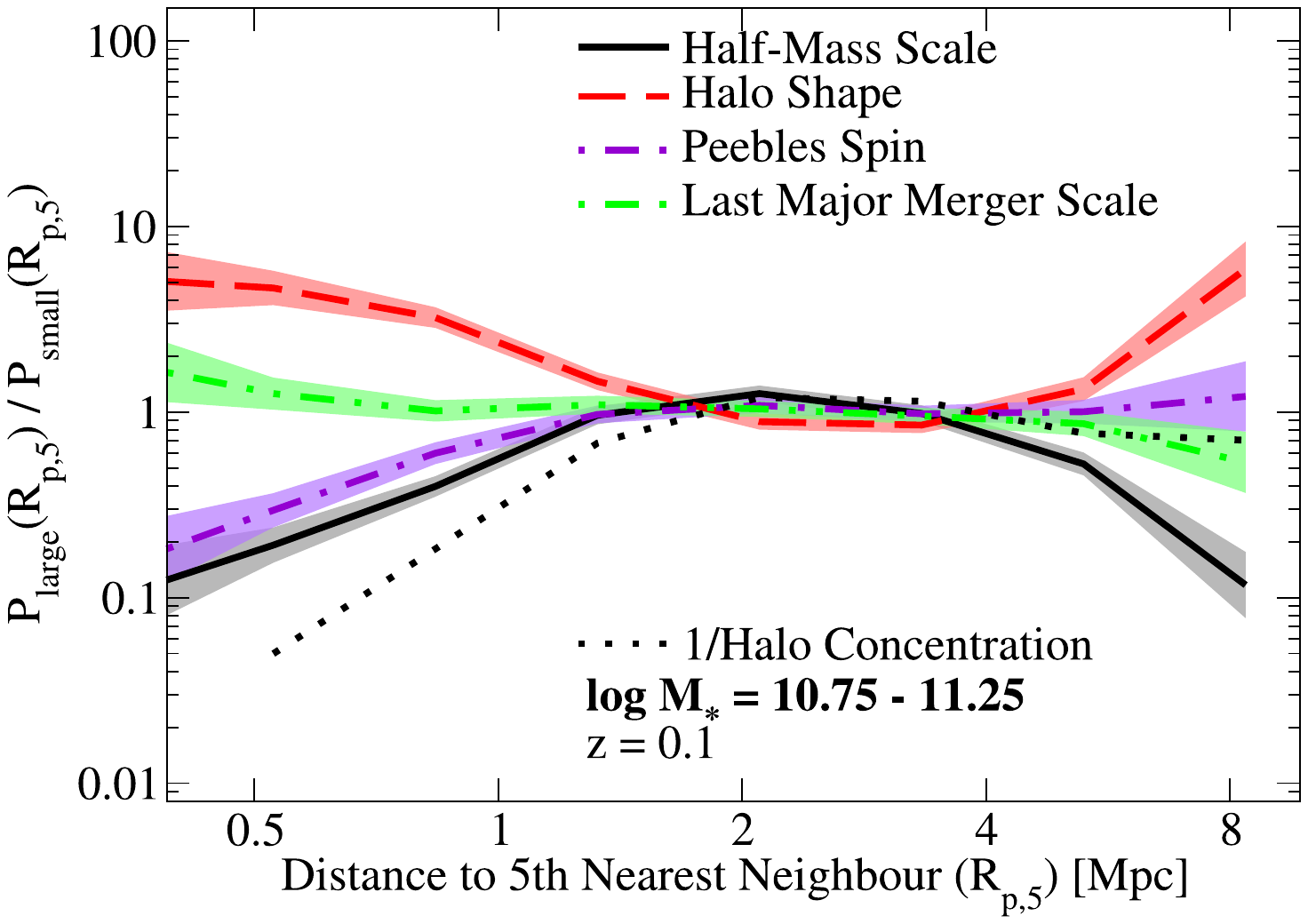}\\[-5ex]
\includegraphics[width=\columnwidth]{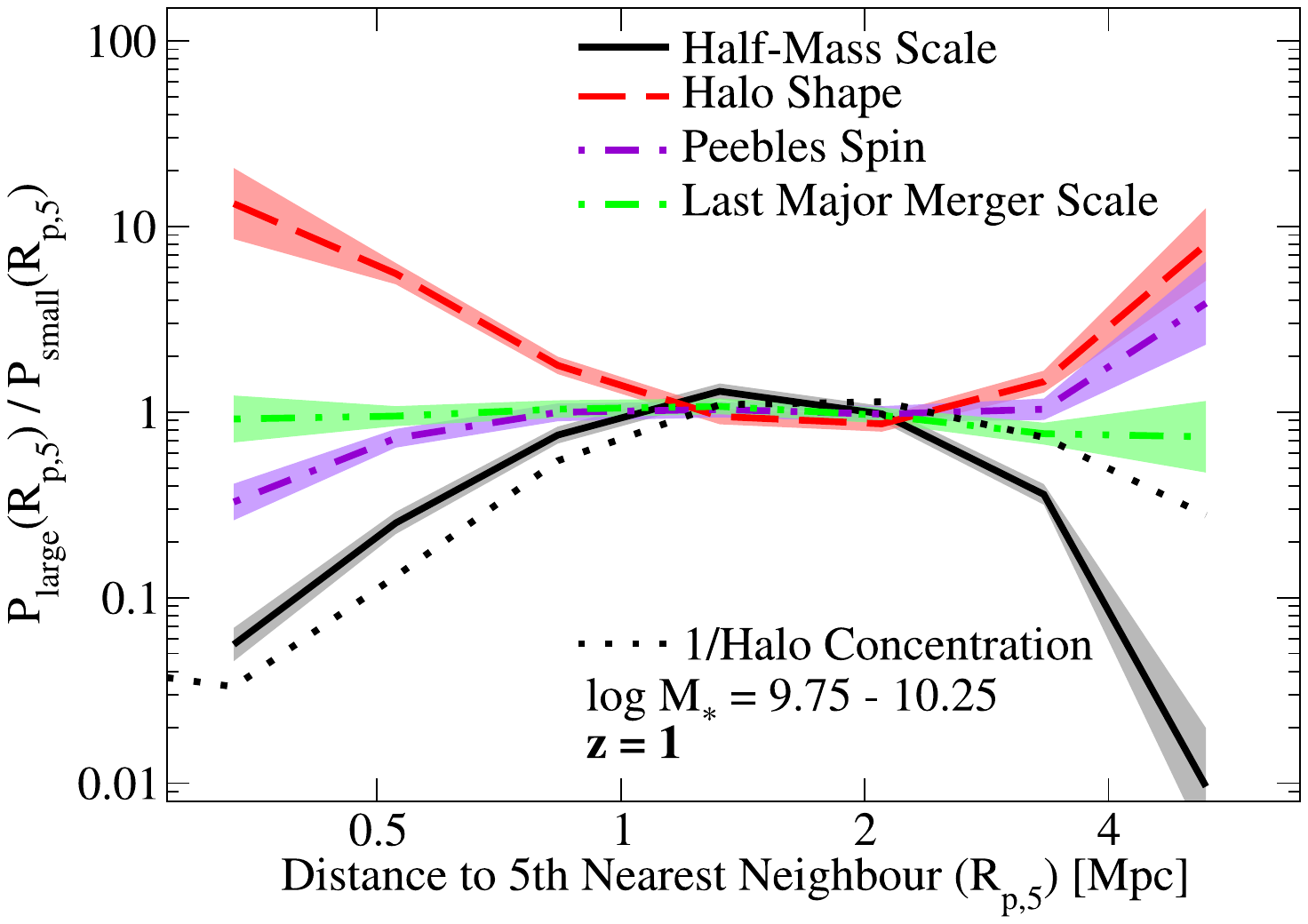}\hspace{-3ex}\includegraphics[width=\columnwidth]{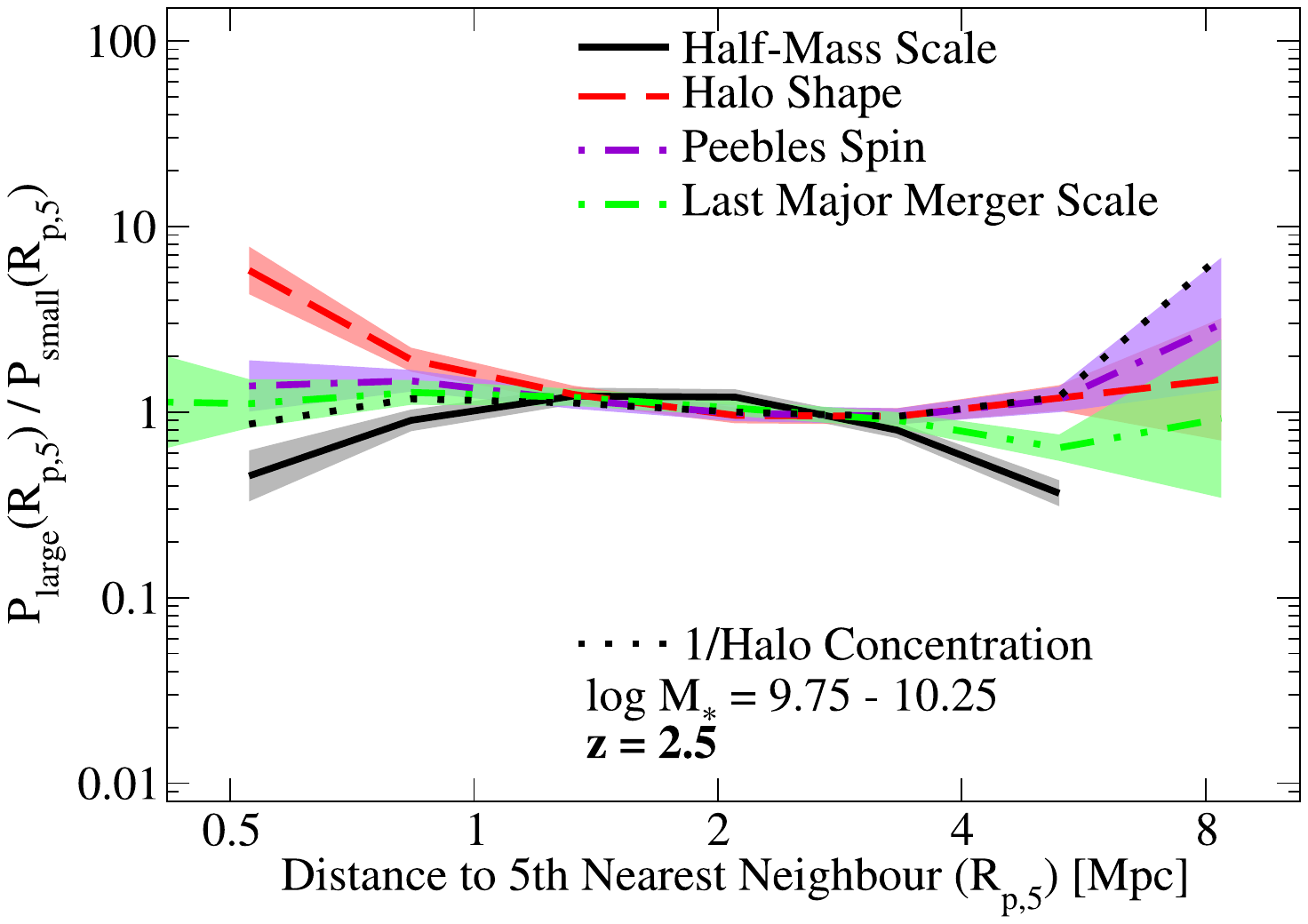}\\[-5ex]
\end{center}
\caption{Ratios of probability distributions of distances to 5$^\mathrm{th}$ nearest within-sample neighbours (5NN PDFs) for galaxies split by host-halo properties.  These halo properties include the scale factor at which the halo reached half its peak mass (\textit{Half-Mass Scale}), the shape ratio $\frac{c}{a}$ (\textit{Halo Shape}), \citet{Peebles69} spin, and the scale factor at which the halo had its last major merger (mass ratio of 1:3 or greater; \textit{Last Major Merger Scale}).  The \textit{dotted line} shows the reciprocal of the clustering ratio for halo concentration for ease of comparison with Fig.\ \ref{f:5nn_ratios}.  There are scale-dependencies for each halo property except for the last major merger scale.   The \textbf{top panel} shows simulated 5NN PDF ratios for galaxies with $10^{9.75} <  M_\ast/\Msun < 10^{10.25}$ from an abundance-matched catalogue at $z=0.1$ (Section \ref{s:mock}).  The \textbf{middle panels} show 5NN PDF ratios at $z=0.1$ for higher-mass galaxies.  The \textbf{bottom panels} show 5NN PDF ratios at $z=1$ and $z=2.5$ for galaxies with $10^{9.75} <  M_\ast/\Msun < 10^{10.25}$.  In all panels, shaded regions correspond to $1\sigma$ jackknife uncertainties convolved with 20\% assumed systematic errors; all panels assume $\pi_\mathrm{max}=$125 Mpc/h.}
\label{f:alt_5nn_ratios}
\end{figure*}

Here, we test the scale-dependence of alternate halo properties including the half-mass scale (i.e., the scale factor at which the halo reached 50\% of $M
_\mathrm{peak}$), the \citet{Peebles69} spin, the shape ratio $\frac{c}{a}$ (i.e., the ratio of smallest to largest principal axes), and the scale factor of the last major halo merger (i.e., with a mass ratio of $1:3$ or larger).  As in the main body, we split haloes into those with above-median and below-median values for these properties.  Fig.\ \ref{f:alt_wp_ratios} shows the resulting ratios for the two-point correlation functions (2PCFs), and Fig.\ \ref{f:alt_5nn_ratios} shows the same for the probability distributions for the distance to the 5$^\mathrm{th}$ nearest neighbour (5NN PDFs).

Halo shape, like $T/|U|$, shows a strong difference between clustering ratios on two-halo scales and on one-halo scales.  Overall, satellites have rounder shapes (higher $\frac{c}{a}$) than central haloes, but the same is not true for satellites that have had recent interactions (i.e., are physically close together).  As a result, haloes with elongated shapes have relatively boosted clustering on very small scales (Fig.\ \ref{f:alt_wp_ratios}).  Overall, the fraction of such interactions is relatively small, so the effect on the 5$^\mathrm{th}$ nearest neighbour distance is also small, and the same effect is not seen in the 5NN PDF ratios (Fig.\ \ref{f:alt_5nn_ratios}).

Interactions also help explain different one-halo vs.\ two-halo clustering ratios for the half-mass scale factor ($a_\mathrm{1/2}$).  The half-mass scale factor is highly anti-correlated with concentration at $z\le 1$  \citep{Wechsler02}, and so they have inverse clustering on two-halo scales.  However, satellite orbits with close passages will lead to eventual disruption of the satellite, imposing a selection effect against haloes with early formation times.  As a result, small-scale clustering is relatively boosted for haloes with larger half-mass scale factors (Fig.\ \ref{f:alt_wp_ratios}).  As with halo shapes, such close interactions between satellites are infrequent, so less of an effect is seen in the 5NN PDF ratios (Fig.\ \ref{f:alt_5nn_ratios}).

The scale factor of the last major merger has relatively little correlation with the 2PCF or the 5NN PDF.  This may seem puzzling, as merger rates overall have strong environmental dependencies \citep{Fakhouri09,Hester10}.  However, major mergers are rare enough that the environment at the time the last merger happened may have nothing to do with the current environment of the halo.  For example, the last major merger for most current satellite haloes occurred when the satellite was more than four virial radii away from its current host halo \citep{BehrooziMergers}.  As a result, detecting haloes with recent major mergers is best done with a different observational statistic, such as close pairs \citep[e.g.,][]{BehrooziMM}.

Lastly, the environmental dependence of the Peebles spin is weaker but broadly similar to that of the Bullock spin.  Although the angular momentum $J_h$ is boosted by close interactions (leading to twin bias), so also is the ratio of kinetic to potential energy, $T/|U|$.  As a result, $|E_h| = |U|-T$ drops, partially compensating for the increase in $J_h$.  The compensation is not perfect, so the same U-shaped clustering ratios result for the Peebles spin (compare Fig.\ \ref{f:alt_wp_ratios} to Fig.\ \ref{f:wp_ratios}), and similar shapes are seen in the 5NN PDF ratios (compare Fig.\ \ref{f:alt_5nn_ratios} to Fig.\ \ref{f:5nn_ratios}).

\section{Observed Probability Distributions for Distances to 5$^\mathrm{th}$ Nearest Neighbours}

\label{a:obs_5nn}

\begin{figure}
\begin{center}
\vspace{-8ex}
\phantom{\hspace{-5ex}}\includegraphics[width=1.1\columnwidth]{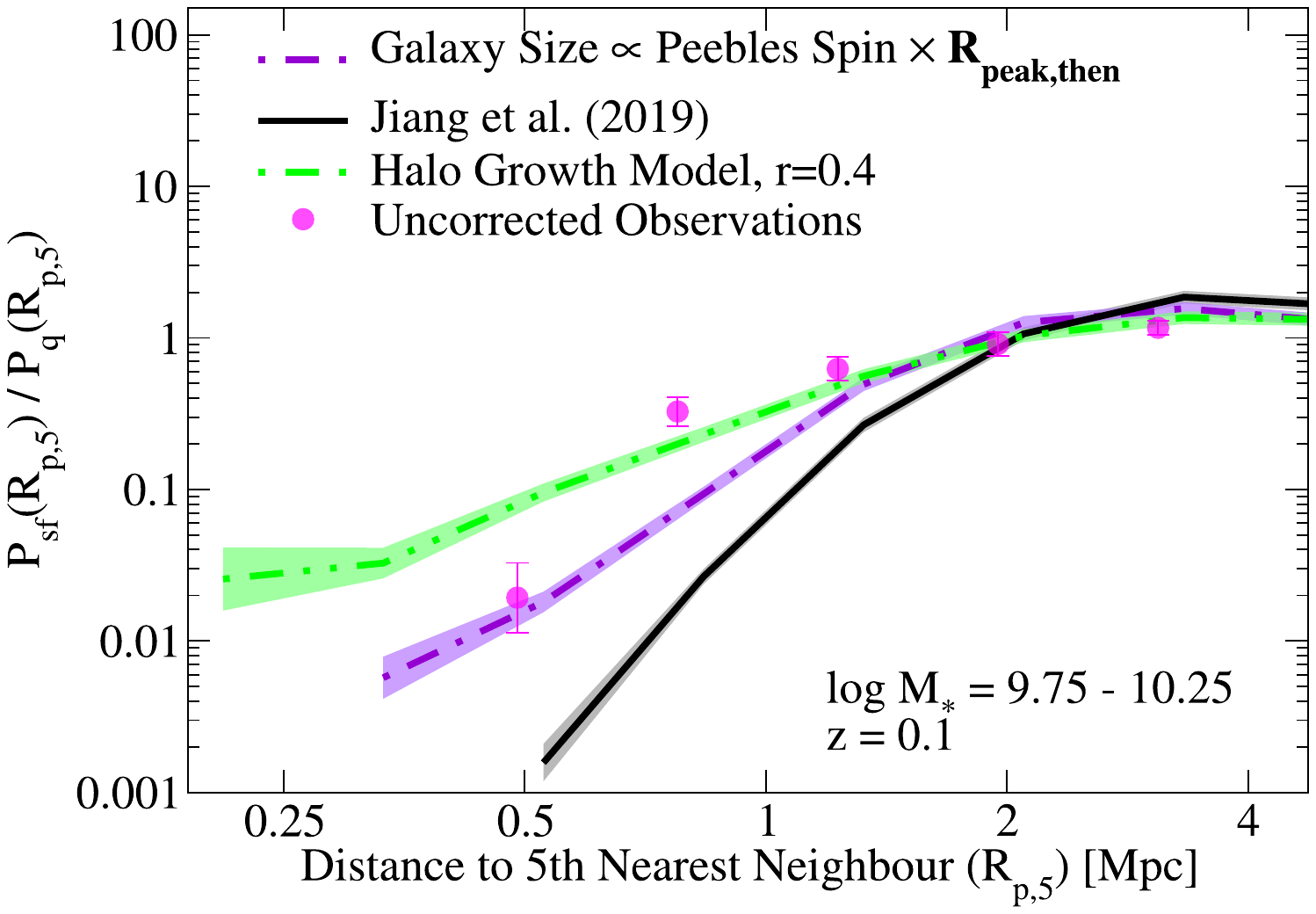}\\[-7ex]
\phantom{\hspace{-5ex}}\includegraphics[width=1.1\columnwidth]{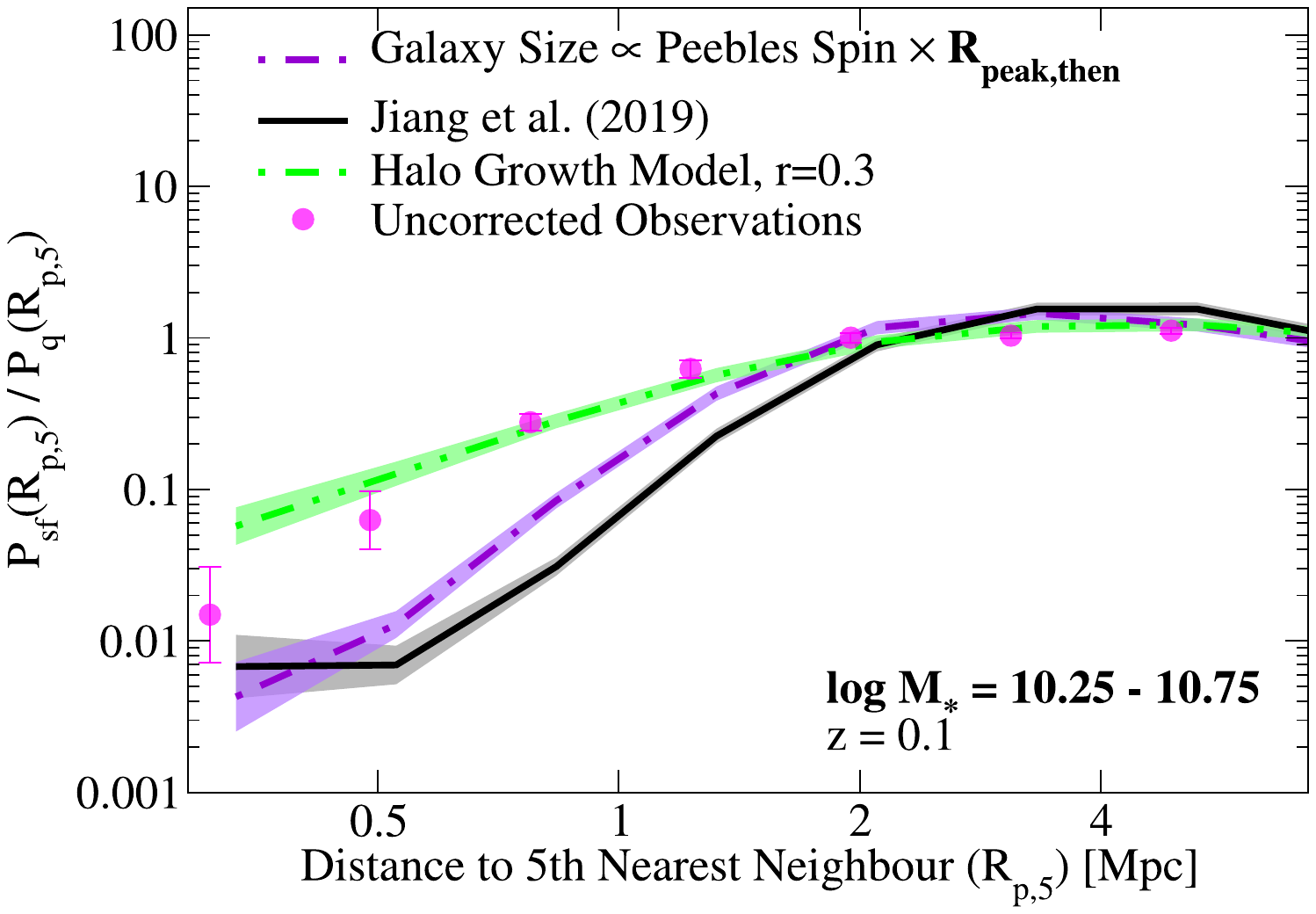}\\[-7ex]
\phantom{\hspace{-5ex}}\includegraphics[width=1.1\columnwidth]{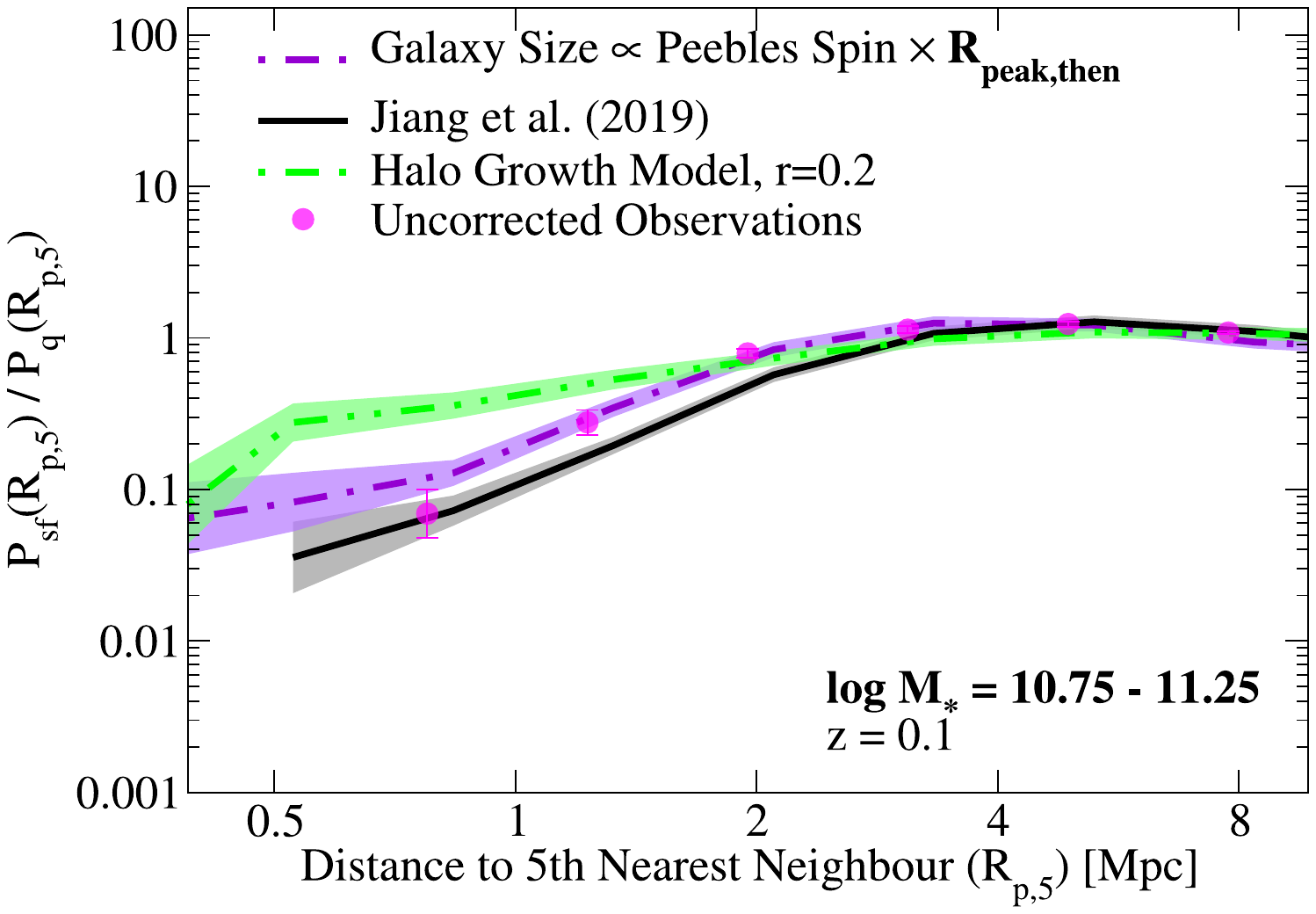}\\[-7ex]
\end{center}
\caption{Ratios of probability distributions of distances to 5$^\mathrm{th}$ nearest within-sample neighbours (5NN PDFs) for large vs.\ small galaxies in both observations and simulations, in several bins of stellar mass.  Galaxy projected half-light sizes in observations have had corrections applied from Section \ref{s:size_3d} to approximate 3D half-mass sizes.  All panels show models where galaxy sizes are correlated with the halo growth rate (\textit{green line}), a model where galaxy size is proportional to the Peebles spin (\textit{purple line}) as well as the \citet{Jiang19} concentration-based model (\textit{black line}).  \textbf{Note that observational results have not been corrected for either fibre collisions or survey masking} (see Appendix \ref{a:obs_5nn}).  In all cases, error bars and shaded regions correspond to jackknife uncertainties.  Results for both observations and simulations are computed with $\pi_\mathrm{max}=13.6$ Mpc $h^{-1}$.}
\label{f:obs_5nn_ratios}
\end{figure}

Given the complexity of building an accurate mock catalogue accounting for both fibre collisions and masking for the SDSS, we do not present observed probability distributions for 5$^\mathrm{th}$ nearest neighbours in the main paper body.  We can nonetheless qualitatively compare the (uncorrected) observed neighbour distributions to the model predictions to demonstrate that: 1) our assumed errors are reasonable, and 2) the observations contain sufficient information to motivate more careful modeling of observational systematics in future work.

To calculate 5$^\mathrm{th}$ nearest neighbour distance distributions (5NN PDFs) in the SDSS, we use a modified version of the \texttt{correl} routine from the \textsc{UniverseMachine}.  This code calculates the $5^\mathrm{th}$ nearest within-sample neighbours for a specified galaxy sample, using the same definition of projected distance as in Section \ref{s:correlation_functions}.  As for observed galaxy 2PCFs, we divide galaxies into ``large'' and ``small'' based on their deprojected 3D size (Section \ref{s:size_3d}), and we use $\pi_\mathrm{max}=13.6$ Mpc/h as the maximum redshift-space separation for neighbours.  We use the same jackknife sampling for 5NN PDFs as for 2PCFs.

As mentioned, we do not attempt to make corrections for survey masking or for fibre collisions for the observed 5NN PDFs.  Not correcting for sky masking will mean that galaxies near boundaries will only have half or fewer of their neighbours within the survey footprint.  This most significantly increases the probabilities of very large 5NN separations for galaxies in voids (i.e., it increases $P(R_{p,5})$ for large $R_{p,5}$).  Not correcting for fibre collisions implies that galaxies in dense environments will have artificially boosted 5NN separations.  As a result, the probability of having very low 5NN separations will decrease, and $P(R_{p,5})$ will steepen at small $R_{p,5}$.

Fig. \ref{f:obs_5nn_ratios} shows results for observed 5NN PDFs.  We find that the observational data show trends that are similar in magnitude and shape to multiple models' predictions.  In particular, differences of over 1 dex in the 5NN PDFs for large and small galaxies are seen in all mass bins, which are well beyond the statistical error bars.  No single model matches all the data well, at least in part because of unmodeled observational systematics.  Nonetheless, these results suggest that it is promising to attempt more systematic forward modeling of 5NN PDFs so that quantitative comparisons can be made.

\section{Disc Galaxies}

\label{a:discs}

\begin{figure*}
\begin{center}
\vspace{-8ex}
\phantom{\hspace{-5ex}}\includegraphics[width=1.1\columnwidth]{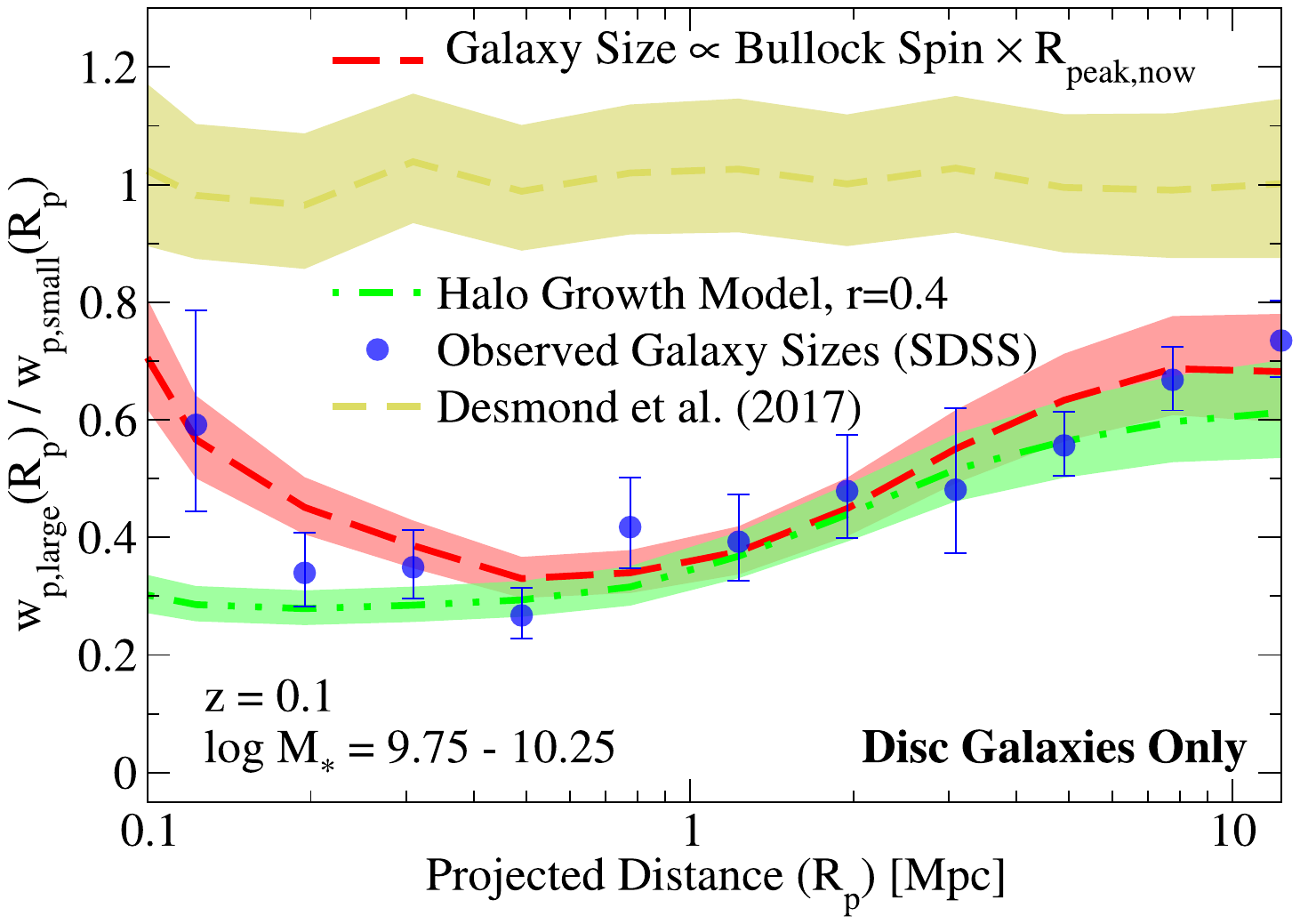}\hspace{-6ex}\includegraphics[width=1.1\columnwidth]{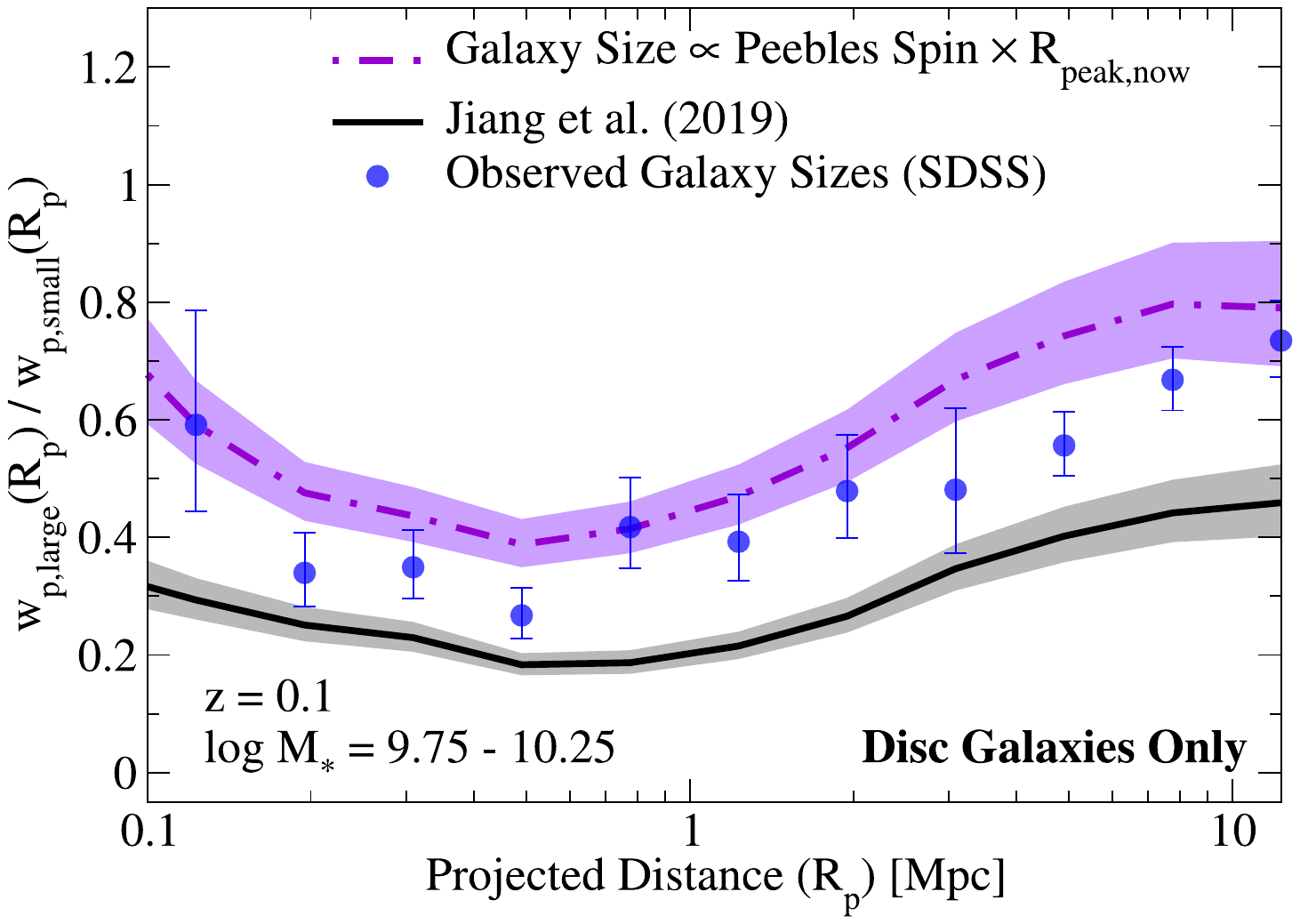}
\vspace{-3ex}
\phantom{\hspace{-5ex}}\includegraphics[width=1.1\columnwidth]{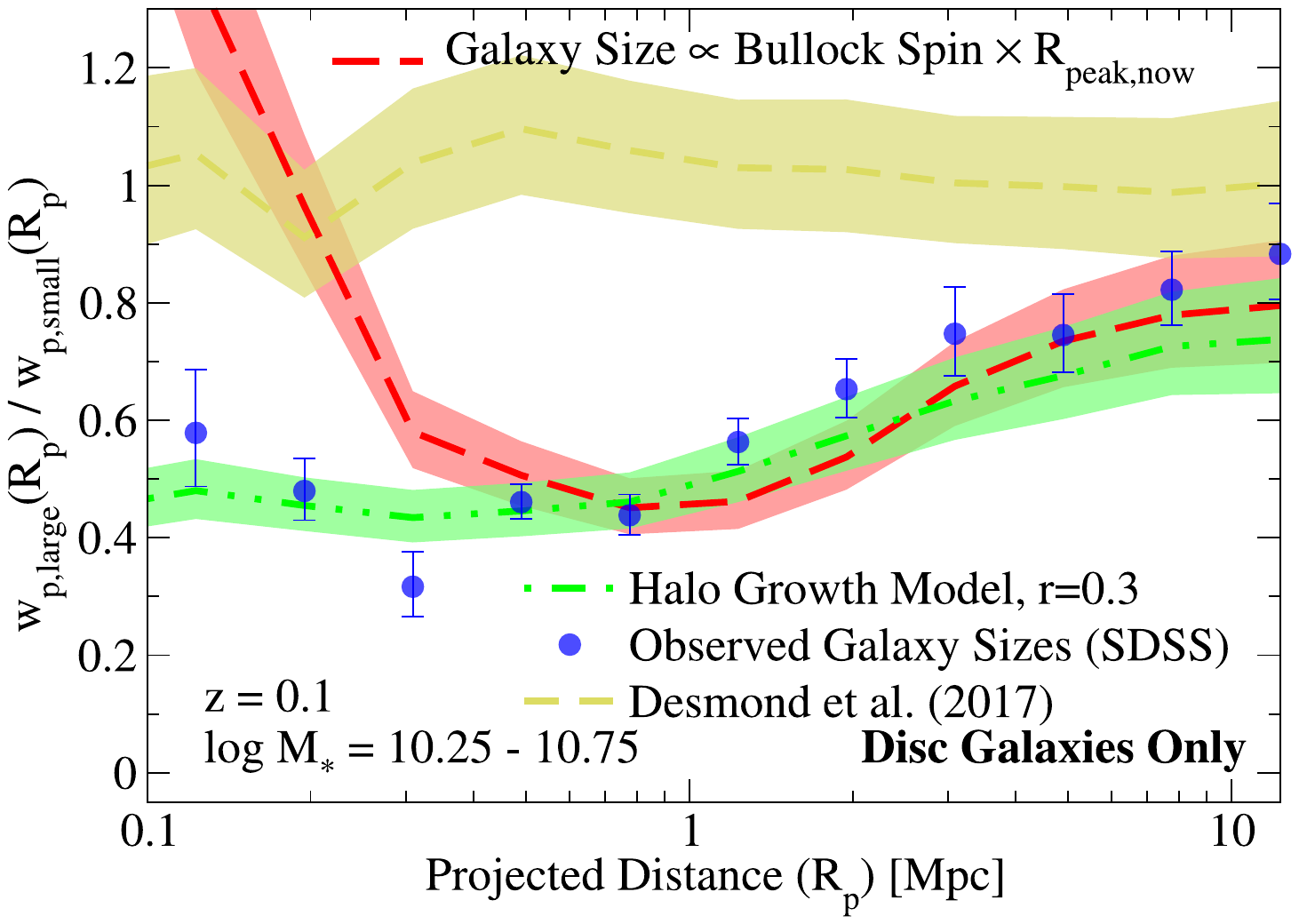}\hspace{-6ex}\includegraphics[width=1.1\columnwidth]{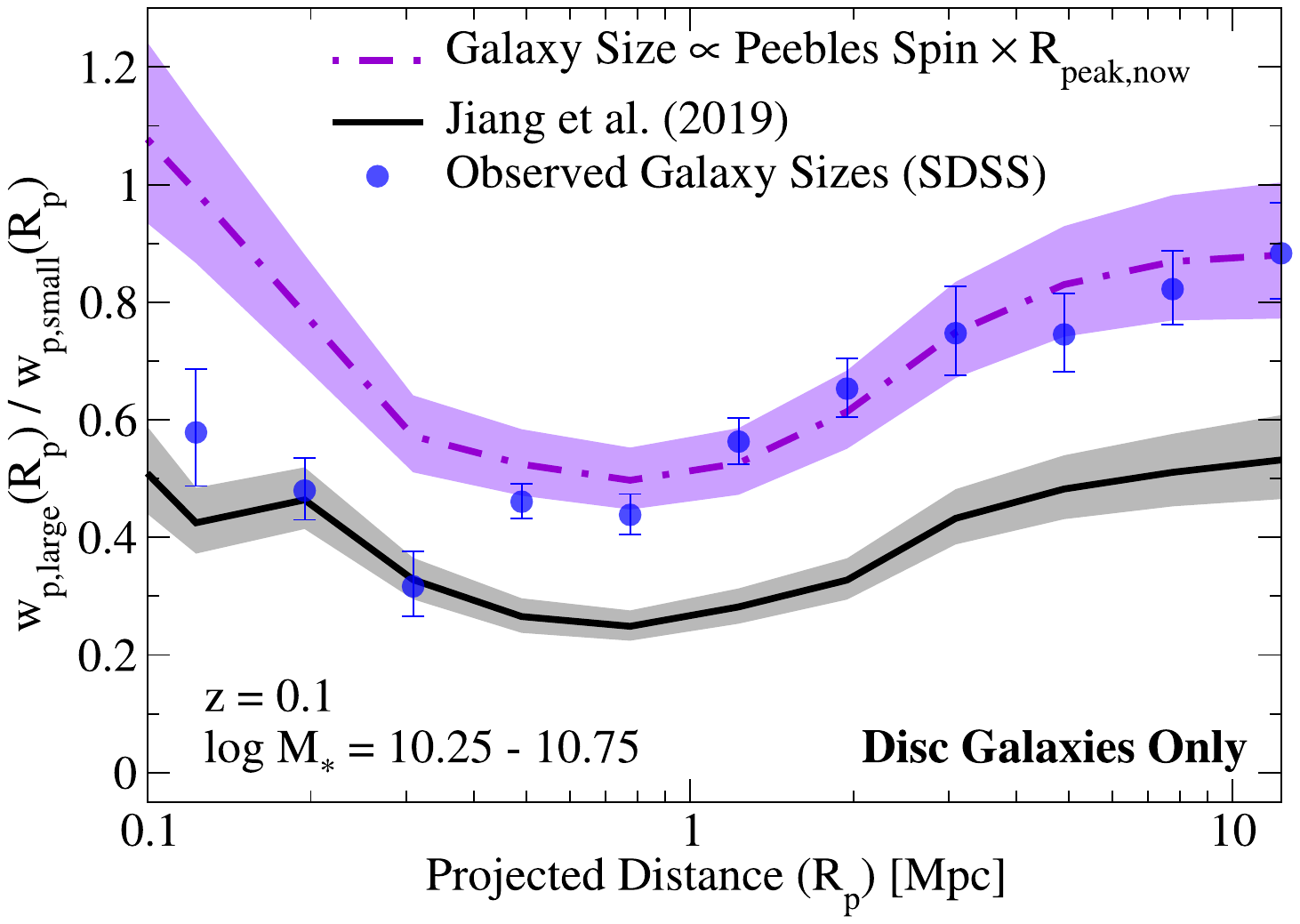}
\vspace{-3ex}
\phantom{\hspace{-5ex}}\includegraphics[width=1.1\columnwidth]{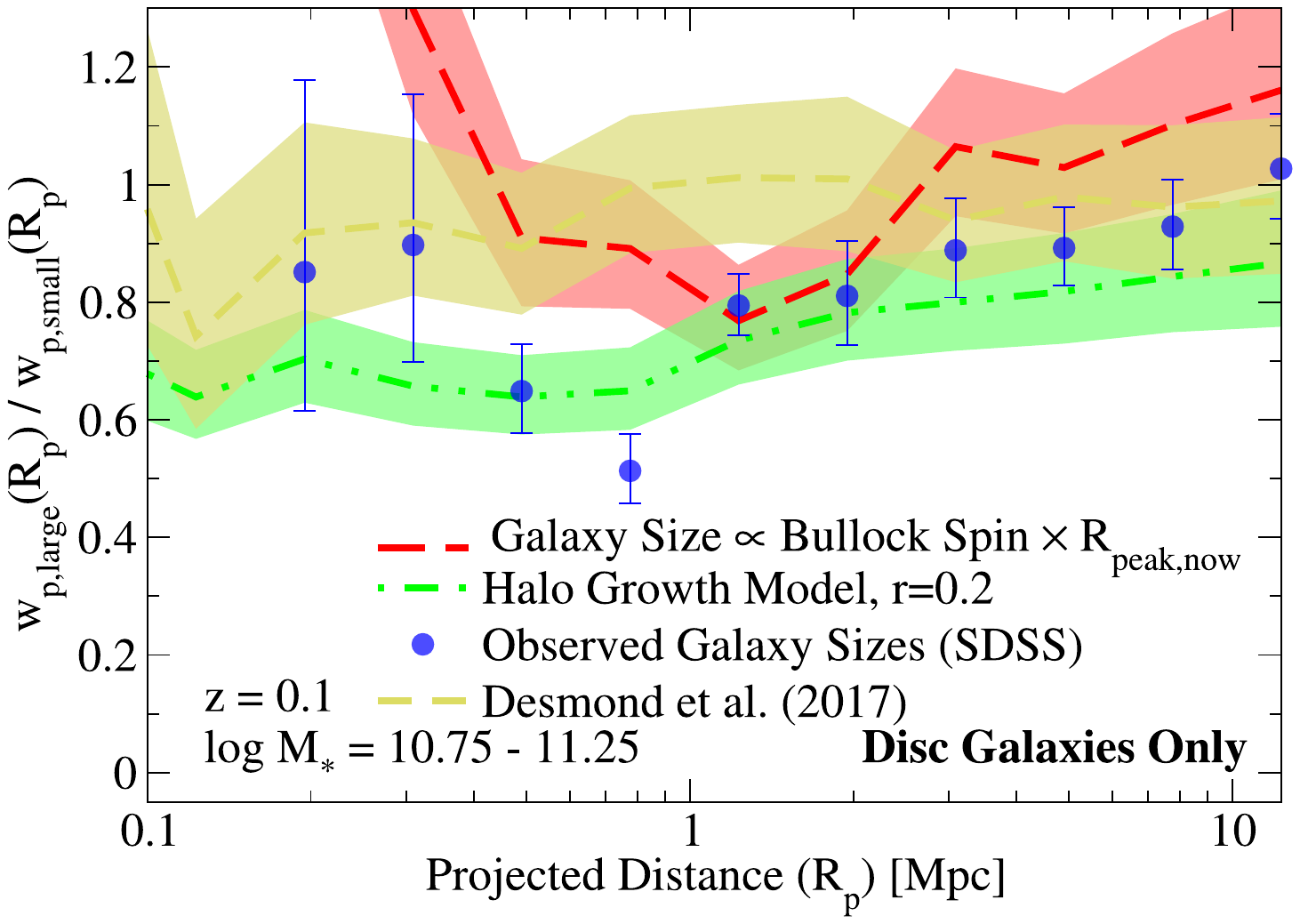}\hspace{-6ex}\includegraphics[width=1.1\columnwidth]{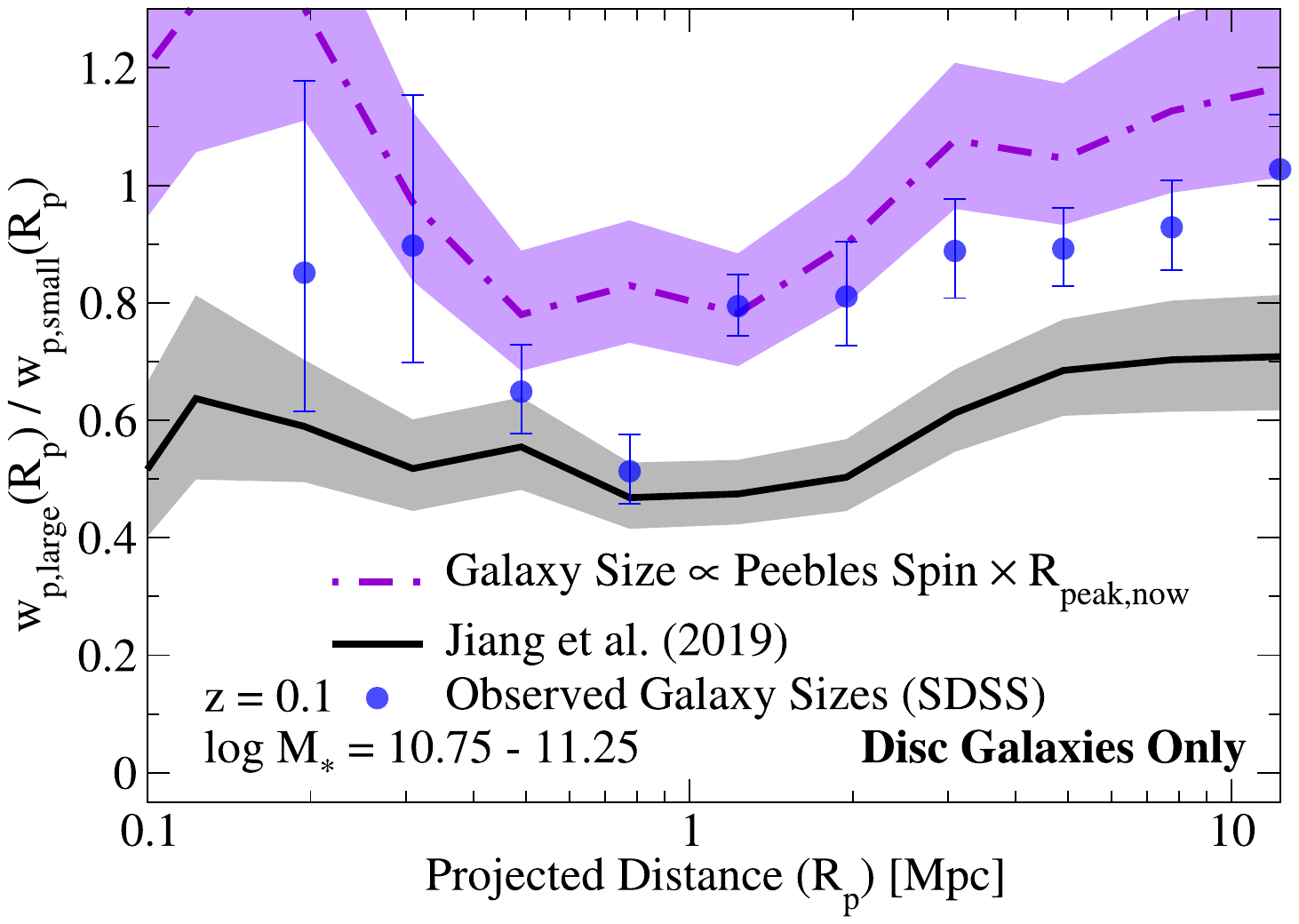}
\vspace{-3ex}
\end{center}
\caption{Ratios of projected two-point correlation functions for large vs.\ small \textbf{disc galaxies} in both observations and simulations, in several bins of stellar mass.  Galaxy projected half-light sizes in observations have had corrections applied from Section \ref{s:size_3d} to approximate 3D half-mass sizes.  \textbf{Left} panels show models where galaxy sizes are proportional to the Bullock spin (\textit{red} line) as well as to the halo accretion rate (\textit{green line}); \textbf{Right} panels show a model where galaxy size is proportional to the Peebles spin (\textit{purple line}) as well as the \citet{Jiang19} concentration-based model (\textit{black line}).  The halo growth model gives the closest predictions to the SDSS observations (\textit{blue points}).  For both spin-based models, the halo radius used is the present-day radius of a halo with the same peak mass.  In all cases, error bars and shaded regions correspond to jackknife uncertainties.  In contrast to previous plots, two-point correlation functions here for both observations and simulations are computed with $\pi_\mathrm{max}=13.6$ Mpc $h^{-1}$.}
\label{f:disc_corrs}
\end{figure*}

For models in which sizes are only assigned to disc galaxies, there is a dilemma: these models do not predict \textit{which} galaxies are disc galaxies, and hence represent incomplete---and therefore untestable---descriptions of galaxy formation.  We can evaluate these models only by choosing how to label haloes as hosting disc or elliptical galaxies.  It is beyond the scope of this paper to evaluate all such choices, so we adopt one such choice and leave the task of evaluating other choices to interested readers.  Specifically, we assume that the probability of being a disc galaxy depends only on the galaxy's stellar mass and specific star formation rate (SSFR).

In our mock catalogues, we assign ``disc'' vs.\ ``elliptical'' labels in narrow bins of stellar mass via conditional abundance matching.  Within each stellar mass bin, haloes are sorted in rank order of observed SSFR, which is assigned by the \textsc{UniverseMachine} \citep{BWHC19}.  Observed galaxies are similarly sorted in rank order of descending observed SSFR as measured in the SDSS \citep{Brinchmann04}.  Observed galaxies are then assigned to haloes in this same rank order, so that every halo receives an SDSS SSFR as well as a \cite{Meert15} bulge/total ratio.  By definition, this preserves the fraction of disc galaxies as a function of observed SSFR and stellar mass.  We classify galaxies (both observed and simulated) as ``discs'' if their bulge fractions are less than 50\%.

We generate sizes for galaxies within the disc population exactly as in Section \ref{s:size_sims}.  In addition, for the \textbf{Desmond et al. (2017) Model}, we test the hypothesis in \cite{Desmond17} that galaxy sizes do not correlate with any property beyond galaxy stellar mass; we do so by randomly assigning labels of ``large'' or ``small'' to haloes.  We classify observed galaxies as ``large'' or ``small'' based on whether they are above or below the following median relation that we fitted to the distribution of observed disc galaxy sizes:
\begin{equation}
    R_\mathrm{3D,discs}(M_*) = 2.67\left[\left(\frac{M_*}{10^{10.95} \Msun}\right)^{0.822} + \left(\frac{M_*}{10^{10.95} \Msun}\right)^{0.037}\right]\;\mathrm{kpc}. \label{e:size_fit_discs}
\end{equation}

Comparisons between models and observations are shown in Fig.\ \ref{f:disc_corrs}.  Most notably, the observations show a strong clustering difference between large and small galaxies at $M_\ast < 10^{10.75}\Msun$.  This rules out the null hypothesis in \cite{Desmond17} that galaxy sizes correlate only with galaxy stellar mass and do not correlate with any other halo properties.  As with Fig.\ \ref{f:size_obs}, there is a strong upturn in the clustering ratio at small scales for all spin-based models.  This prevents any spin-based model from accurately describing clustering for Milky Way-mass galaxies ($10^{10.25}<M_\ast < 10^{10.75}\Msun$).  However, the Bullock spin model does provide a good description of clustering for disc galaxies in our lowest mass bin ($10^{9.75}-10^{10.25}\Msun$).  As with Fig.\ \ref{f:size_obs}, the \cite{Jiang19} model continues to predict much larger clustering differences than actually observed.  The proposed empirical model in Appendix \ref{a:halo_growth} continues to provide a good match to the observed clustering for all mass bins.

\section{Halo Radii at the Time of Peak Mass}

\label{a:rthen}

As described in Section \ref{s:size_sims}, there is no universally-accepted definition for satellite radii.  Here, we use an alternate halo radius definition, $R_\mathrm{peak,then}$, which is the physical radius at the time a halo reaches its peak mass.  This radius can be much smaller than $R_\mathrm{peak,now}$ because the average halo density $\rho_\mathrm{vir}$ increases strongly with redshift.  As a result, using $\lambda\cdot R_\mathrm{peak,then}$ gives smaller satellite radii than using $\lambda \cdot R_\mathrm{peak, now}$.  When splitting galaxies by size, there will hence be a larger proportion of satellites among the galaxies with smaller sizes.  This causes smaller galaxies to have increased clustering both in the one-halo and two-halo regimes (e.g., in the two-halo regime, finding a galaxy in a given location means that there are likely to be satellites nearby in surrounding haloes).

\begin{figure*}
\begin{center}
\vspace{-8ex}
\phantom{\hspace{-5ex}}\includegraphics[width=1.1\columnwidth]{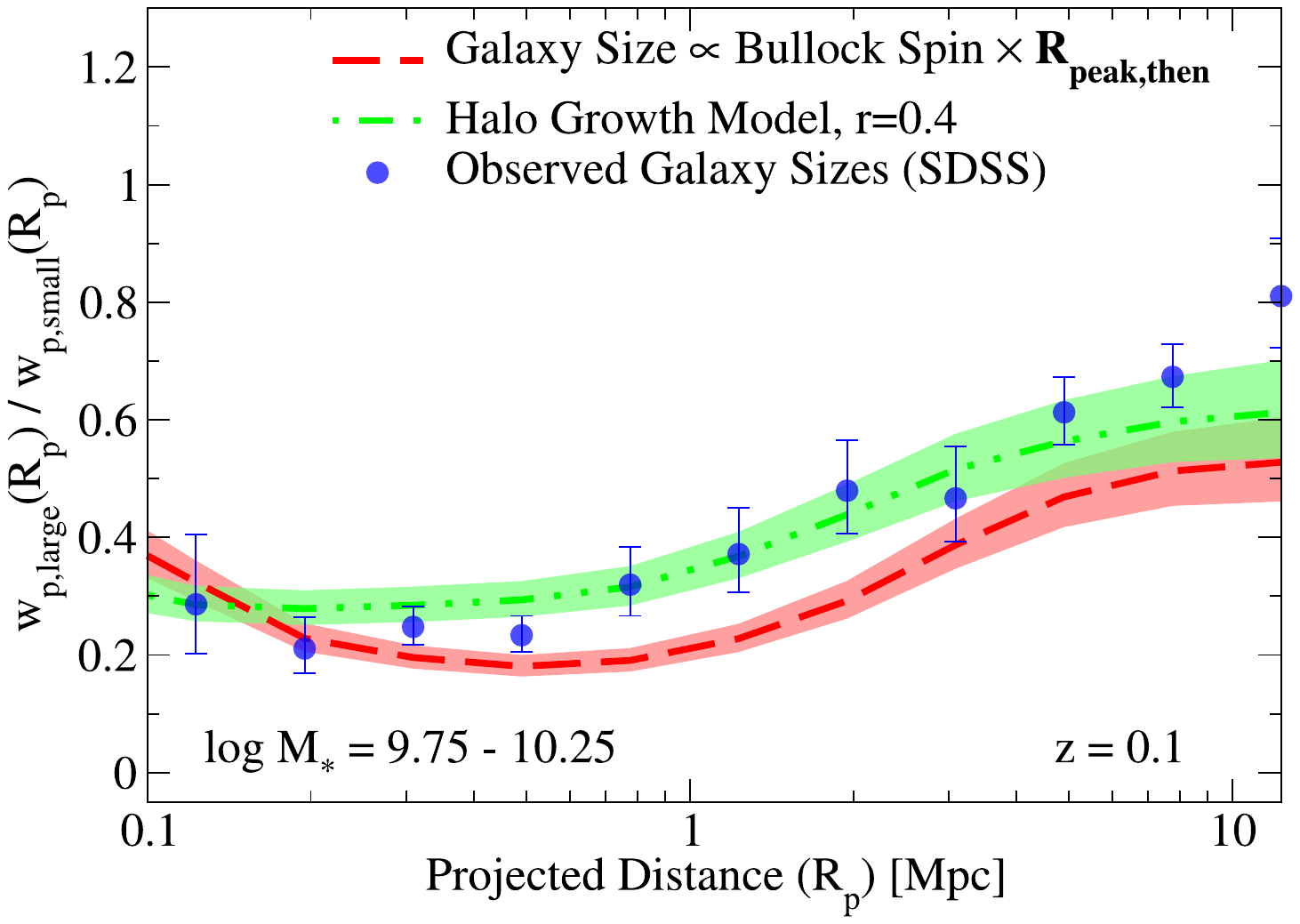}\hspace{-6ex}\includegraphics[width=1.1\columnwidth]{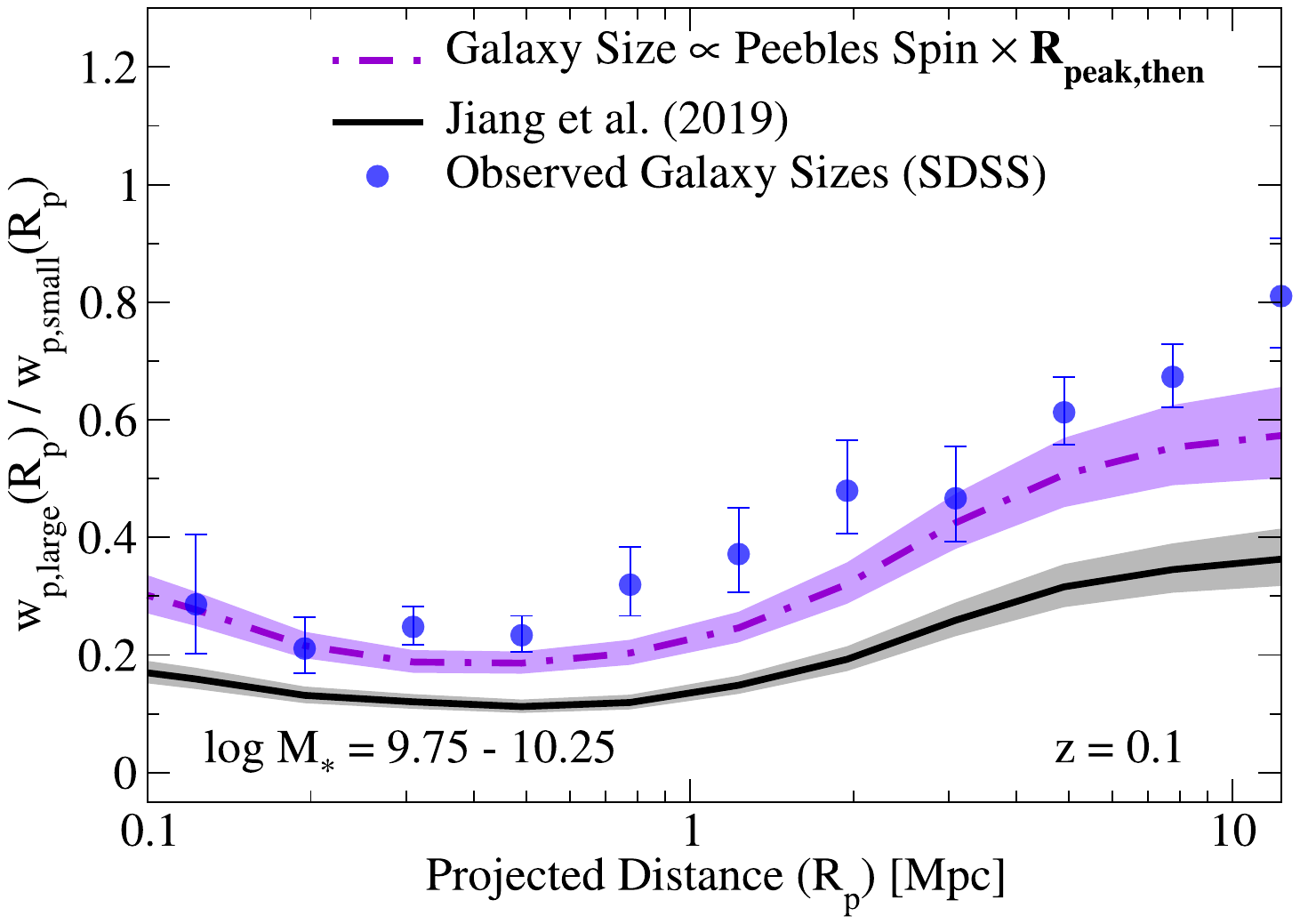}
\vspace{-3ex}
\phantom{\hspace{-5ex}}\includegraphics[width=1.1\columnwidth]{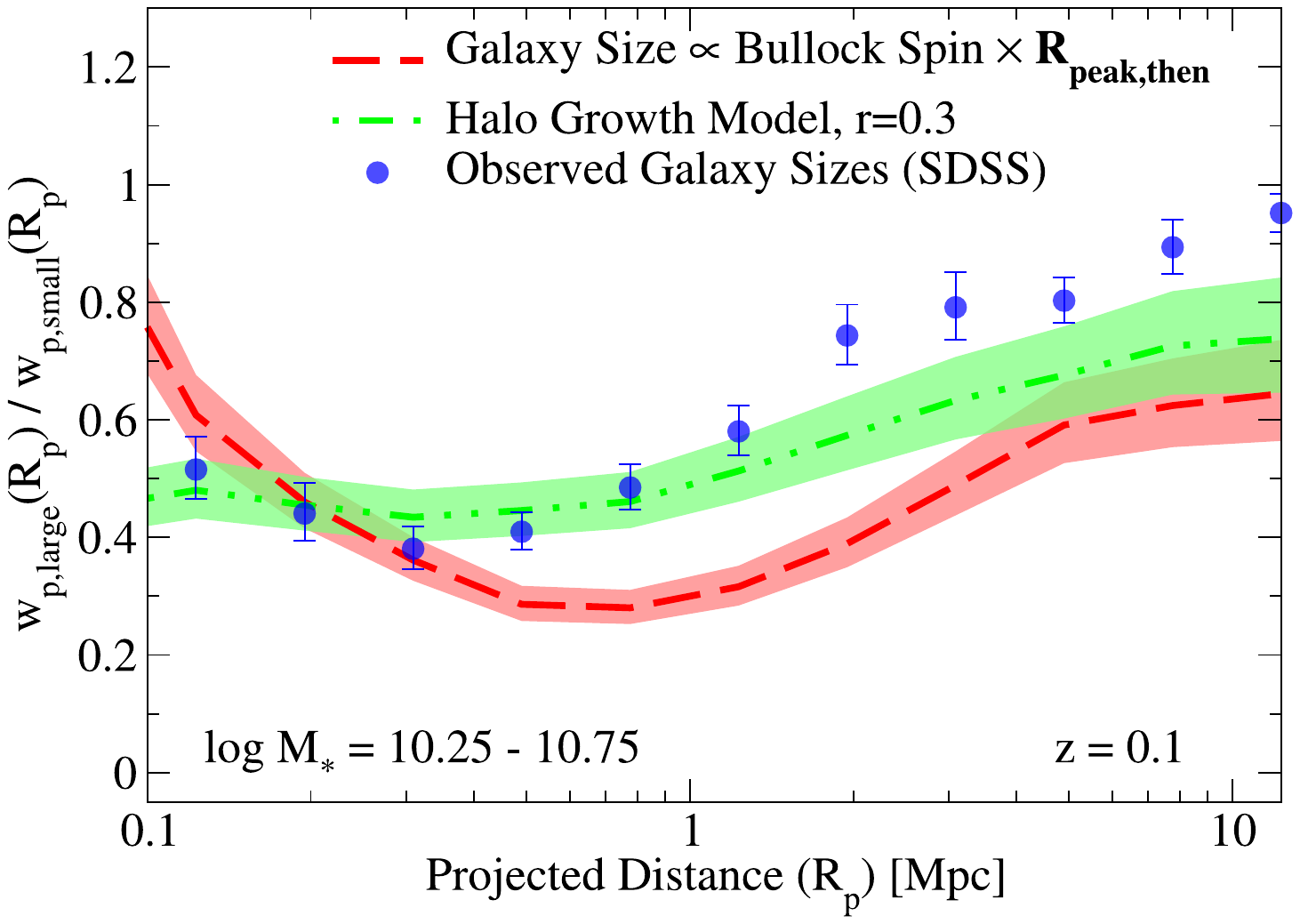}\hspace{-6ex}\includegraphics[width=1.1\columnwidth]{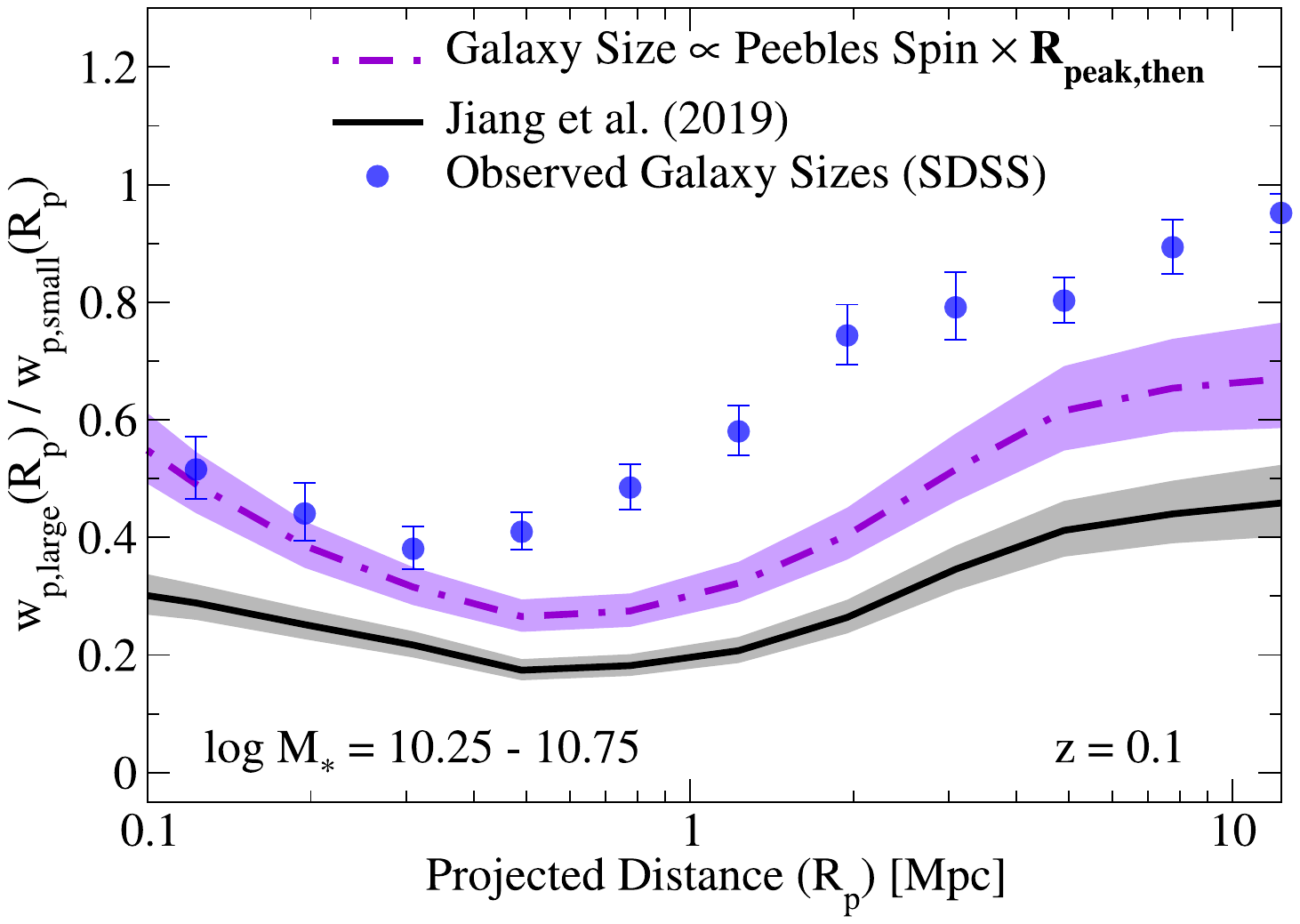}
\vspace{-3ex}
\phantom{\hspace{-5ex}}\includegraphics[width=1.1\columnwidth]{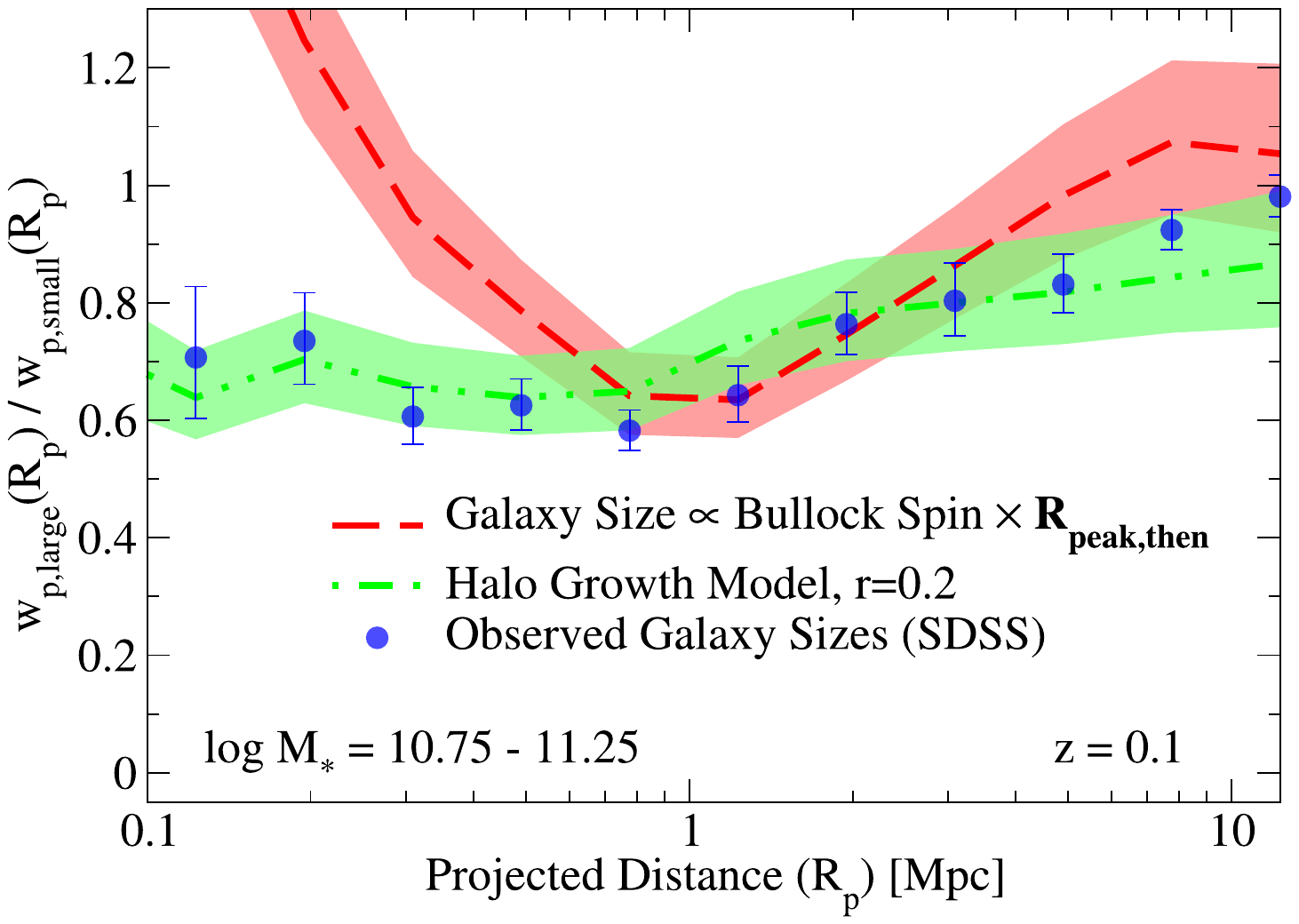}\hspace{-6ex}\includegraphics[width=1.1\columnwidth]{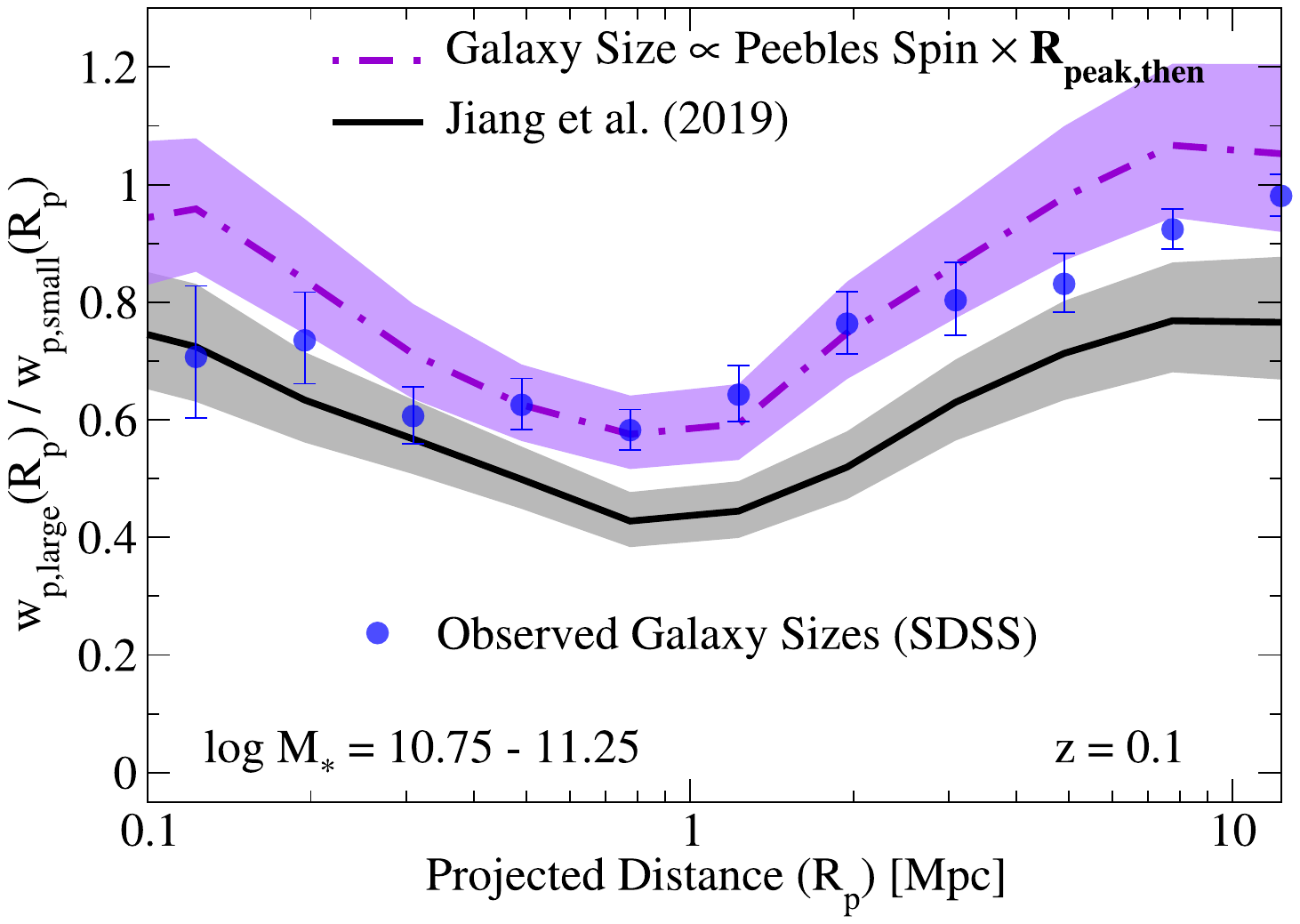}
\vspace{-3ex}
\end{center}
\caption{Ratios of projected two-point correlation functions for large vs.\ small galaxies in both observations and simulations, in several bins of stellar mass.  This figure is equivalent to Fig.\ \ref{f:size_obs} but with the spin-based models using $\mathbf{R_\textbf{peak,then}}$ as the halo radius instead of $R_\mathrm{peak,now}$ (see Appendix \ref{a:rthen}).   Galaxy projected half-light sizes in observations have had corrections applied from Section \ref{s:size_3d} to approximate 3D half-mass sizes.  \textbf{Left} panels show models where galaxy sizes are proportional to the Bullock spin (\textit{red} line) as well as to the halo accretion rate (\textit{green line}); \textbf{Right} panels show a model where galaxy size is proportional to the Peebles spin (\textit{purple line}) as well as the \citet{Jiang19} concentration-based model (\textit{black line}).  The halo growth model gives the closest predictions to the SDSS observations (\textit{blue points}).  For both spin-based models, the halo radius used is the present-day radius of a halo with the same peak mass.  In all cases, error bars and shaded regions correspond to jackknife uncertainties.  As in Fig.\ \ref{f:size_obs}, two-point correlation functions for both observations and simulations are computed with $\pi_\mathrm{max}=13.6$ Mpc $h^{-1}$.}
\label{f:rthen}
\end{figure*}

This effect is shown in Fig.\ \ref{f:rthen}, in which the spin-based size models show increased clustering differences on both one-halo and two-halo scales compared to Fig.\ \ref{f:size_obs}.  Notably, the characteristic shape of the two-point clustering differences predicted by the spin-based size models remains the same: i.e., an upturn in the clustering ratios on small scales that is not present in the observed data for $M_\ast > 10^{10.25}\Msun$.  We find that $R_\mathrm{peak,then}$ generally gives a worse match to the observed data than $R_\mathrm{peak,now}$, with the sole exception of the Peebles spin model for one mass bin.

\section{Halo Growth Model}

\label{a:halo_growth}

We follow the general approach to correlating galaxy and halo properties in \cite{BWHC19}.  As in Section \ref{s:mock}, stellar masses are assigned to haloes via abundance matching.  In narrow bins of assigned stellar mass, we rank-order haloes by their average growth rates over the past dynamical time, $\langle \dot{M}_h \rangle_\mathrm{dyn}$.  We Gaussianize the distribution of halo ranks (i.e., we compute a $Z$-score) using the percentile rank of each halo ($p\in[0,1]$) as follows:
\begin{equation}
    Z_\mathrm{growth} = \sqrt{2}\cdot \mathrm{erf}^{-1}(2p-1),
\end{equation}
where $\mathrm{erf}^{-1}$ is the inverse error function.  The quantity $Z_\mathrm{growth}$ then has a Gaussian distribution with standard deviation $1$, and has perfect rank correlation with halo growth rate.  We then compute the equivalent $Z$-score for galaxy size by:
\begin{equation}
    Z_\mathrm{size} = r Z_\mathrm{growth} + \sqrt{1-r^2} G(1),
\end{equation}
where $G(1)$ is a random number drawn from a Gaussian distribution with standard deviation 1.  By construction, $Z_\mathrm{size}$ is still a Gaussian distribution with standard deviation 1, and has a rank correlation of $r$ with halo growth rate.  One may then assign observed galaxy sizes to haloes in rank order of $Z_\mathrm{size}$ if creating a mock catalogue with stellar masses and sizes.  For the analysis here, we need only split haloes into above-median and below-median galaxy sizes, which is most simply done by splitting haloes on $Z_\mathrm{size}$ being above or below 0, respectively.

\section{Comparison of Overall Galaxy Clustering}

\label{a:clustering}

\begin{figure*}
\begin{center}
\vspace{-8ex}
\phantom{\hspace{-5ex}}\includegraphics[width=1.1\columnwidth]{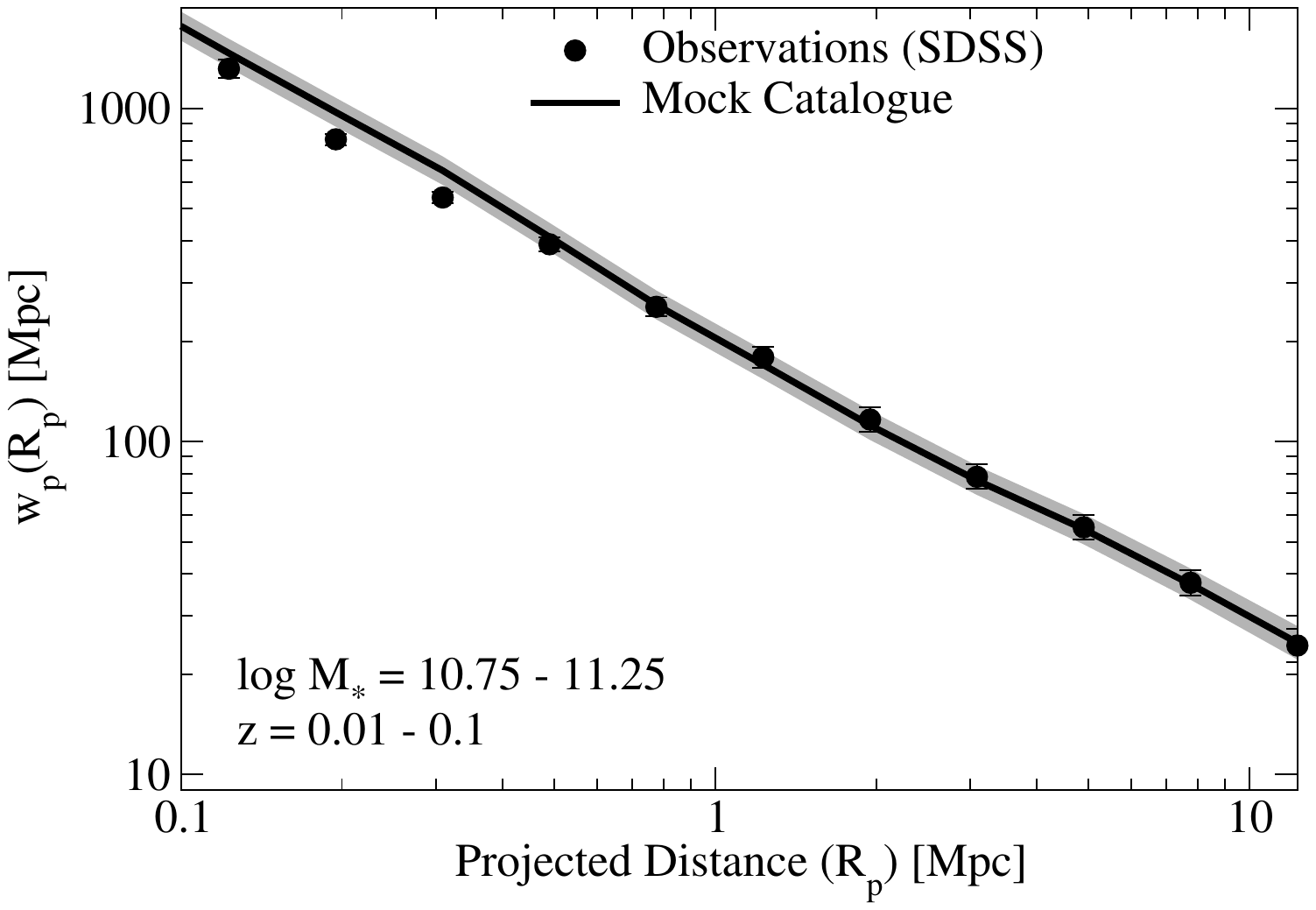}\hspace{-6ex}\includegraphics[width=1.1\columnwidth]{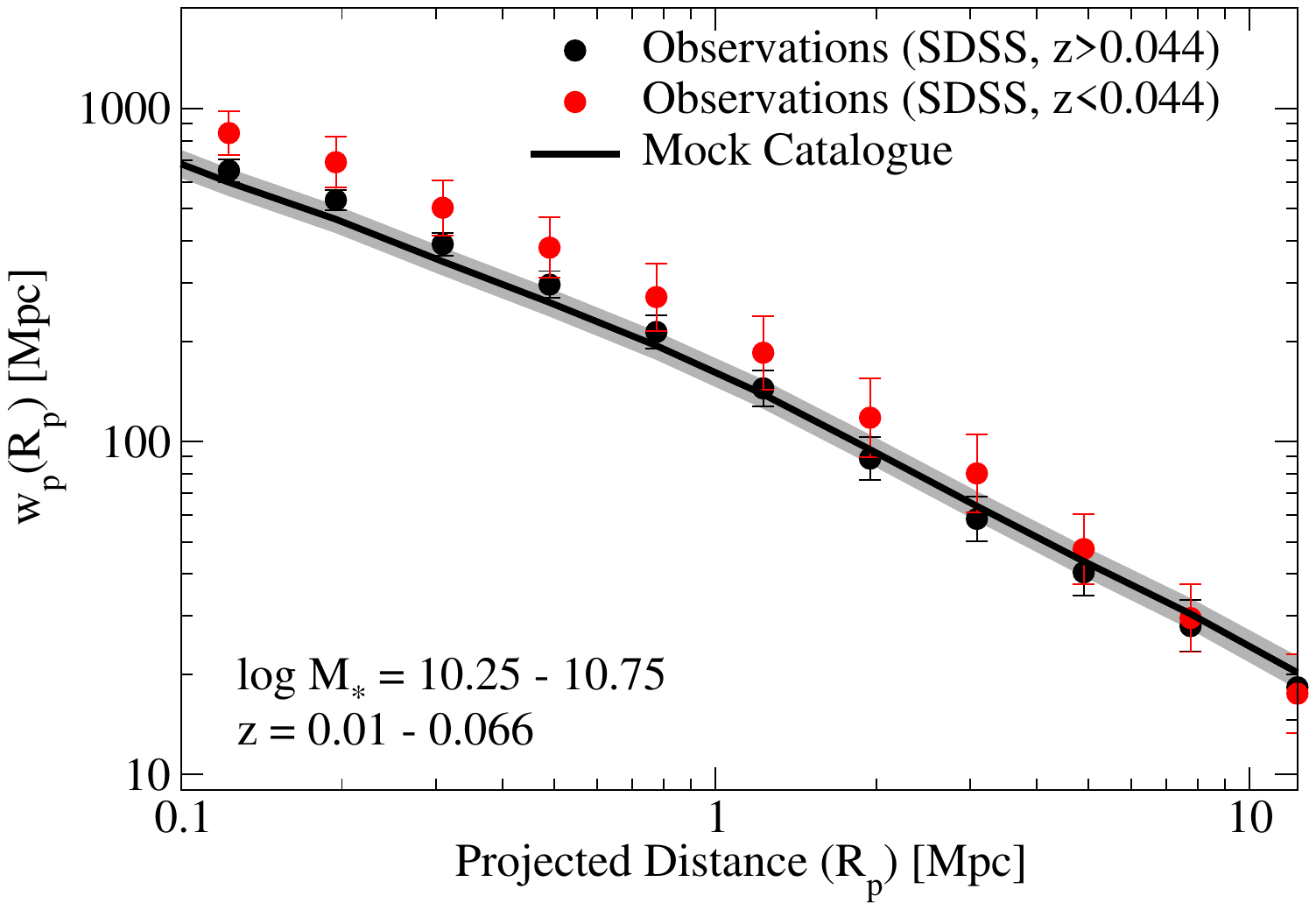}\\[-6ex]
\phantom{\hspace{-5ex}}\includegraphics[width=1.1\columnwidth]{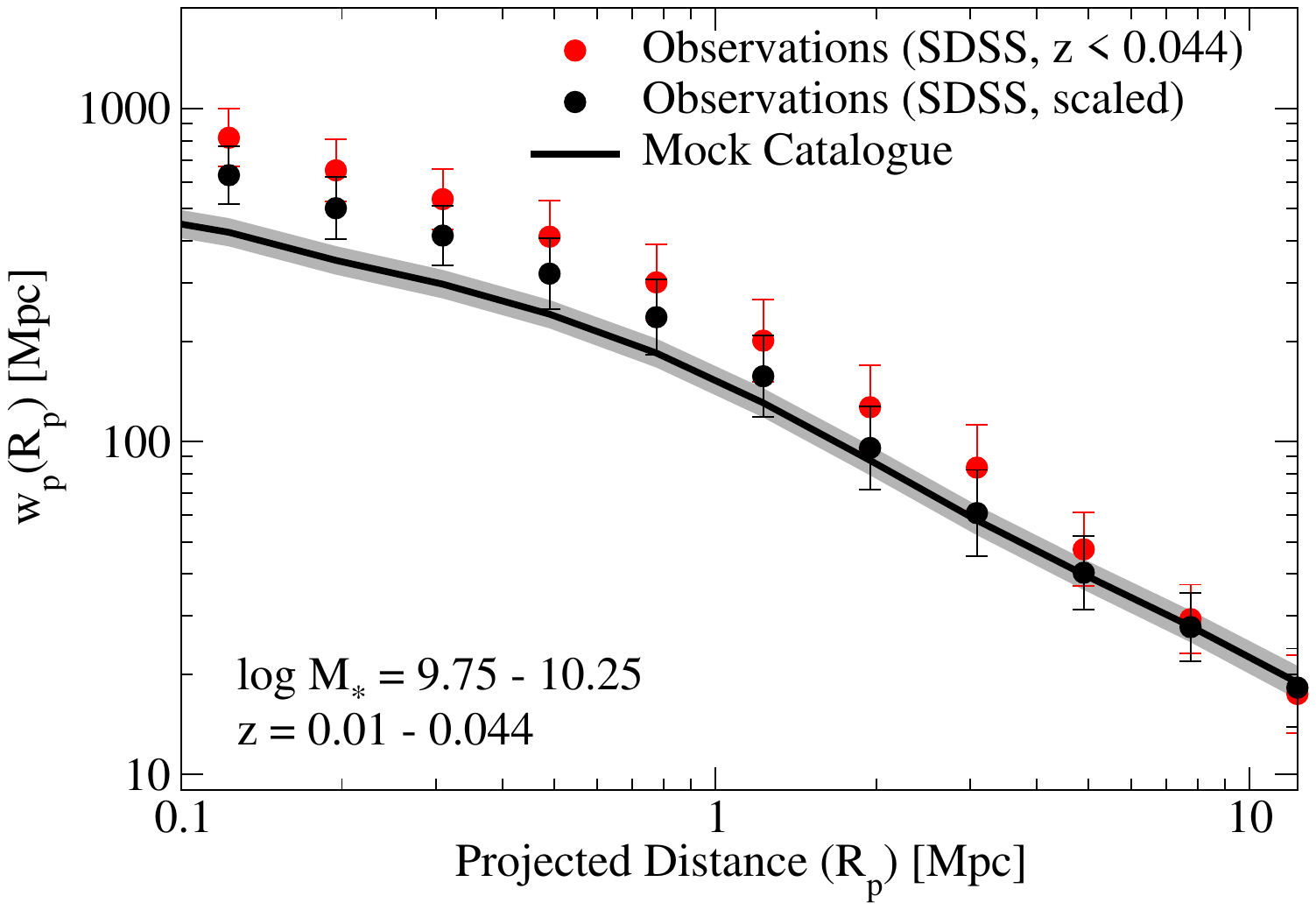}\hspace{-6ex}\includegraphics[width=1.1\columnwidth]{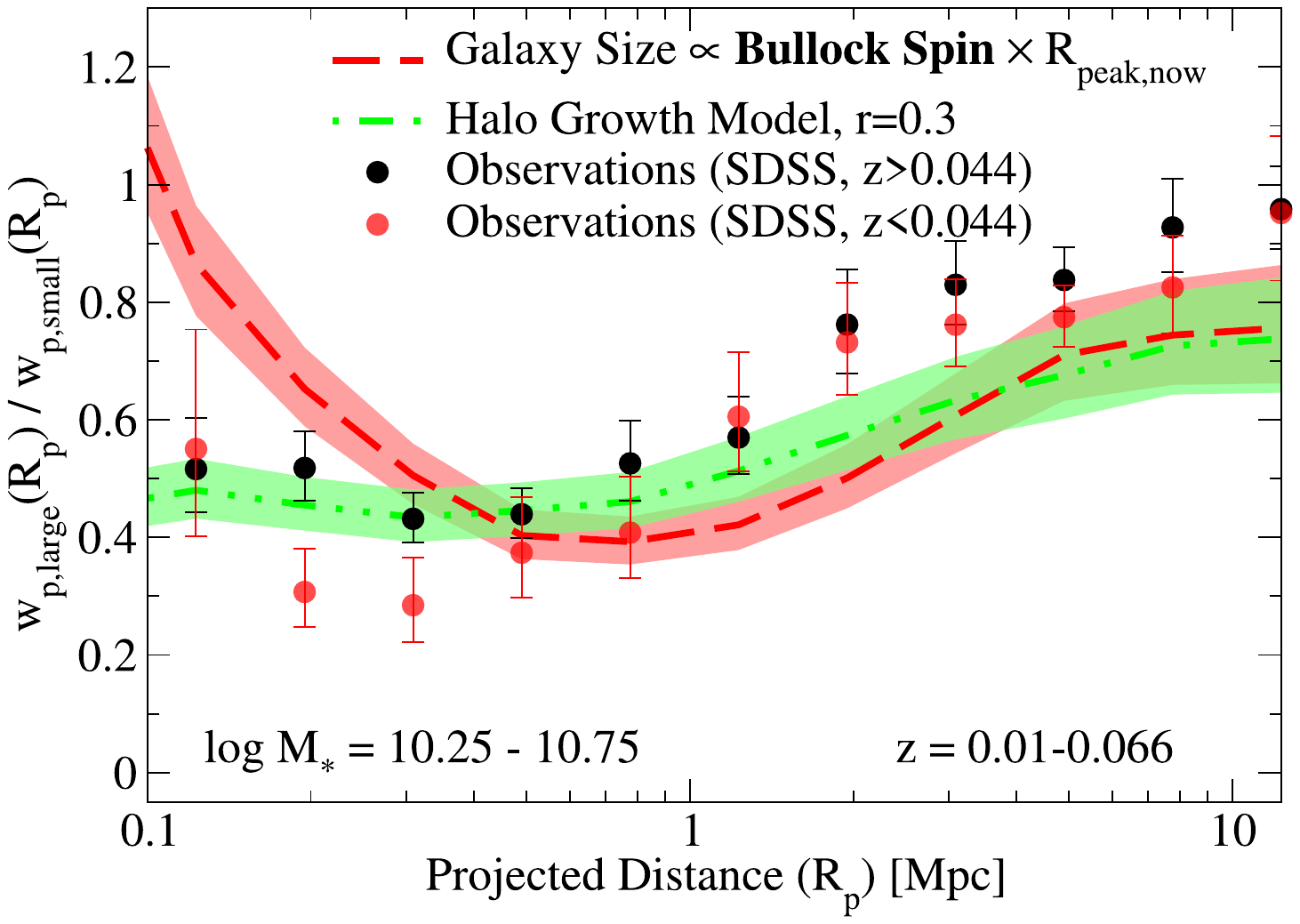}
\vspace{-3ex}
\end{center}
\caption{\textbf{Top and bottom-left} panels: Comparison of overall two-point correlation functions from both observations and simulations, in several bins of stellar mass.  Generally very good agreement is seen, with the exception of the lowest mass bin, due in part to significant sample variance in the SDSS at $z<0.044$ \citep{BWHC19}.  As shown in the \textbf{bottom-right} panel, this sample variance does not affect conclusions from comparing observed clustering ratios for large vs.\ small galaxies.  In all cases, error bars and shaded regions correspond to jackknife uncertainties.  As in Fig.\ \ref{f:size_obs}, two-point correlation functions for both observations and simulations are computed with $\pi_\mathrm{max}=13.6$ Mpc $h^{-1}$.}
\label{f:clustering}
\end{figure*}

Fig.\ \ref{f:clustering} shows the overall clustering normalizations for all galaxies in bins of stellar mass, both in observations and in our mock catalogues.  As discussed in Appendix C of \cite{BWHC19}, there is significant sample variance within the Sloan Digital Sky Survey (SDSS) at $z<0.044$ that results in excess clustering on scales less than 3 Mpc.  We find excellent agreement between our mock catalogues and all clustering measurements at $z>0.044$ (Fig.\ \ref{f:clustering}, top panels).  At $z<0.044$, we reproduce the finding in \cite{BWHC19} that all mass bins at $z<0.044$ have significantly boosted clustering.  For example, at $R_p = 1$ Mpc, the measured clustering of $10^{9.75}-10^{10.25}\Msun$ galaxies is 20\% larger than that of $10^{10.75}-10^{11.25}\Msun$ galaxies (compare Fig.\ \ref{f:clustering} bottom-left panel to top-left panel).  We also show the $z<0.044$ vs.\ $z>0.044$ clustering of $10^{10.25}-10^{10.75}\Msun$ galaxies in the top-right panel of Fig.\ \ref{f:clustering}.  The black points in the bottom-left panel of Fig.\ \ref{f:clustering} show a first-order correction for sample variance in our lowest-mass sample, in which the clustering of $10^{9.75}-10^{10.25}\Msun$ galaxies has been scaled by the ratio between the $z>0.044$ and $z<0.044$ clustering of $10^{10.25}-10^{10.75}\Msun$ galaxies.  This correction leads to significantly better agreement between our mock catalogues and observations.

In the bottom-right panel of Fig.\ \ref{f:clustering}, we show that the clustering ratios of large vs.\ small galaxies are affected very modestly by such sample variance.  We find that the differences between the $z<0.044$ and $z>0.044$ samples are largely confined within the 1-$\sigma$ jackknife errors; in comparison, the departures of, e.g., a Bullock spin-based model tend to be much larger.

\section{Effects of Orphan Galaxies}

\label{a:orphans}

\begin{figure*}
\begin{center}
\vspace{-8ex}
\phantom{\hspace{-5ex}}\includegraphics[width=1.1\columnwidth]{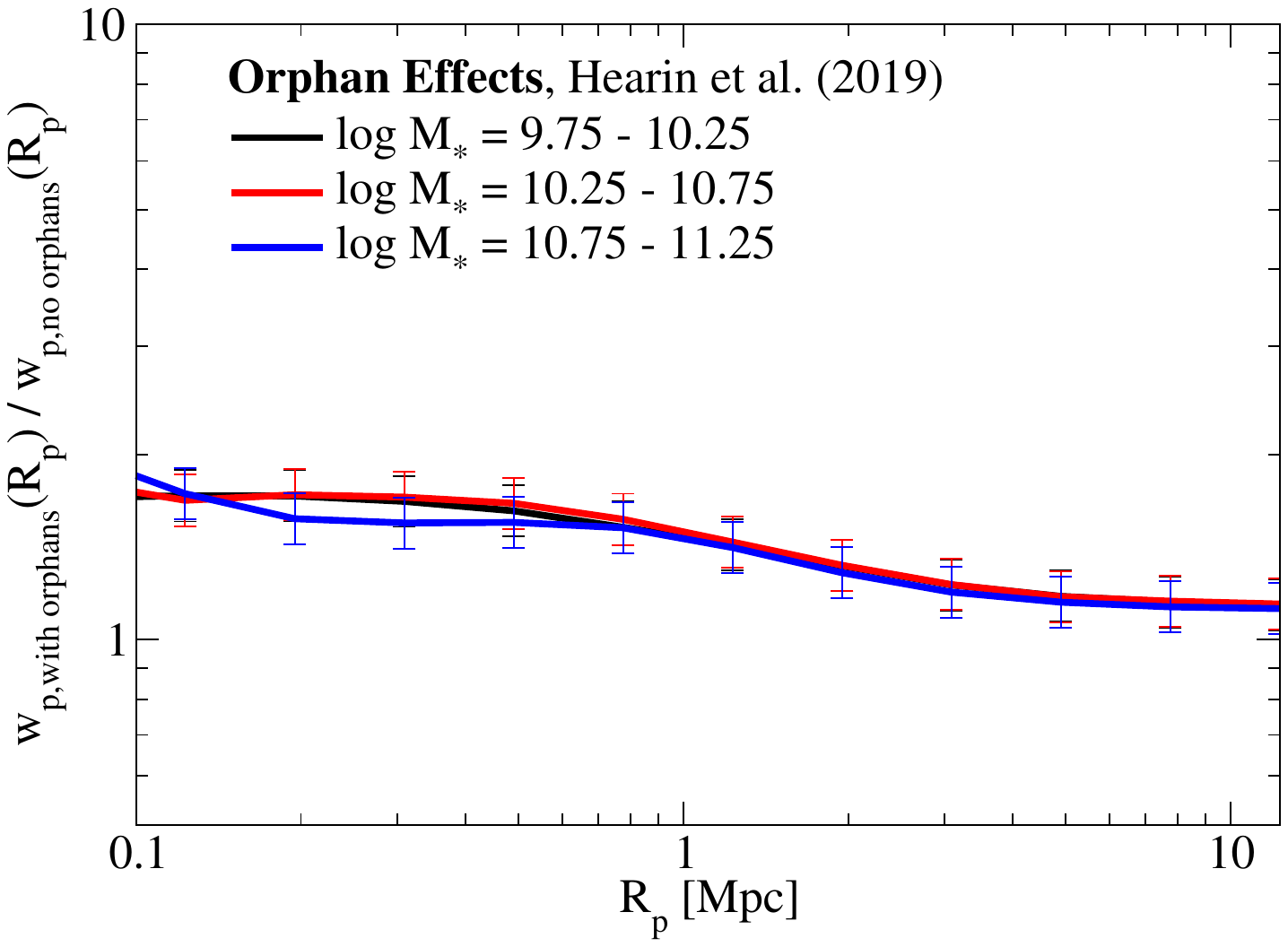}\hspace{-6ex}\includegraphics[width=1.1\columnwidth]{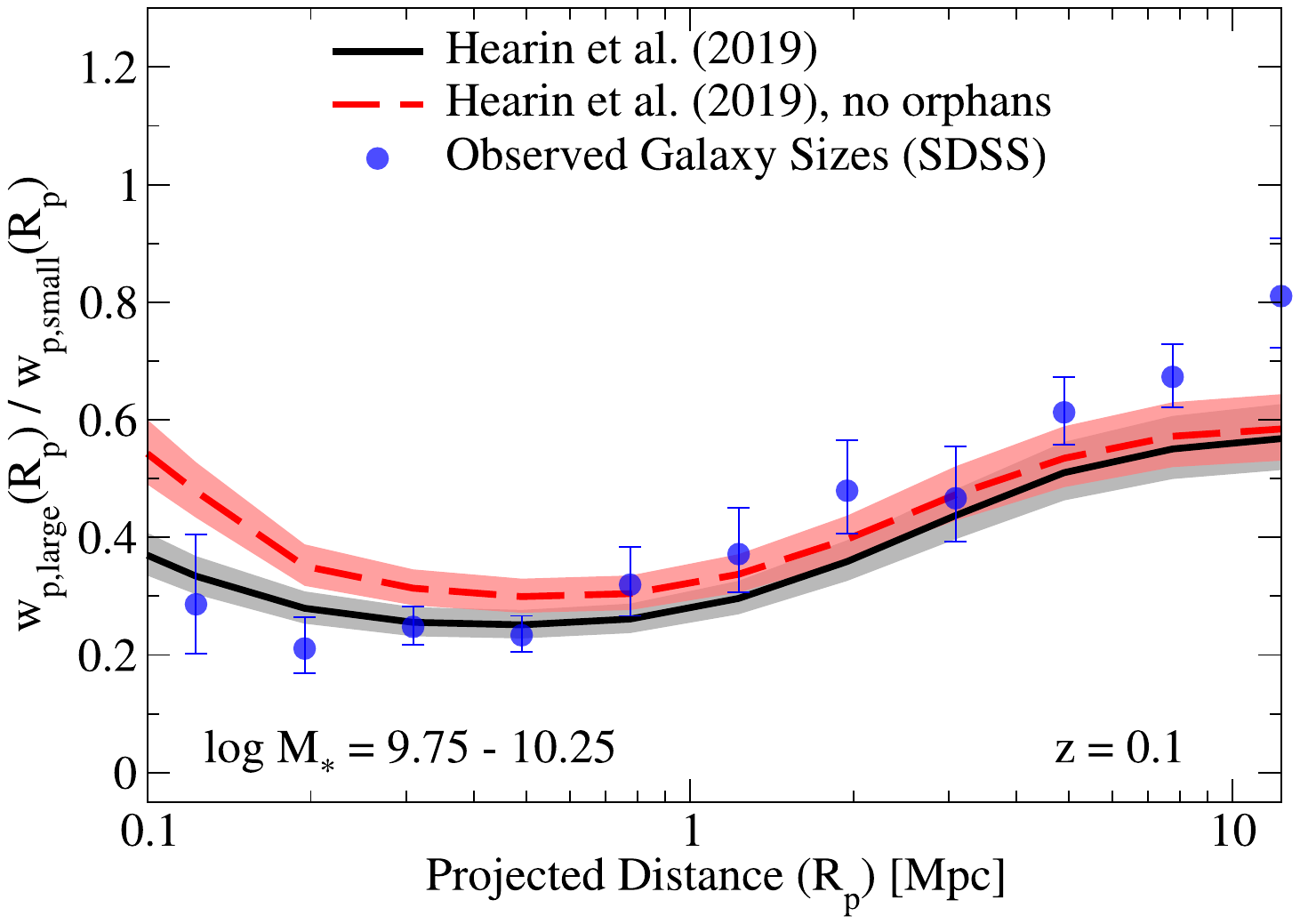}\\[-6ex]
\phantom{\hspace{-5ex}}\includegraphics[width=1.1\columnwidth]{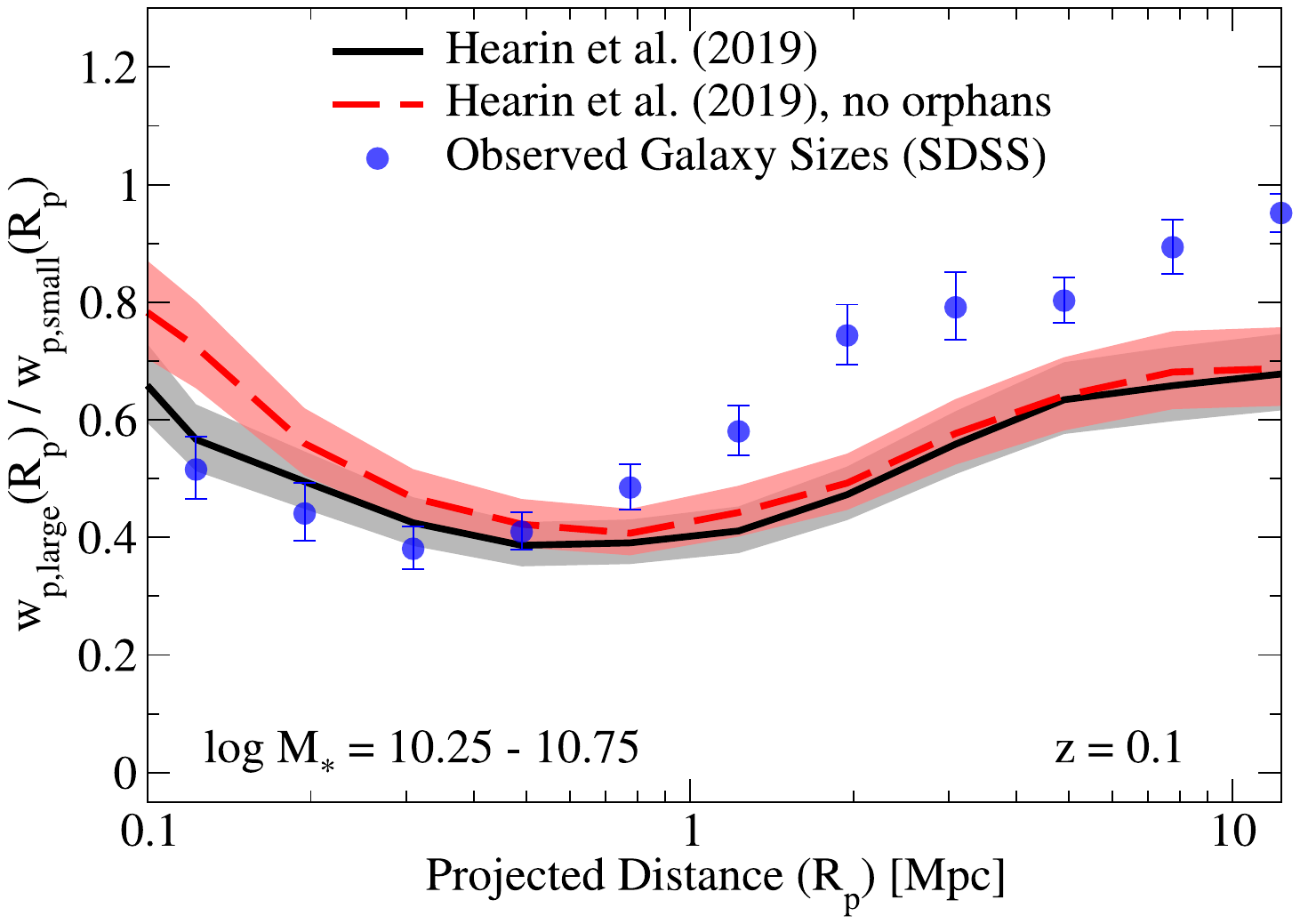}\hspace{-6ex}\includegraphics[width=1.1\columnwidth]{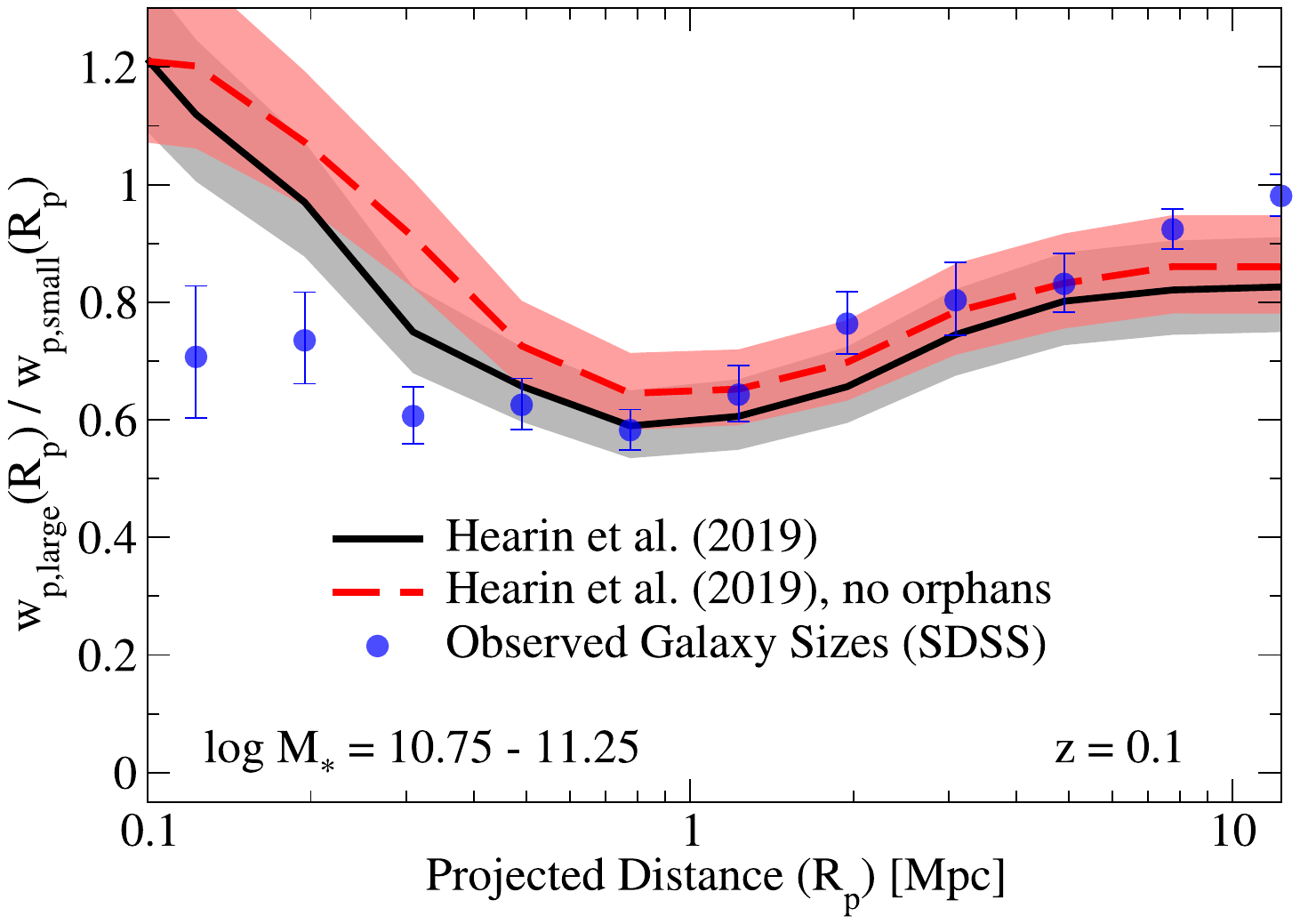}
\vspace{-3ex}
\end{center}
\caption{\textbf{Top-left} panel: ratios of two-point projected correlation functions for all galaxies in the \citet{Hearin19} abundance matching+orphans model, compared to those for non-orphan galaxies from the same model.  \textbf{Top-right} and \textbf{Bottom} panels: Ratios of projected two-point correlation functions for large vs.\ small galaxies in both observations and the \citet{Hearin19} model, in several bins of stellar mass.  Galaxy projected half-light sizes in observations have had corrections applied from Section \ref{s:size_3d} to approximate 3D half-mass sizes.  For the \citet{Hearin19} model, ratios both with and without orphan galaxies are shown.  In all cases, error bars and shaded regions correspond to jackknife uncertainties.  Two-point correlation functions here for both observations and simulations are computed with $\pi_\mathrm{max}=13.6$ Mpc $h^{-1}$.}
\label{f:orphans}
\end{figure*}

Here, we show the effects of orphan galaxies on the clustering ratios considered in this paper.  Although there are a wide variety of orphan models, we limit tests here to the single orphan+galaxy size model developed in \cite{Hearin19}.  Briefly, this model makes a jolly simple assumption that the present-day size of a galaxy is linearly proportional to the size of the virial radius at the time the parent subhalo reached its maximum mass. An additional modeling ingredient is included to incorporate galaxies residing in subhalos that have been disrupted and no longer appear in the standard catalog of subhalos that survive to $z=0.$ All orphan subhalos with $V_{\rm max}/V_{\rm peak}<0.1$ are discarded from consideration; half of the remaining candidates are randomly selected to host satellite galaxies. We refer the reader to the Appendix of \cite{Hearin19} for further details.

The top-left panel in Fig.\ \ref{f:orphans} compares the overall two-point projected correlation functions (2PCFs) in the \citet{Hearin19} model with and without orphan galaxies.  The effects of orphans on the overall clustering mimic the effect of changing the satellite fraction (Fig.\ \ref{f:sat_ratios}), as is naturally expected: an asymptotically flat effect on one-halo scales ($\lesssim$ 0.25 dex for the \citealt{Hearin19} model), which is larger than the asymptotically flat effect on two-halo scales.

Taking the ratio of 2PCFs for large and small galaxies will normalize out any overall boost shared by all galaxies.  As shown in Fig.\ \ref{f:orphans}, the effect of adding orphans to the \citet{Hearin19} size model changes the overall 2PCF but does not have a very large impact on clustering ratios.  The most significant effects are seen on relatively small scales; the error bars no longer overlap only on scales of $300$ kpc and below.  Effect sizes that are this small are not enough to remedy differences between predictions for spin-based models and the SDSS observations, although this simple test does not rule out arbitrary combinations of orphan+spin-based size models.

Overall, the \citet{Hearin19} model provides a better match to the observed data than the spin-based size models (Fig.\ \ref{f:wp_ratios}).  Although the growth-based empirical model in this paper is formally an even closer match to the observed data, mocks using the \citet{Hearin19} model still provide very reasonable results.  We have also found that predictions for $5^\mathrm{th}$ nearest neighbour distances are very similar between the model in this paper and \citet{Hearin19}.  This is reasonable because average halo growth rate is highly correlated with satellite infall time (and hence, the halo radius at peak mass); hence, we would expect a similar satellite population to be classified as ``small'' vs. ``large.''  Differences between the models would be more evident if only central, non-flyby haloes were considered, and only then if a measure of halo accretion rate were available, such as that in \cite{ODonnell20}.

\bsp	
\label{lastpage}
\end{document}